\documentclass[12pt]{article}
\makeatletter
\newcommand{\lambdabar}{{\mathchoice
  {\smash@bar\textfont\displaystyle{0.25}{1.2}\lambda}
  {\smash@bar\textfont\textstyle{0.25}{1.2}\lambda}
  {\smash@bar\scriptfont\scriptstyle{0.25}{1.2}\lambda}
  {\smash@bar\scriptscriptfont\scriptscriptstyle{0.25}{1.2}\lambda}
}}
\newcommand{\smash@bar}[4]{%
  \smash{\rlap{\raisebox{-#3\fontdimen5#10}{$\m@th#2\mkern#4mu\mathchar'26$}}}%
}
\makeatother

\RequirePackage[a4paper,top=30.6mm,bottom=38.6mm,left=26mm,right=26mm,footskip=1.3cm]{geometry}
\usepackage{setspace}
\usepackage{amsmath}
\usepackage{amsfonts} 
\usepackage{amssymb}
\usepackage{graphicx}
\usepackage{hyperref}
\usepackage{tensor}
\usepackage{siunitx}
\usepackage{float}
\usepackage{bm}
\usepackage{slashed}
\usepackage{booktabs}
\usepackage{cancel}
\usepackage{multirow}
\usepackage{comment}
\usepackage{bbold}
\usepackage{caption}
\usepackage{amsmath}
\usepackage{enumerate}
\usepackage{cite}
\usepackage{tensor}
\usepackage{slashed}
\usepackage{inputenc}
\usepackage{rotating}
\usepackage{bigfoot}
\usepackage{cancel}
\usepackage{mathtools}
\usepackage{xcolor}

\DeclarePairedDelimiter{\floor}{\lfloor}{\rfloor}

\usepackage[textsize=footnotesize]{todonotes}

\numberwithin{equation}{section}

\def \< {\left<}
\def \> {\right>}

\newcommand{\be}{\begin{equation}} \newcommand{\ee}{\end{equation}}
\newcommand{\bea}{\begin{eqnarray}}  \newcommand{\eea}{\end{eqnarray}}

\DeclareMathOperator{\Tr}{Tr}

\usepackage{amsfonts, amsthm}
\usepackage[english]{babel}
\usepackage{slashed}
\usepackage{mathrsfs}
\usepackage{amssymb}
\usepackage{color}

\def\Tr{{\mathrm{Tr}}}

\begin{document}
		
	\begin{center}        % Main title
		\LARGE Notes on integrating out M2 branes
	\end{center}
	
	\vspace{0.7cm}
	\begin{center}        % Authors
		{\large Jarod Hattab and Eran Palti }
	\end{center}
	
	\vspace{0.15cm}
	\begin{center}  
		\emph{Department of Physics, Ben-Gurion University of the Negev,}\\
		\emph{Be'er-Sheva 84105, Israel}\\[.3cm]
%		\emph{}\\[.2cm]
%		e-mails:  \tt palti@bgu.ac.il, \;petri@post.bgu.ac.il
	\end{center}
	
	\vspace{1cm}
	
	%%%%%%%%%%%%%%%%%%%%%%%%%%%%%%%%%%%%%%%%%%%%%%%
	%%%%%%%%%%%%%%%%%%%%%%%%%%%%%%%%%%%%%%%%%%%%%%%
	%%%%%%%%%%%%%%%%%%%%%%%%%%%%%%%%%%%%%%%%%%%%%%%
	%%%%%%%%%%%%%%%%%%%%%%%%%%%%%%%%%%%%%%%%%%%%%%%
	%%%%%%%%%%%%%%%%%%%%%%%%%%%%%%%%%%%%%%%%%%%%%%%
	%%%%%%%%%%%%%%%%%%%%%%%%%%%%%%%%%%%%%%%%%%%%%%%
	%%%%%%%%%%%%%%%%%%%%%%%%%%%%%%%%%%%%%%%%%%%%%%%
	%%%%%%%%%%%%%%%%%%%%%%%%%%%%%%%%%%%%%%%%%%%%%%%
	
	\begin{abstract}
	\noindent  
	Integrating out supersymmetric M2 branes wrapped on two-cycles in Calabi-Yau manifolds is an important calculation: it allows the determination of, and in some ways defines, the free energy of topological strings. In these notes, based on a short course aimed at graduate students, we go through various aspects of this calculation in detail. The end result is a recently proposed new formula for the topological string free energy.  
	\end{abstract}
	
	\thispagestyle{empty}
	\clearpage
	
	\tableofcontents
	
%%%%%%%%%%%%%%%%%%%%%%%%%%%%%%%%%%%%%%%%%%%%%%%%%%%%%%%%%%%%%%%%%%%%%%%%%%%%%%%%%%%%
%%%%%%%%%%%%%%%%%%%%%%%%%%%%%%%%%%%%%%%%%%%%%%%%%%%%%%%%%%%%%%%%%%%%%%%%%%%%%%%%%%%%

%%%%%%%%%%%%%%%%%%%%%%%%%%%%%%%%%%%%%%%%%%%%%%%%%%%%%%%%%%%%%%%%%%%%%%%%%%%%%%%%%%%%
\section{Introduction}
\label{sec:int}
%%%%%%%%%%%%%%%%%%%%%%%%%%%%%%%%%%%%%%%%%%%%%%%%%%%%%%%%%%%%%%%%%%%%%%%%%%%%%%%%%%%%

%%%%%%%%%%%%%%%%%%%%%%%%%%%%%%%%%%%%%%%%%%%%%%%%%%%%%%%%%%%%%%%%%%%%%%%%%%%%%%%%%%%%
\subsection{What is String theory?}
\label{sec:whatop}
%%%%%%%%%%%%%%%%%%%%%%%%%%%%%%%%%%%%%%%%%%%%%%%%%%%%%%%%%%%%%%%%%%%%%%%%%%%%%%%%%%%%

In 1994 Polchinski wrote up his Les Houches lecture notes with the title: ``What is String Theory?" \cite{Polchinski:1994mb}. The notes gave two answers to that question: the first half dealt with a technical introduction to the basics of string theory, and the second half attacked the question from a deeper perspective. He struggled with questions that would, less than a year later, lead to his discovery of D-branes \cite{Polchinski:1995mt}. 

The issue that was raised is that in a typical quantum field theory, with coupling constant $\epsilon$, perturbation theory in $\epsilon$, at order $h$, behaves as \cite{LeGuillou:1990nq} 
\be
\sum_{h=0}^{\infty} A_h \;h! \;\epsilon^{2h}  \;,
\label{qftase}
\ee 
with $A_h$ being the contribution from amplitudes at order $h$ but with the $h!$ taken out. In terms of $h$, the $h!$ is the important factor so we can practically take the $A_h$ as constants. We therefore see that the ratio $R_h$ of successive terms in the perturbative expansion behaves as
\be 
R_h \sim h \;\epsilon^2\;.
\label{rhoo}
\ee 
When we reach sufficiently large $h$ that $R_h$ becomes of order one, successive terms begin to grow, and ultimately the series diverges. So perturbation theory does not yield a function in terms of the coupling $\epsilon$, but only an {\it asymptotic series}. That is, if we treat the expansion as over infinite $h$, then the radius of convergence of the series is zero. Of course, an asymptotic expansion is just fine for most practical purposes, and we can calculate in Quantum Electron Dynamics (QED) to great precision at low orders. But, we cannot define the theory through the perturbative expansion.

In string theory, we have a similar problem \cite{PhysRevLett.60.2105}. One way we might try to define string theory is by writing down its partition function ${\cal Z}$, or more conveniently its free energy $F$ defined as
\be
F = \log {\cal Z} \;.
\ee
The free energy $F$ should be a function of the parameters of the theory, and so at least depend on the string coupling $g_s$. So we could try to define string theory through a function $F\left(g_s\right)$. We could consider expanding it order-by-order in the string coupling as a sum of different genus string diagrams
\be 
F\left(g_s\right) = \sum_{g=0}^{\infty} F_g \;g_s^{2g-2} \;,
\label{fgscl}
\ee 
where $g$ is the genus expansion of string theory, and we included an extra overall factor of $g_s^{-2}$ for conventions. But now we face the same problems as in field theory, at each genus we have a large number of diagrams contributing. Indeed, it is expected that the $F_g$ behave as
\be 
F_g \sim \left(2g\right)! \;,
\ee 
so even worse than in quantum field theory. Therefore, the definition (\ref{fgscl}) would only be an asymptotic series (with zero radius of convergence). There are many functions which share the same asymptotic series, which is the correct one? A very explicit and sharp part of the question "What is string theory?" is to ask: "What is the full function $F\left(g_s\right)$"? 

We can gain some clues as to how the full function $F\left(g_s\right)$ should behave by studying more carefully when the perturbative approximation breaks down. Consider the quantum field theory perturbative expansion (\ref{qftase}), which breaks down when $R_h \sim 1$, so at $h_{\mathrm{max}} \sim \epsilon^{-2}$. The smallest term of the series, representing its maximum accuracy, is therefore of the size
\be 
\left(h_{\mathrm{max}}\right)!\; \epsilon^{2h_{\mathrm{max}}} \sim  \left(e^{\epsilon^{-2}\log \epsilon^{-2} - \epsilon^{-2}}\right)\left(e^{2\epsilon^{-2} \log \epsilon}\right) \sim e^{-\frac{1}{\epsilon^2}}\;,
\label{ftl2e}
\ee 
where we used Stirling's approximation $h! \sim e^{h \log h-h}$. 

For string theory, we can consider open string field theory \cite{Witten:1985cc}, which behaves much like field theory in terms of the diverging number of diagrams. The open string coupling $g_o$ is related to the closed-string coupling as $g_o^2 = g_s$. So replacing $\epsilon^2 \rightarrow g_o^2 = g_s$ in (\ref{ftl2e}) we find that string perturbation theory breaks down at terms of order $e^{-\frac{1}{g_s}}$ \cite{alvarez1991random}. We therefore can estimate that completing $F(g_s)$ to a full function would require adding terms of that magnitude, so we could modify (\ref{fgscl}) to a full function which has an expansion of the form
\be 
F\left(g_s\right) = \sum_{g=0}^{g_{\mathrm{max}}} F_g \;g_s^{2g-2} + {\cal O}\left(e^{-\frac{1}{g_s}}\right)\;,
\label{fgsclff}
\ee 
where $g_{\mathrm{max}}$ is the genus order at which the perturbative approximation breaks down.
This full function would then, to some extent, define string theory for us. 

What is the physics which would lead to contributions of order $e^{-\frac{1}{g_s}}$? It would need to be something whose action scaled as $\frac{1}{g_s}$, so an object which is non-perturbative. These are exactly the D-branes of string theory, whose tension behaves as $\frac{1}{g_s}$. It is natural to speculate that Polchinski's discovery of D-branes, one year after his notes were published, was somewhat motivated by looking for these contributions.

%%%%%%%%%%%%%%%%%%%%%%%%%%%%%%%%%%%%%%%%%%%%%%%%%%%%%%%%%%%%%%%%%%%%%%%%%%%%%%%%%%%%
\subsection{What is Topological String theory?}
\label{sec:whatop}
%%%%%%%%%%%%%%%%%%%%%%%%%%%%%%%%%%%%%%%%%%%%%%%%%%%%%%%%%%%%%%%%%%%%%%%%%%%%%%%%%%%%

The discovery of D-branes, associated to the $e^{-\frac{1}{g_s}}$ effects in string theory, was revolutionary for the subject. It played a central role in the so-called second superstring revolution. Nonetheless, to this day we still only have partial information about these effects, and still very far from being able to write down a function $F(g_s)$ for string theory. 

A useful way to proceed is to study simplified versions of string theory. One such theory is topological string theory \cite{Witten:1988xj,Witten:1991zz}. Topological strings are constructed by taking the theory on the string worldsheet to be topological. That is, physical quantities in the theory do not depend on the worldsheet metric. This means that the embedding of the string into the spacetime it is propagating in, the target space, is very much simplified. In particular, we will consider the topological string A-model (there is also a B-model which will not be discussed here). The embedding of the A-model turns out to require holomorphic maps, so the spacetime coordinates $X$ depend on the worldsheet coordinates only holomorphically $X(z)$. So one can define perturbative topological string theory (A-model) as counting holomorphic maps from a genus $g$ worldsheet to the target space. It turns out that, at least for any non-trivial cases, the target space must be a Calabi-Yau (CY) threefold. We will not discuss the worldsheet approach to topological strings in these lectures, and refer to many reviews on the field. In particular, \cite{Marino:2004uf,marino2005chern,Vonk:2005yv,Ooguri:2009qk} are excellent examples. 

We can revisit the question "What is String Theory?" in the context of topological strings and ask "What is Topological String Theory?". Following the discussion above, we can aim to phrase this sharply as the determination of the topological string free energy function. Topological strings are much simpler than full strings, and the only continuous parameters they depend on are the topological string coupling $\lambda$, the Kahler moduli of the Calabi-Yau manifold $v^i$, and the components of the Neveu-Schwarz two-form $B$ along the two-cycles associated to the Kahler moduli, which we label as $b^i$. The index $i$ runs over the range $i=1,...,h^{(1,1)}(\mathrm{CY})$, where $h^{(1,1)}(\mathrm{CY})$ is the $(1,1)$ Hodge number of the Calabi-Yau. We can combine $v^i$ and $b^i$ into complex Kahler moduli $t^i$ as
\be 
t^i = v^i + i b^i \;.
\ee 
Then the topological string free energy would be some function, as in (\ref{fgsclff}), of the form
\be 
F\left(t,\lambda\right) = \sum_{g=0}^{\infty} F_g(t) \;\lambda^{2g-2} + {\cal O}\left(e^{-\frac{t}{\lambda}}\right)\;.
\label{tsfe}
\ee 
Determining the function (\ref{tsfe}) is a simplified version of the problem of determining the function (\ref{fgsclff}). To be clear, we should understand that the function (\ref{tsfe}) is for a given topological background, so for a fixed Calabi-Yau topology.\footnote{As yet, topological strings do not have a background independent formulation, though finding such a formulation was one of the motivations of their introduction.} 

Topological strings pick out a certain sub-sector of physical strings. The topological A-model is related to a certain sub-sector of type IIA strings on Calabi-Yau manifolds. Their relation is most usefully, at least for these lectures, understood from the target space perspective. We consider type IIA string theory on the Euclidean background $\mathbb{R}^4 \times \mathrm{CY}$. As a compactification, this leads to an $N=2$ supergravity theory in four-dimensional flat space. In ten dimensions, type IIA supergravity has one gauge field, and restricting it to the four-dimensional space yields a gauge field $A^0_{\mu}$. Further, we also have $h^{(1,1)}(\mathrm{CY})$ gauge fields in four-dimensions coming from expanding the ten-dimensional three-form $C_3$ in a basis of two-forms $\omega_i$ as $C_3 = \sum_i A^i \wedge \omega_i$. So, overall, we have the four-dimensional gauge fields $A^I_{\mu}$, with $I=\left\{0,i\right\}$. 

The four dimensional supergravity has $h^{(1,1)}$ vector multiplets. A vector multiplet contains a four-dimensional gauge field, and a superpartner complex scalar field. The complex scalars correspond to the $t^i$. The one extra gauge field is part of the gravitational mutltiplet, as a superpartner of the graviton and (two) gravitini. The distribution of the $A^I$ into the vector multiplets and the gravitational multiplet is non-trivial, and background dependent. The particular combination which fits in the gravity multiplet is called the {\it graviphoton}, denoted as $V_{\mu}$. The field-strength of the graviphoton, denoted as $W$, with
\be 
W_{\mu\nu} = \partial_{\mu} V_{\nu} - \partial_{\nu} V_{\mu} \;,
\ee 
 plays a central role for topological strings. The map $\left\{A^I_{\mu}\right\} \rightarrow V_{\mu}$ determines a function called the central charge.\footnote{Note that sometimes, confusingly, $A^0_{\mu}$ is denoted the graviphoton. This is a ten-dimensional name, the four-dimensional graviphoton is in general different $A^0_{\mu}\neq V_{\mu}$.}

The four-dimensional effective action can be understood as being determined by type IIA string scattering amplitudes. Topological string amplitudes are a sub-sector of these amplitudes, and so map to a certain sub-sector of the supergravity action. This map was determined in \cite{Bershadsky:1993cx,Antoniadis:1993ze}. It is explained in more detail in section \ref{sec:map2top}, but schematically the map is such that the topological string coupling is identified with the (Einstein frame) graviphoton field strength as
\be 
\lambda^2 \sim W^2 \;,\;\; \;W^2 \equiv W_{\mu\nu} W^{\mu\nu}\;.
\label{latow}
\ee 
Then the coefficients in the perturbative expansion of the free energy (\ref{tsfe}) are mapped to the terms in the effective Lagrangian of the form
\be 
{\cal L}_{\mathrm{eff}}\left(t,W^2\right) = \sum_{g=1}^{\infty} F_g(t)\; W^{2g-2}\; \left(R^{-}\right)^2 \;,
\label{seffiiawr}
\ee  
where $R^{-}$ is the anti self-dual part of the Riemann tensor.\footnote{Note that in our conventions the factor $\left(R^-\right)^2$ is also associated to a factor of $\lambda^2$. This is the appropriate assignment from a supersymmetric perspective, because $R^-$ and $W$ are in the same (gravitational) multiplet. In these conventions, the topological string free energy (\ref{tsfe}) is defined with an extra factor of $\lambda^{-2}$ relative to the action (\ref{seffiiawr}). This also implies that the $F_0$ part maps to the tree-level kinetic terms in the action, as described in section \ref{sec:4dsugra}.} 

From this perspective, we can think of topological strings as (higher-derivative) terms in the four-dimensional supergravity action, involving higher powers of the graviphoton field-strength. This is a very useful way to think of topological strings. Indeed, we may consider the Lagrangian ${\cal L}_{\mathrm{eff}}\left(t,W^2\right)$ as a definition for the topological string free energy $F\left(t,\lambda^2\right)$. If we do that, then determining the exponential terms in $\lambda$ in the expression (\ref{tsfe}), is mapped to determining such exponential terms in $W^2$ in the ${\cal L}_{\mathrm{eff}}\left(t,W^2\right)$. So to understand topological string theory, we need to understand the four-dimensional effective action.

%%%%%%%%%%%%%%%%%%%%%%%%%%%%%%%%%%%%%%%%%%%%%%%%%%%%%%%%%%%%%%%%%%%%%%%%%%%%%%%%%%%%
\subsection{Integrating out states}
\label{sec:intout}
%%%%%%%%%%%%%%%%%%%%%%%%%%%%%%%%%%%%%%%%%%%%%%%%%%%%%%%%%%%%%%%%%%%%%%%%%%%%%%%%%%%%

In his review \cite{Polchinski:1994mb}, Polchinski refers to Wilson, who was struggling with the question ``What is Quantum Field Theory?" \cite{RevModPhys55583}. Polchinski says Wilson realized that ``{\it the theory is to be organized scale-by-scale rather than graph-by-graph}". It is precisely the number of graphs that led to the problematic factorials in perturbation theory, and so maybe we should think instead about scales. Of course, Wilson was the discoverer of the renormalization group equations framework. He taught us how effective actions should be understood as arising from {\it integrating out} smaller distance scales. For topological strings, we found that we can define the theory through the effective supergravity action. But the integrating out formalism of Wilson applies exactly to the idea of an effective action. So perhaps it is possible to define the theory through an integrating out procedure. One aim of these lecture notes is to argue that, for topological strings, this is indeed possible \cite{Hattab:2024ewk}.    

The idea of integrating out states to generate an effective action splits the degrees of freedom of the theory into ``fast'', or ultraviolet degrees of freedom $\phi_{UV}$, and ``slow'', or infrared degrees of freedom $\phi_{IR}$. One starts from an ultraviolet action $S_{UV}\left[\phi_{UV},\phi_{IR}\right]$ containing both types of fields, and performs a partial path integral over the ultraviolet degrees of freedom. What remains then defines the infrared effective action $S_{IR}\left[\phi_{IR}\right]$, which depends only on the infrared degrees of freedom:
\be 
\int {\cal D}\phi_{IR}{\cal D}\phi_{UV}\; e^{-S_{UV}\left[\phi_{UV},\phi_{IR}\right]} = \int {\cal D}\phi_{IR}\;e^{-S_{IR}\left[\phi_{IR}\right]} \;.
\label{intoutgen}
\ee

A classic, and very useful for us, example of this is the case of QED. In this case we can treat the electron as the ultraviolet degree of freedom $\phi_{UV}=e^-$, and the photon as the infrared degree of freedom $\phi_{IR}=\gamma$. We can then consider integrating out the electron to yield an effective action for the photon. More precisely, we are interested in calculating an effective action for the photon field-strength. This is known as a background field calculation. To do this, we take the photon field, $A_{\mu}$, to have a constant background field-strength $F_{\mu\nu}$, so we take $A_{\nu} = \frac12 x^{\mu}F_{\mu\nu}$. We then perform the partial path integral over the electron field. Importantly, it is possible to do this exactly, because the QED action is quadratic in the electron field and so this is just a Gaussian integral. This calculation was first performed by Euler and Heisenberg in 1936 \cite{Heisenberg:1936nmg}, and is textbook material, see in particular \cite{schwartz2014quantum,Dunne:2004nc} for excellent expositions. The resulting effective Lagrangian, associated to $S_{IR}\left[\phi_{IR}\right]$ above, the Euler-Heisenberg Lagrangian $\mathcal{L}_{EH}$, takes the form (in Minkowski signature) \cite{schwartz2014quantum}
\be 
\mathcal{L}_{EH} = -\frac14 F_{\mu\nu}F^{\mu\nu} - \frac{e^2}{32\pi^2}\int_{0}^{\infty} \frac{ds}{s} e^{-s m_e^2} \; \frac{\mathrm{Re} \cosh \left(e s X \right)}{\mathrm{Im} \cosh \left(e s X \right)} \;F_{\mu\nu} \tilde{F}^{\mu\nu}\;,
\label{ehlag}
\ee 
where $e$ is the electric coupling, $m_e$ is the electron mass, $s$ is the Schwinger proper time parameter, $\tilde{F}_{\mu\nu}$ is the magnetic field-strength, and $X=\sqrt{\frac12 F_{\mu\nu}F^{\mu\nu}+ \frac{i}{2}F_{\mu\nu}\tilde{F}^{\mu\nu}}$. We have dropped here various terms, and cutoffs, which are not important for our purposes but are needed to regulate the result.  

The Euler-Heisenberg Lagrangian manifests some of the key properties of our analysis. We can expand it for small coupling $e \rightarrow 0$, as an asymptotic series, yielding an expansion in powers of $F^2$ which arises from the expansion
\be 
\frac{\mathrm{Re} \cosh \left(e s X \right)}{\mathrm{Im} \cosh \left(e s X \right)} \;F_{\mu\nu} \tilde{F}^{\mu\nu} = \frac{4}{e^2 s^2} + \frac23 F^2 - \frac{e^2s^2}{45} \left(F^4 + \frac74 \left(F\tilde{F}\right)^2 \right)+...\;.
\ee 
This series in powers of the field strength should be compared, for our purposes, to the series expansion of the effective supergravity action in the graviphoton field strength (\ref{seffiiawr}). We learn from this that integrating out a charged particle, at one loop, can generate an infinite series of higher derivative terms in the effective action for the background field strength. So it is possible that we can think of the effective action (\ref{seffiiawr}) as arising from integrating out states charged under the graviphoton. 

This was precisely the idea of Gopakumar and Vafa \cite{Gopakumar:1998ii,Gopakumar:1998jq}. They posited that the effective action (\ref{seffiiawr}) arises from integrating out states charged under the graviphoton. Then by performing such an integrating out calculation, it is possible to calculate the $F_g(t)$. In the type IIA setting, the only states charged under the graviphoton are D-branes, specifically D2 branes wrapping the two-cycles inside the Calabi-Yau that are associated to the moduli $t^i$. One can also consider bound states of D0 branes with the D2 branes. Since the branes are wrapping cycles in the Calabi-Yau, they appear as particles in the four-dimensional action, whose mass (and charge) depends on the $t^i$. So the action (\ref{seffiiawr}) is claimed to arise from integrating out D2-D0 bound states.

But string theory is not as simple as QED, and the D-branes are non-perturbative states in the theory. This means that at weak string coupling, they are heavier than the fundamental strings. So it is not clear how to integrate them out, physics above the string scale is not like a quantum field theory. However, we could consider going to strong type IIA string coupling $g_s \rightarrow \infty$. In this limit, the D-branes become the lightest states, lighter than the fundamental string. We could then consider integrating them out. Strongly-coupled type IIA string theory is M-theory, and lives in eleven, rather than ten, dimensions. The extra dimension is a circle, whose size is set by the string coupling, and which becomes large at strong coupling. The D2 states are uplifted to M2 states, and the D0 charge becomes the Kaluza-Klein momentum along the M-theory circle. Therefore, the calculation we really need to perform is an integrating out calculation of M2 states with Kaluza-Klein momentum. 

There is also a very useful dual perspective on this. We can exchange the momentum modes along the circle for winding modes. This is then the particle picture utilized in \cite{Dedushenko:2014nya}, which we will also use often. Now the M2 branes are wrapping two-cycles in the Calabi-Yau but also the M-theory circle. The strong coupling limit of IIA means a very large M-theory circle. This means that we can treat the M2 brane effectively as a particle with a very long worldline, wrapping the M-theory circle. This is the dual way to see why the states can be integrated out: in the Kaluza-Klein picture they are light, and in the winding mode picture they have long worldlines, so are infrared physics. 

A problem with taking $g_s \rightarrow \infty$ is that the perturbative expansion (\ref{tsfe}) in the topological string coupling $\lambda$  may break down. However, recall that the coupling is not controlled only by the IIA string coupling, but also by the graviphoton field strength $W$. In terms of the string frame graviphoton $W^{S}$, one has a relation of type $\lambda^2 \sim g^2_s \left(W^{S}\right)^2$. So we can consider the double scaling limit $g_s^2 \rightarrow \infty$ but $\left(W^{S}\right)^2 \rightarrow 0$, such that $\lambda \rightarrow 0$. This way we can calculate in strongly-coupled IIA string theory, but in the weakly-coupled topological string. The perturbative topological string only has contributions at fixed genus to each term in $\lambda$, which ensures that the relation with $g_sW^{S}$ is exact and remains the same also at strong type IIA coupling.

The result is therefore a derivation of the perturbative asymptotic series for the topological string free energy, so (\ref{tsfe}) without the exponential terms, from integrating out M2 states. This is the beautiful result of Gopakumar and Vafa, which we will reproduce here. However, it does not solve our problem of finding the full function for the free energy, so calculating the ${\cal O}\left(e^{-\frac{t}{\lambda}}\right)$ terms in (\ref{tsfe}). 

There is significant literature trying to calculate the ${\cal O}\left(e^{-\frac{t}{\lambda}}\right)$ terms in (\ref{tsfe}). We cannot do justice to all this work in these notes, and will not attempt to. Let us mention two particularly important approaches: geometric transitions and resurgence analysis. The former allows to calculate the terms through open-closed string dualities which work in some very special cases of non-compact Calabi-Yau manifolds. A good account of these can be found in \cite{Marino:2004uf}, and they build on the seminal work in \cite{Gopakumar:1998ki}. Resurgence analysis is a general methodology of trying to extract non-perturbative terms from perturbative series. It is a very involved field, and has produced significant success in this context. We refer to \cite{Marino:2012zq,Aniceto:2018bis} for reviews, and to \cite{Pasquetti:2010bps,Hatsuda:2015owa,Hatsuda:2015oaa,Alim:2021mhp,Gu:2023mgf} for a small selection of some of the central relevant and recent results. We return to a discussion and comparison with the resurgence analysis in section \ref{sec:sum}. 

In \cite{Hattab:2024ewk}, it was proposed that the ${\cal O}\left(e^{-\frac{t}{\lambda}}\right)$ terms in  (\ref{tsfe}) are in fact already included in the integrating out calculation of M2 branes! In the context of our discussion so far, the Wilsonian integrating out approach to the effective action thereby defines topological strings non-perturbatively. In terms of our postulated question: what is topological string theory? We could then answer: it is the effective action after integrating out M2 branes. The raison d'etre of these lecture notes is to show this result in detail.  

Actually, we can motivate already at this stage why it is expected that the integrating out calculation should indeed yield the ${\cal O}\left(e^{-\frac{t}{\lambda}}\right)$ terms, rather than just an asymptotic series. We can see this already in QED. Let us return to the Euler-Heisenberg Lagrangian (\ref{ehlag}). Consider writing the background field strength in terms of (parallel) electric $E$ and magnetic $B$ fields through 
\be 
F_{\mu\nu} F^{\mu\nu} = 2 \left(B^2 - E^2\right) \;,\;\; F_{\mu\nu} \tilde{F}^{\mu\nu} = 4 E B\;.
\ee  
We consider the purely electric field limit $B \rightarrow 0$, and note that the Euler-Heisenberg Lagrangian then takes the form \cite{schwartz2014quantum}
\be 
\mathcal{L}_{EH} = \frac12 E^2 - \frac{1}{8\pi^2} \int^{\infty}_0 \frac{ds}{s^3} e^{-sm_e^2} \left( e E s \frac{\cos \left(e E s \right)}{\sin \left(e E s \right)}\right) + ... \;,
\ee 
where we dropped unimportant terms. This has pole divergences at 
\be 
s = \left(\frac{n \pi}{e E}\right) \;, \;\; n \in \mathbb{N}^*\;. 
\ee 
Dealing with the poles can be done through contour integration, which localises onto them and yields an imaginary component to the Lagrangian of
\be 
2 \mathrm{Im}(\mathcal{L}_{EH}) = \frac{\left(e E\right)^2}{8\pi^4} \sum_{n=1}^{\infty} \frac{1}{n^2} e^{-\left(\frac{n \pi}{e E}\right)m_e^2} \;.
\label{schweff}
\ee 
An imaginary component of the action describes a decay in time of the electric field. This is the famous Schwinger effect, where the field is converted into electron-positron pairs. Since the electron has been integrated out, this conversion manifests as a decay of the electric field in time. 

Let us map the parameters to the topological string setting. The electric field-strength $E$ is naturally mapped to the topological string coupling $\lambda$, as in (\ref{latow}). The mass of the electron is mapped to the mass of the wrapped M2 branes, which behaves as $t$ (representing a generic $t^i$). The electron charge is mapped to the M2 charge, but this is equal to its mass because the state is a so-called BPS state, and so we should equate $e$ with $t$ also. Then applying these maps we have that the Schwinger effect contributions (\ref{schweff}) map to terms of the form
\be 
t^2 \lambda^2 \sum_{n=1}^{\infty} \frac{1}{n^2} e^{-n\frac{t}{\lambda}} \;.
\ee 
These are precisely of the form ${\cal O}\left(e^{-\frac{t}{\lambda}}\right)$ as in (\ref{tsfe}). So the proposal in \cite{Hattab:2024ewk}, and of these lecture notes, that the integrating out calculation yields the non-perturbative contributions to the topological string free energy are just the analogues of the Schwinger effect. Note that the relation of the non-perturbative effects to the Schwinger effect was noted already in \cite{Gopakumar:1998vy,Pasquetti:2010bps}, although the actual realization described in these notes is very different.

This is a long introduction, but hopefully it has set the stage, and motivation, for the calculation that these notes describe. We need to integrate out M2 states in an M-theory background of the form $S^1 \times \mathbb{R}^4 \times \mathrm{CY}$, which wrap two-cycles in the Calabi-Yau and have Kaluza-Klein momentum along the $S^1$. The claim is that performing this yields the full, non-perturbative, topological string free energy. So as a function, rather than an asymptotic series. 

The calculation is presented as follows. The main calculation is performed in a four-dimensional supergravity setting, and so this is the first thing we discuss, in section \ref{sec:4dsugra}. However, already before doing the integrating out calculation, it is important to embed this in the five-dimensional background $S^1 \times \mathbb{R}^4$. We do this in section \ref{sec:5dsugra}. We then perform the integrating out calculation in section \ref{sec:intm2}. The result is then embedded into M-theory, and into type IIA string theory, in section \ref{sec:embiia}. Through the type IIA string theory context, we describe the map to the topological string in section \ref{sec:map2top}. The result is an expression for the topological string free energy. However, this expression still requires some evaluation to bring it to a useful form. We do this calculation in section \ref{sec:evalfor}. Finally, we summarise these notes, and discuss some interesting future directions, in section \ref{sec:sum}.   

%%%%%%%%%%%%%%%%%%%%%%%%%%%%%%%%%%%%%%%%%%%%%%%%%%%%%%%%%%%%%%%%%%%%%%%%%%%%%%%%%%%%
\section{The four-dimensional supergravity}
\label{sec:4dsugra}
%%%%%%%%%%%%%%%%%%%%%%%%%%%%%%%%%%%%%%%%%%%%%%%%%%%%%%%%%%%%%%%%%%%%%%%%%%%%%%%%%%%%

Recall that, from the type IIA perspective, we are interested in evaluating the action in the background $\mathbb{R}^4 \times \mathrm{CY}$. However, in this section we focus on the supergravity formalism, which is usually described in Minkowski signature. We therefore take a metric signature $\left(-,+,+,+\right)$, and will switch to Euclidean signature in section \ref{sec:5dsugra}. We work with a flat background metric $g^E_{\mu\nu}$ , where the superscript $E$ labels that the metric is in the Einstein frame (see sections \ref{sec:5dsugra} and \ref{sec:embiia} for discussions). However, we keep track of factors of it due to an overall constant factor, which we denote as $e^{\sigma}$, that will be important when matching to the higher dimensional settings.\footnote{Note that $e^\sigma$ will depend on the supergravity fields, but we are only interested in evaluating the action on a specific constant background, so that $e^\sigma$ can be taken spatially constant $\partial_{\mu} \sigma=0$.} So we write
\be 
g^E_{\mu\nu} = e^{\sigma} \eta_{\mu\nu} \;\;,\;\; \eta_{\mu\nu} = \mathrm{diag\;}\left(-1,+1,+1,+1\right) \;.
\label{mingesig}
\ee 

The formulation of four-dimensional ${\cal N}=2$ supergravity is very well-known. See, for example, \cite{Andrianopoli:1996cm,sugra2012} for excellent overviews. Throughout these notes we follow closely the fantastic paper \cite{Dedushenko:2014nya}. We therefore use its supergravity conventions, which are also those of \cite{sugra2012,DEROO1980175}. 
%We also eventually would like to match the supergravity description for compactifications of type IIA string theory, which are described in detail for example in \cite{Gurrieri:2003st,Grimm:2004ua,Palti:2008mg}. 

\subsubsection*{The supergravity field content}

The supergravity has three types of multiplets, whose component fields transform into each other under supersymmetry. There is the gravitational multiplet, whose bosonic components include the graviton $h_{\mu\nu}$ and the graviphoton gauge field $V_{\mu}$ :
\be 
\mathrm{Gravitational\;multiplet\;}:\; \left\{h_{\mu\nu},V_{\mu}\right\} \;.
\ee 
We label the field strength of the graviphoton as $W_{\mu\nu}$ with
\be 
W_{\mu\nu} = \partial_{\mu} V_{\nu} - \partial_{\nu} V_{\mu} \;.
\ee 

The main multiplets of interest for us are the vector multiplets. Their bosonic component fields are a gauge field and a complex scalar fields. The vector multiplets are usually described in a form that exhibits a certain redundancy. We use one more vector multiplet than there actually are in the theory. The gauge field of this additional fictitious multiplet is actually identified with the graviphoton $V_{\mu}$, and the scalar component can be gauged away by a symmetry of the action. So if there are $h^V$ vector multiplets, labelled by $i=1,2,...,h^V$, we write them as multiplets with index $I=\{0,i\}$. We label them as
\be 
\mathrm{Vector\;multiplets\;}:\; \left\{A^I_{\mu},X^I\right\} \;.
\ee 
We label the field strengths of the gauge fields as
\be 
F^I_{\mu\nu} = \partial_{\mu} A^I_{\nu} - \partial_{\nu} A^I_{\mu} \;.
\ee 

As stated, the scalar fields $X^I$ have one redundant component. This is fixed by imposing the (conformal gauge) constraint
\begin{equation}
\label{constraint}    
i(\overline{X}^I f_I - \overline{f}_IX^I) = 1 \;.
\end{equation}
This constrain ensures that the supergravity is in the Einstein frame. 
The $f_I$ are the supergravity periods, and take the form
\be 
f_{I} =\frac{\partial F^{(S)}_0}{\partial X^I} \;.
\ee 
Here $F^{(S)}_0$ is the supergravity prepotential, which is a holomorphic function of the $X^I$. It is a function of homogeneous degree two, which means that if we rescale 
\be 
X^I \rightarrow \lambda\; X^I \;,
\label{resX}
\ee 
then the prepotential scales as $F^{(S)}_0 \rightarrow \lambda^2 F^{(S)}_0$.  

We can therefore think of the vector multiplet scalar fields $X^I$ as respecting a rescaling symmetry (\ref{resX}), which is then fixed by the constraint (\ref{constraint}). It is useful to therefore work with scalar fields that are manifestly invariant under the rescaling. These are the $Z^I$, which are defined as
\be 
Z^I = \frac{X^I}{X^0} \;.
\label{ZtoXX0}
\ee  
Physically, the $Z^I$ are the relevant fields, but the action is best formulated in terms of the $X^I$.

It is useful to split the field strengths $F^I_{\mu\nu}$ into self-dual and anti self-dual parts
\be 
F^I_{\mu\nu} = F^{I,-}_{\mu\nu} + F^{I,+}_{\mu\nu} \;,
\ee 
where we defined
\be 
F^{I,\pm}_{\mu\nu} = \frac12 \left(F^I_{\mu\nu} \mp \frac{i}{2}\epsilon_{\mu\nu\rho\lambda}F^{I,\rho\lambda} \right) \;.
\ee
The epsilon tensor $\epsilon_{\mu\nu\rho\lambda}$ is defined by $\epsilon_{1234} = \sqrt{-g^E}$, and the fact that it is anti-symmetric in all its indices. In a similar way, we can define the anti-self dual field strength of the graviphoton $W^-_{\mu\nu}$. 

We can also define the anti self-dual parts of the magnetic fields $G_{I,\mu\nu}^-$ as
\be 
G_{I}^{-,\mu\nu} = 2 i \frac{\partial \mathcal{L}}{\partial F^{I,-}_{\mu\nu}}\;,
\ee 
with $\mathcal{L}$ the Lagrangian associated to the action (\ref{IIA}). With these quantities, we can single out the graviphoton $V_{\mu}$ as the particular combination of the $A^I_{\mu}$ whose anti self-dual part of its field strength $W^{-}_{\mu\nu}$ is given by the combination 
\be 
W^{-}_{\mu\nu} = 2 \left(X^I G_{I,\mu\nu}^- - \bar{f}_I  F^{I,-}_{\mu\nu}\right)\;.
\ee 

The final type of multiplets are hypermultiplets, whose bosonic content is two complex scalar fields. They decouple from the other multiplets, and will not play a role in our analysis. We therefore do not discuss them henceforth. 

\subsubsection*{The supergravity (bosonic) action}

The two-derivative bosonic supergravity action is
\begin{equation}
\label{IIA}
    M_p^2 \int  d^4x \sqrt{-g^E} \left[ \frac12 R -g_{ij} \partial_{\mu}Z^i\partial^{\mu}\overline{Z}^j+\frac12\text{Im}\,\mathcal{N}_{IJ} F^{I}_{\mu\nu} F^{J,\mu\nu}+\frac14\text{Re}\,\mathcal{N}_{IJ} \epsilon_{\mu\nu\rho\lambda} F^{I,\mu\nu}F^{J,\rho\lambda}\right] \;.
\end{equation}
Here $\mathcal{N}_{IJ}$ is called the gauge kinetic matrix and $g_{ij}$ the metric on the moduli space. We have allowed an arbitrary prefactor for the action $M_p^2$, which we will fix in section \ref{sec:embiia}, depending on its embedding in string theory. Note also that we are working in specific units, which are explained in section \ref{sec:map2top}. This unit choice is not important at this point, but we state it here so as not to lead to confusion if certain quantities appear to have the wrong mass dimensions. 

The forms of $\mathcal{N}_{IJ}$ and $g_{ij}$ are determined by the supergravity prepotential  $F^{(S)}_0$ as
\bea 
\mathcal{N}_{IJ}&=& \overline{F}_{IJ}+i\frac{N_{IL}X^{L} N_{JK}X^{K}}{N_{MN}X^{M}X^{N}} \;,
\eea 
with $F_{IJ}=\partial_{X^I}\partial_{X^J}F^{(S)}_0$, and
\be 
N_{IJ} = 2\;\text{Im}\,F_{IJ} \;.
\ee 
The moduli space metric is given in terms of the non-homogenous coordinates as
\be 
g_{ij} = -\frac{\partial^2}{\partial Z^{i}\partial \overline{Z}^j}\log\left(-N_{KL}Z^{K}\overline{Z}^{L}\right) \;.
\label{Kahlmet}
\ee 
The metric on the moduli space $g_{ij}$ has an associated Kahler potential $K$, so $g_{ij} = \partial_i \bar{\partial}_j K$, given by 
\be
4e^{-K} = -\frac{2}{\left|X^0\right|^2} \mathrm{Im} \left(\partial_{X^I} \partial_{X^J} F_0^{(S)} \right) X^I \bar{X}^J = -\frac{1}{i\left|X^0\right|^2} \left(f_I \bar{X}^I - \bar{f}_I X^I \right) = \frac{1}{\left|X^0\right|^2} \;,
\label{kahlpot}
\ee
where in the last equality we used the constraint (\ref{constraint}). This fixes the magnitude of $X^0$ in terms of $K$, and the phase is chosen so that we have
\be 
X^0 = -\frac{i}{2} e^{\frac{K}{2}} \;.
\label{x0tok}
\ee 
We have normalized the Kahler potential in (\ref{kahlpot}) so that (\ref{x0tok}) matches the definition in \cite{Dedushenko:2014nya}.

\subsubsection*{The superfields}

So far we described the supergravity using component fields in the superfields. It is useful to describe how this is formulated in terms of the superfields and superspace. Our conventions are those of \cite{Freedman_VanProeyen_2012,Dedushenko:2014nya}, and of the early works \cite{deRoo:1980mm,deWit:1980lyi,Bergshoeff:1980is,deWit:1984rvr}.
The scalar fields $X^I$ are part of superfields $\mathcal{X}^{I}$ which have expansions
\begin{equation}
    \mathcal{X}^{I} = X^{I}+\cdots+\frac{1}{2}\epsilon_{ij}\overline{\theta}^i\sigma^{\mu\nu}\theta^j\mathcal{F}^{I,-}_{\mu\nu}+\dots-\frac{1}{6}(\epsilon_{ij}\overline{\theta}^i\sigma^{\mu\nu}\theta^j)^2D_{\mu}D^{\mu}\overline{X}^{I} \;,
    \label{n2boschmul}
\end{equation} 
where we explicitly write only the bosonic components. We can also define superfields associated to the $Z^I$, denoted as $\mathcal{Z}^I$.
Here $\theta^i$ are the supersymmetry parameters and 
\be 
\mathcal{F}^{I,-}_{\mu\nu} = F^{I,-}_{\mu\nu}-\frac{1}{2}\overline{X}^{I}W^-_{\mu\nu} \;.
\label{deffmcu}
\ee 
The covariant derivative in (\ref{n2boschmul}) is defined by $D_{\mu}X^{I} = (\partial_{\mu}-i\mathcal{A}_{\mu})X^{I}$ where $\mathcal{A}_{\mu}$ is the auxiliary non-dynamical gauge field associated to the common $U(1)$ of the complex scalar fields rescaling invariance $X^{I} \rightarrow \lambda X^{I}$ leftover after fixing the constraint (\ref{constraint}).  

The anti-self dual graviphoton field strength $W_{\mu\nu}^{-}$ is the bottom component of the chiral superfield $\mathcal{W}_{\mu\nu}$ associated to the Weyl multiplet
\begin{eqnarray}
    \mathcal{W}_{\mu\nu} = W_{\mu\nu}^{-}-R^{-}_{\mu\nu\lambda\rho}\epsilon_{ij}\overline{\theta}^i\sigma^{\lambda\rho}\theta^j+\cdots \;.
    \label{gravsupf}
\end{eqnarray} 
Here $R^{-}_{\mu\nu\lambda\rho}$ is the anti self-dual part of the Riemann tensor.

\subsubsection*{Supersymmetric backgrounds}

A supersymmetric background requires that the supersymmetry variations vanish. The graviphoton variation in (\ref{gravsupf}) vanishes because we are in a flat background $R^{-}_{\mu\nu\rho\lambda}=0$. The vector multiplet variations, given by (\ref{n2boschmul}) and (\ref{deffmcu}), require $\mathcal{F}^{I,-}_{\mu\nu} = 0$, and therefore we find the important relation 
\be 
 F^{I,-}_{\mu\nu}=\frac{1}{2}\overline{X}^{I}W^-_{\mu\nu} \;.
\label{deffmcuvan}
\ee 
We will utilise this relation, and therefore it is worth noting that it remains unchanged also in Euclidean signature.

\subsubsection*{The superspace action}

Using the superfields, we can write the two-derivative vector multiplet part of the action (\ref{IIA}) simply as
\begin{eqnarray}
  -i M_p^2 \int\text{d}^4x\;\text{d}^4\theta\;\sqrt{-g^E} \, F^{(S)}_0\left(\mathcal{X}\right) +\mathrm{c.c} \;.
  \label{f0ftermac}
\end{eqnarray}
Note that this involves only a partial integration of superspace, $d^4\theta$ rather than $d^8 \theta$. It is therefore an F-term. This is rather crucial for the integrating out analysis, because F-terms only receive contributions from integrating out certain supersymmetric states called BPS states. It is a special property of extended ${\cal N}=2$ supersymmetry that we can write the action as an F-term, with ${\cal N}=1$ supersymmetry the kinetic terms are given by D-terms. 

To see that this matches the action (\ref{IIA}), we can integrate over the superspace coordinates to obtain
\be
   M_p^2 \int\text{d}^4x \sqrt{-g^E}\left[-iF_{I}D_{\mu}D^{\mu}\overline{X}^I+\frac{i}{4}F_{IJ}\mathcal{F}^{I,-}_{\mu\nu}\mathcal{F}^{I,-,\mu\nu}+\mathrm{c.c}\right] \;.
   \label{act4supre}
\ee
The scalar field kinetic terms can be re-expressed as:
\begin{eqnarray}
\label{scalar}    
- N_{IJ}D^{\mu}X^ID_{\mu}\overline{X}^J \;.
\label{scalarkinpreA}
\end{eqnarray}
We can eliminate the auxiliary field $\mathcal{A}_{\mu}$ in (\ref{scalar}) which gives, using the constraint (\ref{constraint}),
\begin{eqnarray}
    \mathcal{A}_{\mu} = \frac{i}{2}N_{IJ}\left(\partial_{\mu}X^I\overline{X}^{J}-X^I\partial_{\mu}\overline{X}^J\right)\;.
\end{eqnarray}
Substituting the solution for $\mathcal{A}_{\mu}$ into (\ref{scalarkinpreA}) to yield the kinetic terms for the $Z^i$ in (\ref{IIA}). Similarly, for the gauge fields, we can eliminate the graviphoton field strength $W_{\mu\nu}$ in (\ref{act4supre}), which appears through (\ref{deffmcu}), for the $F^{I}_{\mu\nu}$ to obtain the gauge kinetic terms in (\ref{IIA}).

The effective four-dimensional action also admits higher order corrections in the graviphoton field strength. Similarly to (\ref{f0ftermac}), these can also be written as F-terms, so as chiral superspace integrals of holomorphic functions $F^{(S)}_g(\mathcal{X}^{I})$. We can write the total (relevant) F-terms $I$ as an expansion in the graviphoton superfield
\bea
\label{clstFgv}
    I &=& -i M_p^2\int \text{d}^4x\;\text{d}^4\theta \;\sqrt{-g^E} \;F^{(S)}\left(\mathcal{X}^{I},\mathcal{W} \right)  \nonumber \\
    &=& -i M_p^2 \sum_g\int \text{d}^4x\;\text{d}^4\theta\;\sqrt{-g^E} \;F^{(S)}_g(\mathcal{X}^{I})\;\left(\frac{\pi^2}{16}\mathcal{W}_{\mu\nu}\mathcal{W}^{\mu\nu}\right)^g\;.
\eea
Unlike in (\ref{f0ftermac}), for these terms we are interested in saturating the superspace integration with the  $\mathcal{W}_{\mu\nu}$ superfield using the second terms in (\ref{gravsupf}).\footnote{We can also saturate the integral with the vector superfields, which will generate terms similar to the ones from $F^{(S)}_0$ multiplied by $\left(W_{\mu\nu}^-W^{-,\mu\nu}\right)^g$.} 
We see that these will generate terms, involving the Riemann tensor squared $\left(R^-\right)^2$, of the form 
\be 
M_p^2\int d^{4}x\; \sqrt{-g^E}\;F^{(S)}_g(X^I)\;\left(R^-\right)^2\;\left(W_{\mu\nu}^-W^{-,\mu\nu}\right)^{g-1} \;,
\ee 
for $g\geq 1$. In type IIA string theory, these terms only receive contributions from genus $g$. We will see this in section \ref{sec:embiia} by counting powers of the string coupling. This counting is exact because of the dilaton field being part of the hypermultiplet sector. So all the powers of the string coupling come from switching between the Einstein and string frames in $W^-_{\mu\nu}W^{-,\mu\nu}$.

%%%%%%%%%%%%%%%%%%%%%%%%%%%%%%%%%%%%%%%%%%%%%%%%%%%%%%%%%%%%%%%%%%%%%%%%%%%%%%%%%%%%
\section{The five-dimensional background}
\label{sec:5dsugra}
%%%%%%%%%%%%%%%%%%%%%%%%%%%%%%%%%%%%%%%%%%%%%%%%%%%%%%%%%%%%%%%%%%%%%%%%%%%%%%%%%%%%

As described in the introduction, we are interested in a strongly-coupled limit of type IIA string theory, so that the D-branes being integrated out become lighter than the fundamental string. Such a limit is described by M-theory. We are therefore interested in an M-theory background of the form $\mathbb{R}^4 \times S^1 \times \mathrm{CY}$. The Calabi-Yau factor is the topic of section \ref{sec:embiia}, while here we focus on the $S^1$ part of this background. Note that, as described in section \ref{sec:embiia}, the limit is such that the $S^1$ is much longer than the typical Calabi-Yau radius, so that a five-dimensional approach is appropriate. 

\subsubsection*{Switching to Euclidean signature}

The four-dimensional part of the background is $\mathbb{R}^4$.  We can denote its Euclidean coordinates as $x^{\mu}$ with $\mu \in \left\{1,2,3,4\right\}$. The Euclidean action is determined by the Minkowski one with the transformation to Euclidean time $\tau$ through the identification $t=-i\tau$. In the path integral we then have
\be 
e^{iS} = e^{i \int dt d^3x {\cal L}(t,x)}= e^{\int d\tau d^3x {\cal L}(\tau,x)}=e^{-\int d\tau d^3x {\cal L}_E(\tau,x)}=e^{-S_E} \;,
\ee  
where the last two equalities define the Euclidean Lagrangian ${\cal L}_E=-{\cal L}$ , and the Euclidean action $S_E$. The transformation to Euclidean time does not change the parts of the formalism in section \ref{sec:4dsugra} that we will utilize in this section.\footnote{It should be possible to go through all the steps of section \ref{sec:4dsugra} using the full supergravity formalism in Euclidean signature developed in \cite{deWit:2017cle}.} 

The only relevant change to take note of is that in Euclidean signature the anti self-dual component is defined as\footnote{Note that for the metric (\ref{efmet5d}), the combination $\epsilon_{\mu\nu\rho\lambda}W^{\rho\lambda}$ is independent of $\sigma$ because $\epsilon_{1234}=\sqrt{g^E}$.}
\be 
W^{\pm}_{\mu\nu} = \frac12 \left(W_{\mu\nu} \pm \frac12\epsilon_{\mu\nu\rho\lambda}W^{\rho\lambda} \right) \;.
\label{wmtow12} 
\ee
The central demands of the solution we are interested in is that it is supersymmetric and that there is a constant anti self-dual background field strength for the graviphoton $W^{-}_{\mu\nu}$. We can choose this background to have non-vanishing legs only along the first two, and last two, directions. So we have
\be 
W^-_{12} = -W^-_{34} \;,
\label{gravback}
\ee 
with all other components vanishing. 

%%%%%%%%%%%%%%%%%%%%%%%%%%%%%%%%%%%%%%%%%%%%%%%%%%%%%%%%%%%%%%%%%%%%%%%%%%%%%%%%%%%%
\subsubsection*{The five-dimensional solution}
\label{sec:sweuc}
%%%%%%%%%%%%%%%%%%%%%%%%%%%%%%%%%%%%%%%%%%%%%%%%%%%%%%%%%%%%%%%%%%%%%%%%%%%%%%%%%%%%

Reducing M-theory on a Calabi-Yau will preserve 8 supercharges in five dimensions. But we also need to turn on an anti self-dual graviphoton background. Remarkably, there exists a solution where this background does not break any of the supersymmetries and preserves the full 8 supercharges. This background was found in \cite{Gauntlett:2002nw} and is referred to in \cite{Dedushenko:2014nya} as a type of Godel solution. Actually, \cite{Dedushenko:2014nya} generalized the solution in order to include vector multiplets in the theory, which is necessary for its application in our setting.

It is worth noting why an anti self-dual graviphoton background is special. A self-dual or anti self-dual background for a gauge field ensures that there is no gravitational backreaction induced on the geometry. This helps to keep the spacetime flat. However, it does, in general, induce a scalar backreaction. The graviphoton is special in this respect, because it is part of the gravitational supermultiplet which has no scalar components. So it does not induce a scalar backreaction either, keeping a full solution with flat spacetime.

The five-dimensional background metric takes the form \cite{Gauntlett:2002nw,Dedushenko:2014nya}
\begin{eqnarray}
    \text{d}s^2 = -(\text{d}t -U_{\mu}\text{d}x^{\mu})^2+\sum_{\mu = 1}^{4}(\text{d}x^{\mu})^2 \;,
    \label{5dbackpure}
\end{eqnarray}
where $U_{\mu} = \frac{1}{2}T^{-}_{\nu\mu}x^{\nu}$, with $T^{-}_{\mu\nu}$ the constant anti-self-dual five-dimensional graviphoton background. Crucially, this background is in Minkowski signature $(1,4)$, while we need a Euclidean solution of the form $S^1 \times \mathbb{R}^4$. To reach this we rotate to Euclidean time, and also compactify the time direction to a circle. We denote the circle radius as $R_5$. We can introduce a dimensionless parameter  $e^{\sigma}$ which gives the radius in some appropriate energy units. It is useful to keep these general, so we denote the mass scale as $M_5$, and then write
 \be   
 e^{\sigma} = R_5 M_5 \;.
 \label{esigr5m5de}
 \ee  
 We will relate $M_5$ to higher dimensional energy scales in section \ref{sec:embiia}. The compactifying change of variable is then
 \be 
 t = -ie^{\sigma}y \;,
 \label{5deucro}
 \ee 
 where $y$ is periodic $y \in \left[0,\frac{2\pi}{M_5}\right)$ (so identified under $y \sim y + \frac{2\pi}{M_5}$).

 The metric is then written as
\begin{eqnarray}
\label{5dmetim}
    \text{d}s^2 = e^{2\sigma}\left(\text{d}y -\frac{i}{4}e^{-3\sigma/2}V_{\mu}\text{d}x^{\mu}\right)^2+e^{-\sigma}g^{E}_{\mu\nu}\text{d}x^{\mu}\text{d}x^{\nu} \;,
\end{eqnarray}
where
\be 
V_{\mu} = 4 e^{\frac{\sigma}{2}} U_{\mu} \;,
\label{4dto5dgrrel}
\ee 
is the four-dimensional graviphoton (with field strength $W^-_{\mu\nu}$). 
In (\ref{5dmetim}), $g^{E}_{\mu\nu}$ is the four-dimensional metric in the Einstein frame, and it is important to note that it takes the form
\be 
g^{E}_{\mu\nu} = e^{\sigma}\delta_{\mu\nu} \;.
\label{efmet5d}
\ee 
This is the Euclidean version of (\ref{mingesig}). 

The most important part of the five-dimensional solution for us is that after the rotation to Euclidean signature (\ref{5deucro}), the metric in (\ref{5dmetim}) is complex if the graviphoton background is real. To maintain a real metric, we must take a purely imaginary graviphoton background, so imaginary $V_{\mu}$ and $W^{-}_{\mu\nu}$. We write this as
\be 
W^-_{\mu\nu} \in i \;\mathbb{R} \;.
\label{wimag}
\ee 
We will see that it is possible to make sense of an imaginary graviphoton background, but it will be extremely important to keep track of this.

%%%%%%%%%%%%%%%%%%%%%%%%%%%%%%%%%%%%%%%%%%%%%%%%%%%%%%%%%%%%%%%%%%%%%%%%%%%%%%%%%%%%
\section{Integrating out M2 branes}
\label{sec:intm2}
%%%%%%%%%%%%%%%%%%%%%%%%%%%%%%%%%%%%%%%%%%%%%%%%%%%%%%%%%%%%%%%%%%%%%%%%%%%%%%%%%%%%

We now come to the main part of the integrating out calculation. We are interested in integrating out M2 branes wrapped on two-cycles of the Calabi-Yau in the background $\mathbb{R}^4 \times S^1 \times \mathrm{CY}$. We can perform the integrating out calculation from a four-dimensional perspective. We treat each Kaluza-Klein mode, so at fixed momentum along the $S^1$, of each wrapped M2 brane as a four-dimensional massive charged particle. In terms of the four-dimensional supergravity, the states are specified by their charges $q_I$ under the $A^I_{\mu}$. The relation of these charges to the higher-dimensional physical quantities are given in section \ref{sec:embiia}. 
 Equation (\ref{deffmcuvan}) then determines the charge of the states under the graviphoton $V_{\mu}$ to be 
\be 
q_I A^I_{\mu} = \frac12 q_I \bar{X}^I V_{\mu} = \frac14 \bar{Z}\left(q\right) V_{\mu} \;,
\ee 
where we defined the central charge as
\be 
Z\left(q\right) = 2\; q_I X^I \;.
\label{ccX}
\ee
In general, the central charge is a map from electric $q_I$ and magnetic $p^I$ charges to a complex number. In this case all the charged states are purely electric, so $p^I=0$. 

As discussed in section \ref{sec:4dsugra}, in order to contribute to the effective action F-terms, the states being integrated out must be BPS states. In terms of the higher dimensional data, that means that the wrapped brane preserves half the supersymmetries of the background. From the four-dimensional perspective, it means that the mass $m$ of the states is given by the central charge. We can therefore write for the charge $e$ under the graviphoton, and for the mass $m$, the expressions
\be
e = \frac14 \bar{Z}(q) \;, \;\;\; m = \left|\bar{Z}(q)\right| M_{Y} \;.
\label{emmapse}
\ee 
Here $M_{Y}$ is a (constant) mass scale, which we will relate to higher dimensional mass scales in section \ref{sec:embiia}. Essentially, in the background $\mathbb{R}^4 \times S^1 \times \mathrm{CY}$, we have three mass scales: the scale $M_p$ in (\ref{IIA}), which is an overall scale. The scale $M_5$ in (\ref{esigr5m5de}), which is associated to the $S^1$. And the scale $M_Y$ in (\ref{emmapse}), which is associated to the Calabi-Yau. 

The integrating out calculation is performed in a graviphoton background
\be 
V_{\mu} = -\frac12  W^{-}_{\mu\nu} x^{\nu} \;,
\label{vmapse}
\ee 
where $W^{-}_{\mu\nu}$ is constant and of the form (\ref{gravback}). Importantly, we recall that this background is imaginary (\ref{wimag}). Note also that the charge under this background, $e$ in (\ref{emmapse}), is (in general) complex.

%%%%%%%%%%%%%%%%%%%%%%%%%%%%%%%%%%%%%%%%%%%%%%%%%%%%%%%%%%%%%%%%%%%%%%%%%%%%%%%%%%%%
\subsection{Integrating out a charged scalar}
\label{sec:genchsc}
%%%%%%%%%%%%%%%%%%%%%%%%%%%%%%%%%%%%%%%%%%%%%%%%%%%%%%%%%%%%%%%%%%%%%%%%%%%%%%%%%%%%

We begin by deriving the general expression for integrating out a charged scalar in four Euclidean dimensions. This will be extremely similar to the integrating out calculation of M2 branes, up to a few subtleties which we discuss in sections \ref{sec:worldac} and \ref{sec:trace}. This section is almost textbook material, see for example \cite{schwartz2014quantum}, up to the fact that we are working in Euclidean space rather than Minkowski space. 

Consider the action for four-dimensional Euclidean scalar QED 
\be 
S^E_{QED} = \int d^4x \left[ \frac14 F_{\mu\nu}F^{\mu\nu} + \phi^* \left( -D^2 + m^2 \right) \phi \right] \;,
\label{qedsimact}
\ee
with $m$ the mass of the scalar, $D$ the covariant derivative
\be 
D_{\mu} = \partial_{\mu} - i e A_{\mu} \;,
\label{covder}
\ee 
and $e$ the gauge coupling constant. We would like to integrate out $\phi$ to generate an effective action $\Gamma[A]$ for $A_{\mu}$, so perform the partial path integral
\be 
\int {\cal D}A \;{\cal D}\phi\;{\cal D}\phi^* \;e^{-S^E_{QED}\left[A,\phi\right]} = \int {\cal D}A\; e^{-\Gamma[A]} \;.
\ee 
Because the $S^E_{QED}$ action is only quadratic in $\phi$, we can do this exactly as a Gaussian integral. In general, we have for some operator ${\cal M}$, that
\be 
\int {\cal D}\phi\;{\cal D}\phi^* \;e^{-\int d^4x \;\phi^* {\cal M} \;\phi} = \frac{{\cal N}}{\det {\cal M}} \;,
\ee 
where ${\cal N}$ is some (infinite) normalization constant. In our case, the operator is 
\be
{\cal M}=-D^2+m^2\;.
\ee 
So if we write the effective action as
\be 
\Gamma[A] = \frac14 F_{\mu\nu}F^{\mu\nu} + \Delta\Gamma[A] \;,
\ee 
with $\Delta\Gamma[A]$ being the contribution to the effective action from integrating out the scalar, then we have
\be 
\Delta\Gamma[A] = -\log {\cal N} + \log \det {\cal M} = -\log {\cal N} + \Tr \log {\cal M} \;.
\ee
The trace is over the eigenvalues of the operator. In this case, we can evaluate it by first taking a derivative with respect to $m^2$, 
\be 
\frac{\partial}{\partial m^2} \left[\;\Tr \log \left( -D^2 + m^2\right)\;\right] = \Tr \left( \frac{1}{-D^2 + m^2}\right) \;. 
\ee 
Then, using Schwinger's representation for an operator
\be 
\frac{1}{{\cal M}} = \int^{\infty}_0 ds\; e^{-s {\cal M}} \;,
\ee 
we can write the trace as
\be 
\Tr \left( \frac{1}{-D^2 +m^2}\right) = \int^{\infty}_0 ds \;e^{-s m^2}\; \Tr\left (e^{-sH}\right)\;,
\ee
with a Hamiltonian defined as 
\be 
H=-D^2 \;.
\ee 
Finally, we need to integrate back with respect to $m^2$ to arrive at
\be 
\Delta\Gamma[A] = -\int^{\infty}_{\epsilon} \frac{ds}{s} \;e^{-s m^2}\; \Tr\left (e^{-sH}\right) \;.
\label{gensconac}
\ee 
Here we included an ultraviolet cutoff $\epsilon$ which absorbs the integration constant and the $\log {\cal N}$ factor in it. So they are all packaged into the divergence as $\epsilon \rightarrow 0$, which needs to be regulated.
We can evaluate the trace in any basis we wish. If we do it in terms of position space eigenstates, then we can write
\be 
\Tr\left (e^{-sH}\right) = \int d^4x\;  \langle x | e^{-sH} | x \rangle \;,
\label{trposi}
\ee 
and so the contribution to the effective Euclidean Lagrangian $\Delta{\cal L}^E$ is
\be 
\Delta{\cal L}^E = -\int^{\infty}_{\epsilon} \frac{ds}{s} \;e^{-s m^2} \langle x | e^{-sH} | x \rangle \;. 
\label{contefflag}
\ee
Our main interest is in the expression (\ref{gensconac}), which calculates the contribution to the effective action as a trace over a Hamiltonian. 

%%%%%%%%%%%%%%%%%%%%%%%%%%%%%%%%%%%%%%%%%%%%%%%%%%%%%%%%%%%%%%%%%%%%%%%%%%%%%%%%%%%%
\subsection{Performing the trace over M2 branes}
\label{sec:trace}
%%%%%%%%%%%%%%%%%%%%%%%%%%%%%%%%%%%%%%%%%%%%%%%%%%%%%%%%%%%%%%%%%%%%%%%%%%%%%%%%%%%%

In section \ref{sec:genchsc}, we explained how to integrate out a charged scalar in Euclidean space. In our setting, we need to perform such an integrating out calculation for the fields associated to the wrapped M2 branes. This means we need to replace the general gauge field with the graviphoton $A_{\mu} \rightarrow V_{\mu}$ , and use the values (\ref{emmapse}) and (\ref{vmapse}). 

Starting from (\ref{gensconac}), we need to perform the trace $\Tr \left(e^{s D^2} \right)$. Importantly, the graviphoton background (\ref{gravback}) does not have any components mixing the $x^1$ and $x^2$ directions with the $x^3$ and $x^4$ directions: $W^{-}_{13}=W^{-}_{14}=W^{-}_{23}=W^{-}_{24}=0$. So we have two decoupled systems, each in an $\mathbb{R}^2$ subspace of the four-dimensional theory. The Hamiltonian therefore factorises and we can write
\be 
\Tr \left(e^{s D^2} \right)= \Tr \left(e^{s \left(D_{1}^2+D_{2}^2\right)}\right) \Tr \left(e^{s \left(D_4^2 + D_3^2 \right)} \right) = \left[ \Tr \left(e^{s \left(D_{1}^2+D_{2}^2\right)}\right) \right]^2 \;.
\label{trr4tor2}
\ee 
In the last equality we used the anti self-dual condition $W^-_{12}=W^-_{43}$.

Taking the covariant derivative (\ref{covder}), and evaluating with the values (\ref{emmapse}) and (\ref{vmapse}), we can write
\be
-s \left(D_{1}^2+D_{2}^2\right) = -\sum_{i=1,2} s \left( \frac{\partial}{\partial x^i} + \frac{i}{8} \bar{Z}(q)\; W^-_{ij} x^j \right)^2  = -\pi_1 \pi^1 - \pi_2 \pi^2 \;,
\label{hathdef}
\ee
where we define
\be 
\pi_i \equiv \frac{\partial}{\partial y^i} + \frac{i}{2} B(u)\; \epsilon_{ij} y^j\;,
\ee 
with
\be 
y^i \equiv \frac{x^i}{\sqrt{s}}  \;\;,\;\;\; u \equiv s \bar{Z}(q) \;\;,\;\;\; B(u) \equiv \frac14 u W^-_{12} \;,
\label{defyuB}
\ee
and $\epsilon_{12}=-\epsilon_{21}=1$. Note that the index on the $\pi_i$ and $y^i$ is raised and lowered with the Einstein metric (\ref{efmet5d}), so with $g^E_{ij}=e^{\sigma}\delta_{ij}$.

This rewriting brings the Hamiltonian into a familiar form for performing the trace. Let us calculate the commutation relations
\be
\left[\pi_1,\pi_2\right] = -i B(u)\;.
\label{commrelB}
\ee
Let us assume that $B(u)$ is real and positive
\be 
B(u) \in \mathbb{R}^+ \;.
\label{Bassum}
\ee 
This may seem a strange assumption, given its complex definition (\ref{defyuB}), but we will return to justify it later on. 
So, assuming (\ref{Bassum}), then (\ref{commrelB}) is exactly the commutation relation of momentum generators for a two-dimensional charged particle in a magnetic field. It is simple to solve this system (see, for example \cite{Tong:2016kpv}). We define creation and annihilation operators
\be 
a = \frac{1}{\sqrt{2B(u)}} \left( \pi_1 + i\pi_2\right) \;,\;\; a^{\dagger} = \frac{1}{\sqrt{2B(u)}} \left( -\pi_1 + i\pi_2\right) \;,
\ee
with commutation relations
\be 
\left[a,a^{\dagger} \right]=1 \;.
\ee 
Then taking (\ref{hathdef}) as the Hamiltonian $\hat{H}$, we can write
\be 
\hat{H} = -\pi_1 \pi^1 - \pi_2 \pi^2 = 2B(u) e^{-\sigma}\left(a^{\dagger}a +\frac12 \right) \;.
\ee 
The energy states of this Hilbert space have the well-known spectrum
\be 
E_n = 2 B(u)e^{-\sigma} \left( n+\frac12 \right) \;,\;\; n \in \mathbb{N} \;.
\label{enelvl}
\ee 

It will be useful for later to note the form of the wavefunction solutions. A simple way to do this is by defining the complex combination 
\be 
w = y^1 + i y^2 \;.
\ee 
Then we can write the annihilation operator as
\be 
a = \sqrt{\frac{2}{B}} \left(\partial_{\bar{w}} + \frac{B}{4} w \right) \;. 
\ee 
Since this annihilates the ground state $|0\rangle$ we have
\be 
\left(\partial_{\bar{w}} + \frac{B}{4} w \right) |0\rangle = 0\;,
\ee 
which gives
\be
|0\rangle = f(w)e^{-\frac{B}{4}|w|^2} \;,
\ee 
with $f(w)$ an arbitrary holomorphic function. We can write a basis of wavefunctions as
\be 
\omega^m e^{-\frac{B}{4}|w|^2} \;,
\ee 
with $m$ an integer. We are interested in the location of the localized wavefunction in $\mathbb{R}^2$. This is calculated by the solution to
\be 
\frac{\partial}{\partial |w|^2} \left( |w|^{2m} e^{-\frac{B}{2}|w|^2}\right) = 0 \;,
\ee 
which gives
\be 
|w|_{\mathrm{loc}} = \sqrt{\frac{2m}{B}} \;.
\ee 
Therefore, the number of states we can fit in a unit area is $\frac{B}{2\pi }$.
So the number of states, for a given energy level, in $\mathbb{R}^2$ is given by  
\be 
\int d^2y \;\frac{B(u)}{2\pi} \;,
\label{trace2dsp}
\ee 
where we have taken the same density of states for all the energy levels.\footnote{We have presented a heuristic derivation of (\ref{trace2dsp}) here, but it is an exact result, see for example \cite{Tong:2016kpv}.} 

\subsubsection*{Performing the trace}

We have now characterised the states sufficiently well to be able to calculate the trace. 
From (\ref{enelvl}), we can write the trace over different energy levels as
\be 
\Tr_E \left(e^{-\hat{H}} \right) = \sum_{n \geq 0}^{\infty} e^{-2B(u)e^{-\sigma}\left(n+\frac12\right)}  = \frac{e^{-B(u)e^{-\sigma}}}{1-e^{-2B(u)e^{-\sigma}}} = \frac{1}{2 \sinh \left(B(u)e^{-\sigma}\right)}\;. 
\ee 
The full four-dimensional trace (\ref{trr4tor2}) is the square of this, so we can write 
\be 
\Tr_E \left(e^{s D^2} \right) = \frac{1}{\left( 2 \sinh \left(B(u)e^{-\sigma}\right)\right)^2} \;.
%= -\frac{1}{\left(2 \sin \left( \frac{iuW^-_{12}e^{-\sigma}}{4}\right)\right)^2} \;.
%= -\frac{1}{\left(2 \sin \left( \frac{u \left|W^-_{12}\right|e^{-\sigma}}{4}\right)\right)^2}\;,
\label{energytrace}
\ee
%where in the last equality we used the fact that $W^-_{12}$ is imaginary, and the relation (\ref{radiufor}).
This is not quite the full trace, we have only traced over the different energy levels. We still need to include the trace over states which are degenerate in energy but differ in the the space $\mathbb{R}^4$. For the case of $\mathbb{R}^2$ this gives a factor as in (\ref{trace2dsp}), and so for $\mathbb{R}^4$ we get the square of this
\be 
\int d^4y \;\left(\frac{B(u)}{2\pi}\right)^2\;.
\label{trace4dsp}
\ee 
We can therefore write the full trace as
\be 
\Tr \left(e^{s D^2} \right) = \frac{1}{\left(2\pi\right)^2}\int d^4y \;\frac{B(u)^2}{\left(2 \sinh \left(B(u)e^{-\sigma}\right)\right)^2} \equiv T(u)\;.
\label{full}
\ee
Here we defined $T(u)$, which is a meromorphic function of $u$. 

\subsubsection*{The assumption $B(u)>0$ and the effective action}

At this point we need to return to the assumption (\ref{Bassum}). Given the definition (\ref{defyuB}), we see that this is not true in the physical solution (\ref{5dmetim}), where $W^-_{12}$ is purely imaginary, $\bar{Z}(q)$ is complex, and $s$ is real. However, in the contribution to the effective action (\ref{gensconac}), the result $T(u)$ appears inside an integral
\be 
-\int_{\epsilon}^{\infty} \frac{ds}{s}\; e^{-s \left|Z(q)\right|^2M_Y^2}\;T(u)= -\int^{\infty e^{i\theta}}_{0^+} \frac{du}{u} e^{-u Z(q)M_Y^2} \;T(u) \;,
\label{intstou}
\ee 
where we used $m^2 = \left|Z(q)\right|^2 M_Y^2$ (\ref{emmapse}).
Here the integration is performed along a line in the $u$-plane which starts at the origin and goes out to infinity at an angle 
\be 
\theta = \tan^{-1} \left(\frac{\mathrm{Im\;} \bar{Z}(q)}{\mathrm{Re\;} \bar{Z}(q)}\right) \;,
\ee 
and we recall that $u = s \bar{Z}(q)$ (\ref{defyuB}).
We have taken here the cutoff $\epsilon$ to approach zero, but not including zero,
\be 
\epsilon = 0^+ \;.
\label{preuvepcut}
\ee

\begin{figure}[h]
\centering
\includegraphics[width=0.6\textwidth]{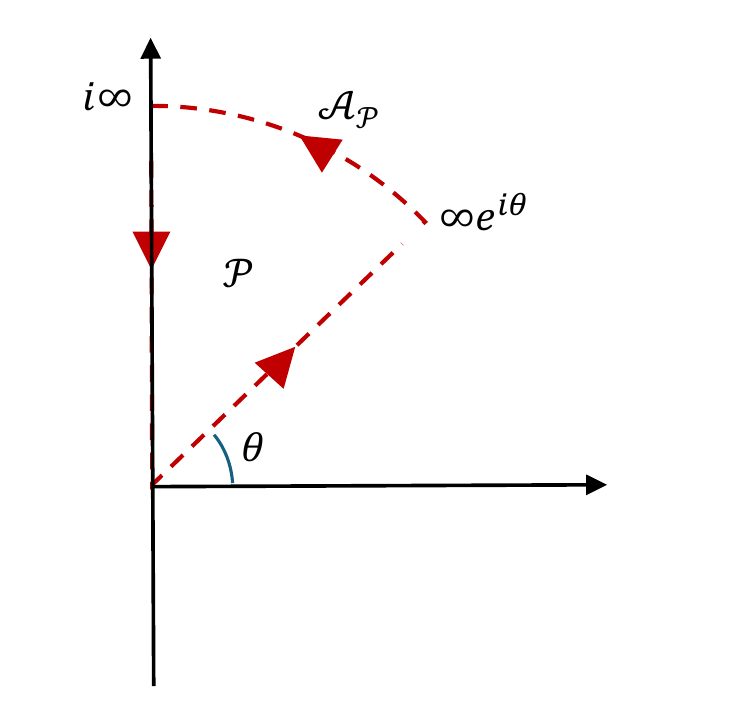}
\caption{Figure illustrating the decomposition of the integration path in (\ref{uintpathcon}).}
\label{fig:ucon}
\end{figure}

Let us take $\theta > 0$ first. In the complex $u$-plane, we have that, for $W^-_{12}$ purely imaginary,  $B(u)=\frac14 u W^-_{12}$ is real when $u$ is on the positive imaginary axis. So for $u \in i \mathbb{R}^+$ we have
\be 
B(u) = \pm \frac{1}{4}  \left| u W^-_{12}\right| \;.
\ee 
This justifies the assumption (\ref{Bassum}), since the sign of $B(u)$ is not important, only that it is real. If $B(u)$ is negative, then we can just interchange $y^1 \leftrightarrow y^2$ and obtain the same result. Therefore, what we would like to do is evaluate (\ref{intstou}) on the half-line of the positive imaginary axis, rather than the half-line along the angle $\theta$. We can relate the two integrals as
\be 
\oint_{{\cal P}} = \int^{\infty e^{i\theta}}_{0^+} - \int^{i\infty }_{0^+} + \int_{{\cal A}_{\cal P}} \;,
\label{uintpathcon}
\ee 
where $\oint_{{\cal P}}$ is a contour integral along a contour composed of the two half-line integrals, plus the arc connecting them at infinity. This is shown in figure \ref{fig:ucon}.
Because $T(u)$ is a meromorphic function, the contribution from the contour integral $\oint_{{\cal P}}$ is given by the residue of the poles of $T(u)$ inside the contour. For $W^-_{12}$ imaginary, the poles of $T(u)$ are on the real axis, and therefore there are no poles inside the contour ${\cal P}$, and so the integral vanishes. Further, as $|u| \rightarrow \infty$, $e^{-u Z(q)}T(u)$ behaves as
\be 
e^{-u Z(q)}T(u) \sim e^{-\mathrm{Re}(u)\;\mathrm{Re}(Z(q))+\mathrm{Im}(u)\;\mathrm{Im}(Z(q))}\; e^{-\frac12 \mathrm{Im}(u) |W^-_{12}|e^{-\sigma}} \;,
\ee  
which vanishes at infinity. Therefore, also the contribution from $\int_{{\cal A}_{\cal P}}$ in (\ref{uintpathcon}) vanishes. This means that the integration along the two half-lines is equal
\be 
\int^{\infty e^{i\theta}}_{0^+} = \int^{i\infty }_{0^+} \;.
\ee 
So we can perform the calculation of the trace along the imaginary axis in the $u$-plane, as we have done in this section, and the result is equal to the physical evaluation along the line at angle $\theta$.

\subsubsection*{Back to the physical effective action}

The physical effective action is given by the left hand side of (\ref{intstou}), so in terms of $s$, $Z(q)$ and $W^{-}_{12}$, rather than $u$ and $B(u)$. We can convert between them using (\ref{defyuB}). It is also useful to write $|W^-_{12}|$ as
\be 
|W^-_{12}| = \frac12 e^{\sigma} \sqrt{-W^-_{\mu\nu}W^{-,\mu\nu}} \equiv \frac12 e^{\sigma} \sqrt{ - W^2 }\;,
\ee 
where we used (\ref{wmtow12}) and (\ref{efmet5d}). We therefore find
\be 
\Delta \Gamma_S = -\frac{\left(\bar{Z}(q)\right)^2}{64\left(2\pi\right)^2} \int d^4x \;\sqrt{g^E}\int^{\infty}_{\epsilon} \frac{ds}{s} \;e^{-s \left|Z(q)\right|^2M_Y^2}\; \frac{\left(-W^2\right) }{\left(2 \sin \left(\frac{s \bar{Z}(q) \sqrt{ - W^2 }}{8}\right)\right)^2} \;,
\label{fullcontscal}
\ee 
where we used that $e^{2\sigma} = \sqrt{g^E}$.

This is an important intermediate result, it gives the contribution from integrating out a single scalar field, with mass and charges of the M2 branes. We have yet to show that this is how integrating out the full M2 branes should be captured, and indeed there will be some small modifications, these are discussed in section \ref{sec:worldac}. 

%%%%%%%%%%%%%%%%%%%%%%%%%%%%%%%%%%%%%%%%%%%%%%%%%%%%%%%%%%%%%%%%%%%%%%%%%%%%%%%%%%%%
\subsection{Supersymmetry and the worldline action}
\label{sec:worldac}
%%%%%%%%%%%%%%%%%%%%%%%%%%%%%%%%%%%%%%%%%%%%%%%%%%%%%%%%%%%%%%%%%%%%%%%%%%%%%%%%%%%%

In this section we study in more detail the description of wrapped M2 branes as four-dimensional charged scalars with a simple QED-type action (\ref{qedsimact}). As explained in \cite{Dedushenko:2014nya}, a very useful way to approach the problem is through the worldline formalism (see, for example, \cite{Schubertlect,Corradini:2015tik} for reviews, and \cite{Strassler:1992zr} for a good introduction). 

Consider performing the Hamiltonian trace in position space, so as in (\ref{trposi}). It is possible to write this in terms of a path integral for the particle worldline, starting at position $x$ and returning to it after $s$ Euclidean time has passed. For example, for an uncharged real scalar we can write
\be 
\langle x|e^{-sH_{S}}|x\rangle = \int_{x(0)}^{x(s)} {\cal D}x(\tau)\;\mathrm{Exp} \left[ -\int_0^{s} d\tau \left( \frac14 \dot{x}^{\mu} \dot{x}_{\mu} \right)\right] \;,
\ee 
where $\dot{x}^{\mu} = \frac{\partial x^{\mu}}{\partial \tau}$ and $H_S = - \partial^{\mu}\partial_{\mu}$. For a charged scalar we have
\be 
\langle x|e^{-sH_{CS}}|x\rangle = \int_{x(0)}^{x(s)} {\cal D}x(\tau)\;\mathrm{Exp} \left[ -\int_0^{s} d\tau \left( \frac14 \dot{x}^{\mu} \dot{x}_{\mu}  - i e \dot{x}^{\mu} A_{\mu}\right)\right] \;,
\label{csworldl}
\ee 
with $H_{CS} = - D^2$, and $D_{\mu}$ as in (\ref{covder}). 

For the M2 brane, we have a three-dimensional world surface, rather than a particle worldline. However, since the brane is wrapping a two-cycle in the Calabi-Yau, we can approximate its action as a worldline as long as the two-cycle size is much smaller than the worldline path. The relevant worldlines for the integrating out calculation turn out to be at least as long as the perimeter of the $S^1$ factor in the background, so of length $2 \pi R_5$. There are two way to see this. The first is directly by inspecting the final result of the integrating out computation (\ref{rcres}). This is given as a contour integral which localises onto poles at fixed worldline path length. For weak topological string coupling $\lambda < 1$, the first pole corresponds to a path length which winds one time around the $S^1$. The second way is through a dual description, where we exchange the Kaluza-Klein momentum modes of the M2 branes for Winding modes. This is the particle picture of \cite{Dedushenko:2014nya}. Exchanging winding and momentum modes is similar to T-duality, but here it is along the M-theory circle. Since the M-theory circle controls the string coupling, it is more similar to S-duality from the type IIA perspective. So we expect it to be some sort of exchange of fundamental string physics and D-brane physics. Indeed, this is precisely the idea behind the Gopakumar-Vafa calculation \cite{Gopakumar:1998ii,Gopakumar:1998jq}. From the type IIA perspective, after integrating out the D2-D0 states, we will generate terms of the form $e^{-t}$ in the effective action (\ref{seffiiawr}). But these type of terms are already understood from the fundamental string description as worldsheet instanton contributions. They arise from the fundamental string wrapping the two cycles. The fundamental string arises in M-theory from an M2 brane wrapping the M-theory circle. So the worldsheet instanton perspective is one where the M2 branes are wrapping both the M-theory circle and the Calabi-Yau two-cycles. That is exactly the winding mode, or particle picture. 

The limit in which the integrating out calculation is performed, is such that 
\be 
\delta \equiv \frac{R_{CY}}{R_{M}} \rightarrow 0 \;,
\label{detldef}
\ee 
 where $R_{CY}$ is the typical Calabi-Yau length scale. We discuss this in more detail in section \ref{sec:embiia}. Therefore, we can say that a worldline description is correct, up to corrections in the parameter $\delta$ in (\ref{detldef}). More accurately, we can perform an expansion of the M2 worldvolume action in $\delta$, and at leading order the action is constant along the Calabi-Yau part and can be integrated over, yielding a worldline action. 

The effective worldline action for the M2 branes was calculated in \cite{Dedushenko:2014nya}. In the particle picture, this is a calculation in the five-dimensional background (\ref{5dbackpure}), and the worldline is along the $t$ direction. We start from the relativistic particle action
\be 
\int dt \left( -\sqrt{-g_{MN}\dot{x}^M \dot{x}^N} + U_M \dot{x}^M \right) \;.
\label{unitwl5dac}
\ee 
Here we set the particle mass and charge equal to 1.\footnote{One can see that in five-dimensions the mass and charge should be set equal from (\ref{emmapse}). As discussed in section \ref{sec:map2top}, we work in units where $M_Y=1$, and the four-dimensional graviphoton is related with a factor of $4$ to the five-dimensional one (\ref{4dto5dgrrel}).} We then expand it out in derivatives, about the background (\ref{5dbackpure}), to obtain the non-relativistic action. The resulting worldline action takes the form\footnote{An important subtlety is that because the graviphoton appears in the four-dimensional metric, the coupling in the particle worldline action to the graviphoton is doubled. We also drop the constant factor of $-1$ in the worldline action which contributes to the mass part of (\ref{gensconac}), rather than the Hamiltonian.} 
\be   
\int dt \left( \frac12 \dot{x}^{\mu} \dot{x}_{\mu}  - T^-_{\mu\nu} x^{\nu}\dot{x}^{\mu}   \right)\;.
\label{wlsusyac}
\ee
The action now needs to be supersymmetrized. The worldline fields are the $x^{\mu}$, which determine the fluctuations of the worldline along the $\mathbb{R}^4$ part of the spacetime. So we need four fermionic degrees of freedom partners. They are written as $\psi_{Ai}$, with $A=1,2$ and $i=1,2$ (the $A$ is the spacetime spinor index, and the $i$ denote the two spacetime spinors in a hypermultiplet). The resulting action is then completely fixed by supersymmetry to be \cite{Dedushenko:2014nya}
\be   
\int dt \left( \frac12 \dot{x}^{\mu} \dot{x}_{\mu}  -  T^-_{\mu\nu} x^{\nu}\dot{x}^{\mu}  + \frac{i}{2} \epsilon^{ij}\epsilon^{AB} \psi_{Ai} \dot{\psi}_{Bj} \right)\;.
\label{wlsusyacpreee}
\ee
%Here we have set $M=1$, since we are only concerned with the form of the worldline action. 
%Note that the supersymmetry algebra is such that the central charge $\xi=2M$ \cite{Dedushenko:2014nya}. So setting $M=\frac12$ implies the central charge is normalised to one $|\xi|=1$.

We can now map this worldline action to the four-dimensional setting, so using the metric (\ref{5dmetim}). The important change is that while we use the same worldline action, we are not integrating the worldline as wrapping the M-theory circle, but from a four-dimensional perspective the worldline (in Euclidean time $\tau$) has a length given by the Schwinger proper time parameter $s$, that we integrate over. So we have for the Euclidean worldline action
\be   
S_{WL} = \int_0^s d\tau \left( \frac12 \dot{x}^{\mu} \dot{x}_{\mu}  - \frac{i}{4} \bar{Z}(q) W^-_{\nu\mu} x^{\nu}\dot{x}^{\mu}  + \frac12 \epsilon^{ij}\epsilon^{AB} \psi_{Ai} \nabla_{\tau} \psi_{Bj} \right)\;,
\label{wlsusyacpre}
\ee
where $\nabla_{\tau}$ is the covariant derivative with the spin-connection (explicitly given in \cite{Dedushenko:2014nya}). We used the relation (\ref{4dto5dgrrel}) between the five-dimensional and four-dimensional graviphotons, and have included the charge $\bar{Z}(q)$ under the four-dimensional graviphoton.\footnote{Note that the five-dimensional and four-dimensional central charges are related as in (\ref{centbps5}) and (\ref{mbpsfrom5dm2}).}

In order to match the worldline action (\ref{wlsusyacpre}) with the form (\ref{csworldl}), we need to introduce coordinates $\tilde{x}^{\mu}$ as 
\be 
x^{\mu} = \frac{\tilde{x}^{\mu}}{\sqrt{2}}  \;\;,\;\; d^4\tilde{x} = 4 \;d^4 x \;. 
\label{xrescaling}
\ee 
Then we reach the action
\be   
S_{WL} = \int_0^s d\tau \left( \frac14 \dot{\tilde{x}}^{\mu} \dot{\tilde{x}}_{\mu}  - \frac{i}{8} \bar{Z}(q) W^-_{\nu\mu} \tilde{x}^{\nu}\dot{\tilde{x}}^{\mu}  + \frac12 \epsilon^{ij}\epsilon^{AB} \psi_{Ai} \nabla_{\tau} \psi_{Bj} \right)\;.
\label{wlsusyac}
\ee
This action is indeed supersymmetric under the four-dimensional supersymmetry algebra \cite{Dedushenko:2014nya}. Actually, the supersymmetry algebra fixes (\ref{wlsusyac}) which is why we needed to perform a rescaling (\ref{xrescaling}), rather than factor out an overall factor of $\frac12$.

The action (\ref{wlsusyac}) coincides with (\ref{csworldl}), up to the fermionic terms. It is useful to split the fermion modes into zero modes, denoted $\psi_{Ai}(0)$ , which are independent of $\tau$, and the higher modes. The zero modes are the superpartners of the collective coordinates of the particle, which move the whole worldline along the $\mathbb{R}^4$. The zero modes are the only ones relevant for the loop worldline, since it has periodic boundary conditions while fermions are anti-periodic.\footnote{More precisely, their determinant contribution in the path integral cancels with an odd component of the bosonic contribution \cite{Dedushenko:2014nya}.} The $\psi_{Ai}(0)$ are in fact identified with the supersymmetry parameters $\theta$. The precise relation is \cite{Dedushenko:2014nya}
\be 
\psi_{Ai}(0) = \sqrt{- i2\bar{Z}(q)} \;\theta_{Ai} \;.
\ee 
Schematically, the path integral behaves as
\be 
\int d^4\psi \;e^{-\psi^2} =-\frac{1}{4\bar{Z}(q)^2} \int \;d^4\theta \;e^{-2\bar{Z}(q) \theta^2}\;,
\ee 
and so the effect of the fermions is to introduce a pre-factor in front of the superspace integral
\be 
-\frac{1}{4\bar{Z}(q)^2} \;d^4\theta \;.
\label{zb2facthepre}
\ee 
The terms quadratic in $\theta$ in the action then combine with the bosonic ones to form the superfields. Finally, we need to undo the change of variable (\ref{xrescaling}), which means that overall the action integration measure is
\be 
-\frac{1}{\bar{Z}(q)^2}\; d^4x  \;d^4\theta \;.
\label{zb2facthe}
\ee 

In summary, we therefore find that we can utilise the worldline (\ref{csworldl}), and therefore the results of section \ref{sec:trace}, in particular (\ref{5deucro}) hold precisely. The only modification is that we must account for the extra prefactor (\ref{zb2facthe}) (which will cancel against the $\bar{Z}(q)^2$ factor in (\ref{5deucro})).

\subsubsection*{Exactness of the action}

There is a very important point, for our purposes, which was explained in \cite{Dedushenko:2014nya}. It is that the action (\ref{wlsusyac}), or (\ref{wlsusyacpreee}), is exact in the sense of its contribution to F-terms. If we added anything else to it, then it would not contribute to F-terms anymore. Because the worldline formalism maps the worldline action to the Hamiltonian $-D^2$ which we used in our calculation, this Hamiltonian is also exact in the sense of calculating the F-terms. We therefore can expect that it yields the full answer for the effective action F-terms, and not only an approximation or an asymptotic series. 

If we define the topological string free energy as the F-terms in the effective action which involve purely the graviphoton background, then we can expect the integrating out calculation to yield the exact answer for the free energy. 

Let us see why adding more terms to the worldline action (\ref{wlsusyac}) would not lead to appropriate F-terms. We have seen that the worldline action yields a factor of $d^4 \theta$ from the fermion zero modes. This is because the action is supersymmetric, and so preserves half the supersymmetries of the background. If we added anything else to it, which is not supersymmetric, then we would break all the supersymmetries, and gain additional fermionic zero modes as goldstone bosons of that breaking. This would lead to terms of type $d^8 \theta$, which are not F-terms. This is physics of the statement that only BPS states can generate F-terms. 

There is another very constraining property beyond supersymmetry of the action. The effective action F-terms cannot depend explicitly, in the Einstein frame, on the circle radius. This is the statement that the string coupling appears only through the combination $W^-_{\mu\nu}W^{-,\mu\nu}$. If we added more terms to the action (\ref{wlsusyac}), then we would generate power of the circle radius $e^{\sigma}$ which are not accompanied by the appropriate powers of $W^-_{\mu\nu}$ \cite{Dedushenko:2014nya}. It would be impossible to interpret such terms as F-terms, because in a supersymmetric framework the dilaton should not mix with the vector multiplets. Also, by definition, they would have no interpretation in terms of the topological string because their dependence on the string coupling would not be through $\lambda$.    

Overall, it is hard to see how to include corrections to the worldline action, and therefore to the Hamiltonian, which would contribute to the restricted F-terms. One possibility is that there are certain local interactions between multi-wrapped branes that could contribute, though these were also dismissed in \cite{Dedushenko:2014nya}. We can therefore hope for an exact result for the free energy, rather than an asymptotic series. We note, however, that the topological string free energy is not defined precisely by the Wilsonian action F-terms. There are certain non-local terms in the effective action which arise from integrating out massless modes associated to degenerate worldsheet geometries \cite{Antoniadis:1993ze,Bershadsky:1993cx}. Such terms are not captured by the integrating out calculation we are performing. They are controlled instead by the holomorphic anomaly equation \cite{Bershadsky:1993cx}. Therefore, we can expect an exact result for the Wilsonian effective action, which is mapped to the topological string free energy in the so-called holomorphic limit \cite{Dedushenko:2014nya}.

%%%%%%%%%%%%%%%%%%%%%%%%%%%%%%%%%%%%%%%%%%%%%%%%%%%%%%%%%%%%%%%%%%%%%%%%%%%%%%%%%%%%
\subsection{Higher genus contributions}
\label{sec:highersp}
%%%%%%%%%%%%%%%%%%%%%%%%%%%%%%%%%%%%%%%%%%%%%%%%%%%%%%%%%%%%%%%%%%%%%%%%%%%%%%%%%%%%

In section \ref{sec:worldac}, when constructing the supersymmetric worldline action (\ref{wlsusyac}) we considered the universal target space embedding bosonic coordinates $x^{\mu}$, and their fermionic superpartners $\psi_{Ai}$. However, in general there may by other fields living on the M-theory worldvolume that should be included in the particle worldline action. Because of supersymmetry, we can count the different fields by studying the fermionic fields only. Further, we know that only fermionic zero modes contribute to the loop integral. So in order to count the fields in the worldline action we need to count the fermionic zero modes. We can do this in type IIA or in M-theory, the number of zero modes is a protected index which cannot change under continuous variation of parameters. 

As in \cite{Dedushenko:2014nya}, we will restrict in these notes to the cases where the cycle wrapped by the M2 brane is rigid. This means that there are no massless deformation modes of the cycle inside the Calabi-Yau. Note that this is a very restrictive condition, and the generic cycle in a Calabi-Yau will have deformations. In the case when there are deformations, the spectrum, and Hamiltonian, of the fermionic zero modes is modified. This modification needs to be accounted for when performing the trace. In \cite{Gopakumar:1998jq} this trace just yields an overall multiplicity, which is part of the Gopakumar-Vafa invariant. However, we will not assume this here, and instead, as in \cite{Dedushenko:2014nya}, will restrict to rigid cycles only.  

In the rigid case, the number of fermionic zero modes is determined by the topology of the two-cycle that the D2 or M2 brane wraps. Let us denote the two-cycle as ${\cal C}_2$. A nice way to count the fermions is by mapping them to differential forms (see, for example, \cite{green1988superstring} for a nice account). In general, we can construct a differential form from spinors by wedging between them anti-symmetric products of gamma matrices $\gamma_{\mu}$. So an $n$-form $\omega^{(n)}$ can be constructed from two spinors $\eta_i$, $i=1,2$, as
\be 
\omega^{(n)}_{\mu_1\mu_2...\mu_n} = \bar{\eta}_1 \gamma_{[\mu_1}\gamma_{\mu_2}...\gamma_{\mu_n]}\eta_2 \;,
\ee 
where the square brackets around the indices denote anti-symmetrization. Depending on the number $n$ being even or odd, the spinors $\eta_1$ and $\eta_2$ are required to have the same or opposite chirality. In the extended supersymmetry setting we are in, there are always the appropriate chirality spinors since both chiralities appear. By mapping the spinors to forms, we map the problem of counting the number of spinors to counting the number of forms. In particular, massless spinors map to harmonic forms. The number of fermionic zero modes is counted by the Betty numbers $b_n$, or Hodge numbers $h^{p,q}$ (with $p+q=n$), of ${\cal C}_2$. Recall that the top and bottom harmonic forms of a compact manifold are unique, so we have
\be 
b_0\left({\cal C}_2\right) = b_2\left({\cal C}_2\right)=1 \;.
\ee 
These are universal and are counting the universal zero modes $\psi(0)_{Ai}$ superpartners of the bosonic coordinates. The possibly additional zero mode fermions that we are interested in here are counted by the middle Betty and Hodge numbers
\be 
b_1\left({\cal C}_2\right) = 2 r \;,\;\; h^{1,0}\left({\cal C}_2\right)=h^{0,1}\left({\cal C}_2\right)=r \;,
\ee 
where $r$ is the genus of ${\cal C}_2$. The $h^{1,0}$ and $h^{0,1}$ give the opposite chirality fermions of a hypermultiplet. We therefore see that if ${\cal C}_2$ has genus $r$, then we find $r$ additional hypermultiplets.\footnote{These should really be understood as composing a single higher spin field \cite{Gopakumar:1998ii,Gopakumar:1998jq,Dedushenko:2014nya}, but this is not relevant for our purposes.} 

While the universal zero modes $\psi(0)_{Ai}$ were not traced over, but rather mapped to the superspace integration factor in the action. The possible additional fermion zero modes associated to the higher genus contributions do need to be traced over. Their contribution to the trace, as in (\ref{full}), can therefore be written as
\be 
\Tr_{{\cal H}_r} \left[\left(-1\right)^F e^{-s \tilde{H}}\right] \;,
\ee 
where $\Tr_{{\cal H}_r}$ denotes tracing over the Hilbert space of the $r$ hypermultiplets, $F$ is the fermion number of the states, and $\tilde{H}$ is the Hamiltonian associated to the states. The best way to determine $\tilde{H}$ is again through the worldline formalism, and we do this in appendix \ref{sec:fertra}. 
The Hamiltonian is identified as
\be 
\tilde{H}= \frac{i}{2} \bar{Z}(q) \; \sqrt{ - W^2 } \;J \;,
\ee 
with $J$ the helicity of the states. In a single hypermultiplet we have two complex scalars with $J=0$, and two opposite chirality Weyl Fermions with helicity $J=\pm \frac12$. We therefore have
\be 
\Tr_{{\cal H}_r} \left[\left(-1\right)^F e^{-s \tilde{H}}\right] = \left(2 - e^{-is\frac14 \bar{Z}(q) \sqrt{ - W^2 }} - e^{+is \frac14\bar{Z}(q) \sqrt{ - W^2 }} \right)^r= \left(2 \sin \left( \frac{s\bar{Z}(q) \sqrt{ - W^2 }}{8}\right) \right)^{2r}\;.
\label{trhyrsp}
\ee 
This factor then multiplies the trace over the universal hypermultiplet modes. The net effect is therefore that the power of the Sine in (\ref{fullcontscal}) is shifted
\be 
\frac{1}{\left(2 \sin \left( \frac{s\bar{Z}(q) \sqrt{ - W^2 }}{8}\right) \right)^2} \rightarrow \frac{1}{\left(2 \sin \left( \frac{s\bar{Z}(q) \sqrt{ - W^2 }}{8}\right) \right)^{2-2r}} \;.
\ee 

%%%%%%%%%%%%%%%%%%%%%%%%%%%%%%%%%%%%%%%%%%%%%%%%%%%%%%%%%%%%%%%%%%%%%%%%%%%%%%%%%%%%
\subsection{The resulting effective action F-terms}
\label{sec:fullferms}
%%%%%%%%%%%%%%%%%%%%%%%%%%%%%%%%%%%%%%%%%%%%%%%%%%%%%%%%%%%%%%%%%%%%%%%%%%%%%%%%%%%%

We are now ready to write the complete effective action F-terms from integrating out a single M2 brane. We will generalise this to all the M2 branes in section \ref{sec:map2top}. We start from the effective action from integrating out a scalar (\ref{fullcontscal}). 
The supersymmetric version of this, so for a hypermultiplet, includes also the fermionic contributions. These were worked out in section \ref{sec:worldac}, and yield a superspace integral factor $d^4\theta$, and a prefactor of $-\left(\bar{Z}(q)\right)^{-2}$, as in (\ref{zb2facthe}), so that we have for the hypermultiplet contribution $\Delta \Gamma_H$ the expression
\be 
\Delta \Gamma_H = \frac{1}{64\left(2\pi\right)^2} \int d^4x \;d^4\theta \;\sqrt{g^E}\int^{\infty}_{\epsilon} \frac{ds}{s} \;e^{-s \left|Z(q)\right|^2M_Y^2}\; \frac{\left(-W^2\right) }{\left(2 \sin \left(\frac{s \bar{Z}(q) \sqrt{ - W^2 }}{8}\right)\right)^{2-2r}} \;,
\label{fullhyper}
\ee 
where we include the generalisation to the genus $r$ case, as in section \ref{sec:highersp}.

The expression (\ref{fullhyper}) requires a little explaining. By writing the integral over superspace, we mean that the fields $X^I$ and $W^{-}_{\mu\nu}$ should really be replaced by their superfields ${\cal X}^I$ and ${\cal W}_{\mu\nu}$, as in (\ref{n2boschmul}) and (\ref{gravsupf}). We have not explicitly derived this replacement, though it follows directly from supersymmetry. It is possible to derive the completion into the graviphoton superfield straightforwardly from the worldline action (\ref{wlsusyac}), see \cite{Dedushenko:2014nya}. The promotion of the $X^I$ to superfields explicitly requires more work, and has been done to some extent in \cite{Dedushenko:2015rga}. Note that if we do not promote the $X^I$ to superfields, the genus-zero contribution $F_0$ vanishes, since there are no $\theta$ parameters to saturate the superspace integral. 

The expression (\ref{fullhyper}) contains two unspecified mass scales, $M_Y$ explicitly, and $M_5$ implicitly through (\ref{esigr5m5de}). In order to match with the supergravity effective action we also need to determine the overall prefactor of $M_p^2$ in (\ref{IIA}). These mass scales are determined by the embedding of the background $\mathbb{R}^4\times S^1$ into the full higher dimensional background  $\mathbb{R}^4\times S^1 \times \mathrm{CY}$. This embedding is the topic of section \ref{sec:embiia}.  

%%%%%%%%%%%%%%%%%%%%%%%%%%%%%%%%%%%%%%%%%%%%%%%%%%%%%%%%%%%%%%%%%%%%%%%%%%%%%%%%%%%%
\section{Embedding in M-theory and string theory}
\label{sec:embiia}
%%%%%%%%%%%%%%%%%%%%%%%%%%%%%%%%%%%%%%%%%%%%%%%%%%%%%%%%%%%%%%%%%%%%%%%%%%%%%%%%%%%%

In section \ref{sec:intm2}, we performed the integrating out calculation in a very general way, within the context of four-dimensional supergravity. In this section we embed this calculation into the higher dimensional M-theory, and type IIA string theory, setting.

%%%%%%%%%%%%%%%%%%%%%%%%%%%%%%%%%%%%%%%%%%%%%%%%%%%%%%%%%%%%%%%%%%%%%%%%%%%%%%%%%%%%
\subsection{Scales, metrics and masses}
\label{sec:scalbps}
%%%%%%%%%%%%%%%%%%%%%%%%%%%%%%%%%%%%%%%%%%%%%%%%%%%%%%%%%%%%%%%%%%%%%%%%%%%%%%%%%%%%

Both string theory and M-theory have the remarkable property that they only contain a single mass scale, which fixes the units of measurement. From the M-theory perspective, the most natural scale is related to the tension of the M2 branes $T_{M2}$. We denote this scale as $M_2$, and write
\be 
M_2 = \left( T_{M2}\right)^{\frac13} \;.
\label{m2def}
\ee
On the other hand, from the string theory perspective the most natural scale is the string scale, which is related to the fundamental string tension $T_{F1}$ as
\be 
M_s = \left( \frac{T_{F1}}{2\pi} \right)^{\frac12} = \frac{1}{2\pi \sqrt{\alpha'}} \;.
\label{msdef}
\ee 
The two mass scales (\ref{m2def}) and (\ref{msdef}) are related by the string coupling $g_s$, which in M-theory has the interpretation of controlling the radius $R_M$ of the extra circle dimension. The simplest way to determine their relation is by noting that, when it does not wrap the M-theory circle, the M2 brane has an interpretation as a D2 brane in type IIA. The tension of a D$p$ brane in string theory is given by
\be 
T_{Dp}=\left(\frac{2\pi}{g_s}\right) M_s^{p+1} \;.
\label{dbranten}
\ee 
Equating the D2 tension with the M2 tension we find the relation
\be 
M_2 = \left(\frac{2\pi}{g_s}\right)^{\frac13} M_s \;.
\label{m2toms}
\ee 

To obtain the relation between the string coupling $g_s$ and the M-theory circle radius $R_M$, we note that when the M2 brane wraps the M-theory circle it is interpreted as a fundamental string. In general, if we have a $p$-brane that is wrapping a $q$-dimensional cycle, the tension of the remaining $(p-q)$-brane is given by 
\be 
T_{p-q} = T_p \;\mathrm{Vol}_q \;,
\label{tenvolcyc}
\ee  
where $\mathrm{Vol}_q$ is the volume of the $q$-cycles. Applying this to an M2 brane wrapping the M-theory circle we obtain
\be 
T_{F1}  = T_{M2}\; 2\pi R_M = T_{D2}\; 2\pi R_M = \frac{\left(2\pi\right)^2}{g_s} R_M M_s^3 \;.
\ee 
Using (\ref{msdef}), we therefore find
\be 
g_s = 2\pi R_M M_s \;.
\label{gsrmrel}
\ee 

We also need to determine the higher dimensional Planck scales, so the scales multiplying the $d$-dimensional Ricci scale $R^{(d)}$. From dimensional reduction we can relate the 11-dimensional Newton's constant $\kappa_{11}$ with the ten-dimensional one $\kappa_{10}$ through
\be 
\frac{1}{2\kappa_{11}^2} \int_{{\cal M}_{11}} d^{11}x\sqrt{-g^{(11)}} R^{(11)} = \frac{2\pi R_M}{2\kappa_{11}^2} \int_{{\cal M}_{10}} d^{10}x\sqrt{-g^{(11)}} R^{(11)} = \frac{1}{2\kappa_{10}^2} \int_{{\cal M}_{10}} d^{10}x\sqrt{-g^{S}} R^{S} \;.
\label{k11k10rel}
\ee 
We can choose to identify the ten-dimensional part of the eleven-dimensional metric, $g^{(11)}_{MN}$ , with the ten-dimensional string frame metric $g^{S}_{MN}$ , 
\be 
g^{(11)}_{MN} = g^{S}_{MN} \;,
\label{g11eqgs}
\ee 
where $M,N$ are ten-dimensional indices. In the string frame, we have the relation
\be 
\frac{1}{2\kappa_{10}^2} = \frac{2\pi M_s^8}{g_s^2}\;.
\label{strfrpre}
\ee 
Therefore, from (\ref{gsrmrel}), (\ref{k11k10rel}) and (\ref{strfrpre}), we find
\be 
\frac{1}{2\kappa_{11}^2} = \frac{1}{2\pi R_M}\frac{1}{2\kappa_{10}^2} = \frac{2\pi M_s^9}{g_s^3}\;.
\label{k11toms}
\ee 
We have now determined the starting points of the higher dimensional theories, writing all the relevant quantities in terms of the string scale $M_s$ and the string coupling $g_s$.

\subsubsection*{The Einstein frame}

Famously, string theory contains no dimensionless parameters. By this we mean that quantities such as the string coupling $g_s$, or $R_M M_s$, are actually promoted to spacetime fields. For example, the string coupling is given by the expectation value of the dilaton field $\phi$ as
\be 
g_s = e^{\phi} \;.
\ee 
The Einstein frame is defined to be a choice of spacetime metric where the coefficient in front of the Ricci scalar in the action is independent of any of these fields. So it is given purely in terms of some constants and dimensionful mass scales. We can call whatever is in front of the Ricci scalar the Planck scale. More precisely, in our conventions we pull out this overall scale from the full action, so we write for the action in $d$-dimensions
\be 
S_{(d)} = \left(M_p^{(d)}\right)^{d-2} \int d^d x\sqrt{-g^E}  \left( \frac12 R^{(d)} + {\cal L}^{(d)}_M \right) \;,
\label{einframaction}
\ee 
where $g_E$ is the Einstein frame metric, $M_p^{(d)}$ is the $d$-dimensional Planck scale, $R^{(d)}$ is the $d$-dimensional Ricci scalar, and ${\cal L}^{(d)}_M $ is the $d$-dimensional matter Lagrangian. The Einstein frame is then such that, if we chose to, we could work in Planck units $M_p=1$.

In order to reach the Einstein frame, we will often need to perform a Weyl rescaling. Let us consider two metrics, in $d$ dimensions, $g^E_{\mu\nu}$ and $g^S_{\mu\nu}$ that are related by a dimensionless parameter $\sigma'$ as
\be 
g^S_{\mu\nu} = e^{2\sigma'}\;g^E_{\mu\nu} \;.
\label{weyres}
\ee   
Then we have the relations
\be 
\sqrt{-g^S} R^S = e^{\left(d-2\right)\sigma'}\sqrt{-g^E} R^E \;,
\label{wetlresctran}
\ee 
where $R^S$ and $R^E$ denote the Ricci scalars constructed from the metrics $g^S_{\mu\nu}$ and $g^E_{\mu\nu}$ respectively. 

\subsubsection*{Measuring masses with metrics}

If we perform a Weyl rescaling (\ref{weyres}), then the Ricci scalar transforms as (\ref{wetlresctran}). However, in (\ref{einframaction}) we have also the matter Lagrangian ${\cal L}^{(d)}_M$ that we need to transform. To see how this behaves, we can consider a massive scalar field
\be 
{\cal L}^{(d)}_M = -g^{S,\mu\nu}\partial_{\mu} \varphi \partial_{\nu} \varphi + m_{(S)}^2 \varphi^2 \;.
\ee  
Here we are using the metric $g^S_{\mu\nu}$, and we label the mass of the scalar as matching that metric $m_{(S)}$. Now if we write it in terms of the metric $g^E_{\mu\nu}$ we have
\be 
{\cal L}^{(d)}_M = e^{-2\sigma'} \left[ -g^{E,\mu\nu}\partial_{\mu} \varphi \partial_{\nu} \varphi + \left(e^{\sigma'}m_{(S)}\right)^2 \varphi^2  \right] \;.
\ee 
The overall factor behaves like the Ricci scalar $R^{(d)}$ in (\ref{einframaction}), so that the relation with the matter Lagrangian is maintained. However, the mass of the scalar field depends on which metric we use to measure it, so with respect the the metric $g^E_{\mu\nu}$ we can write
\be 
m_{(E)} = e^{\sigma'}m_{(S)} \;.
\label{weymasrel}
\ee 
This is an example of the general statement that dimensionful quantities, such as masses $M$ or radii $R$ , depend on the metric we use to measure them. We can see this by noting that a change in the metric can be undone with a change in the coordinates, and rescaling the time coordinate will change how energies and masses are measured, while rescaling space coordinates will change how distances are measured. So when we state the mass of an object, we should specify with respect to which metric it is measured.

%%%%%%%%%%%%%%%%%%%%%%%%%%%%%%%%%%%%%%%%%%%%%%%%%%%%%%%%%%%%%%%%%%%%%%%%%%%%%%%%%%%%
\subsection{Embedding in M-theory}
\label{sec:embmt}
%%%%%%%%%%%%%%%%%%%%%%%%%%%%%%%%%%%%%%%%%%%%%%%%%%%%%%%%%%%%%%%%%%%%%%%%%%%%%%%%%%%%

The most natural setting to perform the integrating out calculation is in M-theory. It is therefore also natural to write everything in units of the M2 tension, so the mass scale $M_2$ in (\ref{m2def}). Indeed, the calculation in \cite{Dedushenko:2014nya} is performed in such units (setting $M_2=1$). In this section, we consider the embedding into M-theory, as in \cite{Dedushenko:2014nya}, and use such units. In section \ref{sec:embiiast} we will consider instead an embedding into type IIA string theory, in that case the natural units will be using $M_s$ in (\ref{msdef}). 

\subsubsection*{Reducing on the Calabi-Yau to five dimensions}

From the M-theory perspective, we first would like to compactify on a Calabi-Yau manifold $Y$ to five dimensions, and then on a circle to four dimensions. Let us expand the Kahler form $J$ of the Calabi-Yau in a two-form basis
\be 
J = s^i \omega_i \;\;,\;\;\; i=1,...,h^{1,1}(Y) \;.
\ee 
The Kahler form integrated over a two-cycle ${\cal C}_i$ gives the volume of the cycle measured in units of $M_2$
\be 
M_2^2\;\mathrm{Vol\;} {\cal C}_i  = \int_{{\cal C}_i} J = s^i\;,
\label{volci}
\ee 
where we have chosen a dual two-cycle basis to the $\omega_i$ such that $\int_{{\cal C}_i} \omega_j=\delta^i_j$ . The volume of the Calabi-Yau manifold $Y$, in units of $M_2$, is given by
\be 
{\cal V}^{M_2}_Y = \frac16 \int_{Y} J \wedge J \wedge J = \frac16 \kappa_{ijk} s^i s^j s^k \;,
\label{vym2def}
\ee 
where we defined the triple intersection numbers
\be 
\kappa_{ijk} =  \int_{Y} \omega_i \wedge \omega_j \wedge \omega_k \;.
\ee

The compactification results in a five-dimensional supergravity. The vector multiplets are composed of a real scalar, and a vector field. The vector fields arise from expanding the M-theory three-form $C_3$ in the same two-form basis
\be 
C_3 = V^i \wedge \omega_i \;. 
\ee 
Similarly to the four dimensional supergravity described in section \ref{sec:4dsugra}, there are only $h^{1,1}-1$ vector multiplets. One combination of the $V^i$ is embedded into the gravity multiplet, and forms the graviphoton. The physical scalars $s^i$ are reduced by one by writing them in terms of $h^i$ , defined as
\be 
h^i = \left({\cal V}^{M_2}_Y \right)^{-\frac13} s^i  \;.
\label{hidef}
\ee 
And then from (\ref{vym2def}) we have the constraint
\be 
\frac16 \kappa_{ijk} h^i h^j h^k = 1 \;, 
\label{5dconstr}
\ee 
which reduces the degrees of freedom by one. So, in other words, varying the $h^i$ under the constraint (\ref{5dconstr}) keeps the volume fixed. So the volume is not part of the vector multiplets, and instead is in a hypermultiplet.

Reducing the eleven-dimensional Ricci scalar gives
\be 
\frac{\left(M_2\right)^9}{\left(2\pi\right)^2} \int_{\mathbb{R}^{1,3}\times S^1 \times Y} d^{11}x\sqrt{-g^{(11)}} R^{(11)} = \frac{\left(M_2\right)^3}{\left(2\pi\right)^2} {\cal V}^{M_2}_Y \int_{\mathbb{R}^{1,3}\times S^1} d^{5}x\sqrt{-g^{(11)}} R^{(11)} \;.
\ee 
To go to the Einstein frame we therefore should identify
\be 
g^{(11)}_{MN} = \left({\cal V}^{M_2}_Y \right)^{-\frac23} g^{E,5}_{MN} \;,
\label{welres115}
\ee 
with $M,N$ five-dimensional indices. In the Einstein frame, using (\ref{wetlresctran}), we then have
\be 
\frac{\left(M_2\right)^3}{\left(2\pi\right)^2} \int_{\mathbb{R}^{1,3}\times S^1} d^{5}x\sqrt{-g^{(E,5)}} R^{(E,5)} \;.
\ee

\subsubsection*{Reducing on the circle to four dimensions}

Next we need to reduce on the circle, with radius $R_5$, to four dimensions. We therefore have 
\be 
\frac{\left(M_2\right)^3}{\left(2\pi\right)^2} \int_{\mathbb{R}^{1,3}\times S^1} d^{5}x\sqrt{-g^{(E,5)}} R^{(E,5)} = \frac{\left(M_2\right)^3 R_5}{2\pi} \int_{\mathbb{R}^{1,3}} d^{4}x\sqrt{-g^{(E,5)}} R^{(E,5)} \;.
\ee 
To go to the Einstein frame we identify
\be 
g^{(E,5)}_{\mu\nu} = \left(R_5 M_2\right)^{-1} g^E_{\mu\nu} \;,
\label{5to4wres}
\ee 
with $\mu,\nu$ four-dimensional indices.
We then arrive at
\be 
\frac{\left(M_2\right)^2}{2\pi} \int_{\mathbb{R}^{1,3}} d^{4}x\sqrt{-g^{E}} R^{(E)}  \;.
\ee

We can now determine some of the parameters of the embedding of the integrating out calculation into M-theory. The Weyl rescaling (\ref{5to4wres}) is nothing but the metric relation (\ref{5dmetim}), which fixes
\be 
e^{\sigma_{M_2}} = R_5 M_2 \;.
\ee 
Here we place an $M_2$ subscript on $\sigma$ to remind us that it is $\sigma$ when evaluated in $M_2$ units.
In terms of the mass scale $M_5$ introduced in (\ref{esigr5m5de}), we therefore have
\be 
M_5 = M_2 \;.
\ee 

We can also read off the overall factor in the four-dimensional action, as in (\ref{IIA}), which gives
\be 
M_p^2 = \frac{\left(M_2\right)^2}{\pi}  \;.
\label{mpinm2unt}
\ee 
 
In terms of the vector multiplets. The five-dimensional vectors $V^i$ directly lead to four-dimensional vectors $A^i$. Each vector in the five-dimensional multiplet also has a Wilson line along the circle direction, and so leads to a pseudo-scalar $\alpha^i$. These pair up with the real five-dimensional scalar $h^i$ to form the complex scalar of a four-dimensional vector multiplet. In terms of the $Z^i$ of section \ref{sec:4dsugra}, we have the relation \cite{Dedushenko:2014nya}
\be 
Z^i = \alpha^i + i e^{\sigma} h^i \;.
\label{zitoah}
\ee 
The constraint (\ref{5dconstr}) then reads
\be 
\frac16 \kappa_{ijk} \;\mathrm{Im\;}Z^i \;\mathrm{Im\;}Z^j \;\mathrm{Im\;}Z^k = e^{3\sigma}\;.
\label{relsigvol}
\ee 
Then we see that the $Z^i$ can be taken as unconstrained, by appropriately adjusting $e^{\sigma}$, leading to the $h^{1,1}$ independent vector multiplets. 

Finally, we should determine the relation between $R_5$ and the M-theory circle radius in eleven dimensions $R_M$. They are related by the metric rescaling (\ref{welres115}), so they are measured relative to different metrics, as discussed in section \ref{sec:scalbps}. The relation is then
\be 
R_5 = \left( {\cal V}^{M_2}_Y \right)^{\frac13} R_M \;,
\ee 
which gives
\be 
e^{\sigma_{M_2}} =  \left({\cal V}^{M_2}_Y  \right)^{\frac13} R_M M_2 \;.
\ee

\subsubsection*{BPS states and central charges}

Wrapping M2 branes on the two-cycles ${\cal C}_i$ leads to five-dimensional BPS states. Their charges under the five-dimensional gauge fields $V^i$ are given by the wrapping numbers, and denoted as $\beta_i$. This is a BPS state, and its mass is then given by the central charge $\xi(\beta) = \beta_i h_i$ of the five-dimensional supersymmetry algebra \cite{Dedushenko:2014nya}
\be 
M^{(5)}_{BPS} = \beta_i h^i M_2 \;.
\label{centbps5}
\ee  
As a check, we can compare this to a direct evaluation of the M2 brane tension wrapping a two-cycle ${\cal C}_i$. The eleven-dimensional mass is
\be 
M^{(11)}_{BPS} = T_{M2} \; \beta_i\mathrm{Vol\;}{\cal C}_i = \beta_i s^i M_2  = \left( {\cal V}^{M_2}_Y \right)^{\frac13}\beta_i h^i M_2 \;.
\ee 
The five-dimensional mass is then given by a Weyl rescaling (\ref{welres115}), which gives precisely (\ref{centbps5}). 

The four-dimensional mass is related to the five-dimensional mass (\ref{centbps5}) by the Weyl rescaling (\ref{5to4wres}), and so reads
\bea 
M^{M_2}_{BPS} &=& e^{-\frac{\sigma_{M_2}}{2}}  \left|\beta_i h^i  - i e^{-\sigma_{M_2}} n - i e^{-\sigma_{M_2}} \beta_i \alpha^i  \right|M_2  \nonumber \\
&=& \frac{1}{\left({\cal V}^{M_2}_Y R_M M_2 \right)^{\frac12} } \left|\beta_i Z^i + n \right| M_2\;.
\label{mbpsfrom5dm2}
\eea  
Here we have introduced the contribution from the Kaluza-Klein momentum $n$, which contributes a mass $\frac{n}{R_5 M_2} = e^{-\sigma_{M_2}}n$.

We can compare this to the BPS mass (\ref{emmapse}) in the four-dimensional supergravity analysis of section \ref{sec:4dsugra}. Using (\ref{x0tok}), the central charge (\ref{ccX}) can be written as
\be 
Z\left(q\right) =-i \left|2X^0\right|\left(\beta_i Z^i + n \right) \;,
\label{cencxomod}
\ee 
where we identified
\be 
q_I = \left(\beta_i,n\right) \;.
\ee 
The BPS mass is therefore given by 
\be 
M^{M_2}_{BPS} = \left|2X^0\right|\left|\beta_i Z^i + n \right| M_Y\;.
\ee 
We then see that they match under the identification
\be 
\left|2X^0 \right| = e^{-\frac{3\sigma_{M_2}}{2}} = \frac{1}{\left({\cal V}^{M_2}_Y R_M M_2 \right)^{\frac12} }  \;,
\ee 
and 
\be 
M_Y = M_2 \;.
\label{myinm2unit}
\ee 

This completes the embedding of the integrating out calculation into M-theory, reduced on a Calabi-Yau, and then again on a circle. For completeness, we can also give the leading order (in Im $Z^i$) form of the prepotential
\be 
F_0^{(S)} = -\frac{\kappa_{ijk}}{6} \frac{X^i X^j X^k}{2X^0} \;.
\ee 
Note that there is an extra factor of 2 in the denominator, relative to usual conventions, because of the factor of 2 in $X^0$ in (\ref{x0tok}). 

%%%%%%%%%%%%%%%%%%%%%%%%%%%%%%%%%%%%%%%%%%%%%%%%%%%%%%%%%%%%%%%%%%%%%%%%%%%%%%%%%%%%
\subsection{Embedding in type IIA string theory}
\label{sec:embiiast}
%%%%%%%%%%%%%%%%%%%%%%%%%%%%%%%%%%%%%%%%%%%%%%%%%%%%%%%%%%%%%%%%%%%%%%%%%%%%%%%%%%%%

In the type IIA setting, we compactify on a Calabi-Yau $Y$ directly to four dimensions. The natural scale is not the M2 tension, but the fundamental string tension. We therefore work in units of $M_s$. 

The steps to follow are similar to the M-theory case in section \ref{sec:embmt}. We expand the Kahler form $J$, and the NS two-form $B_2$, in two-form basis
\be
    J= \sum_{i}\frac{v^i}{2\pi}\omega_i\;\;\;\;,\;\;\;\;\; B_2 = -\sum_{i} \frac{b^i}{2\pi}\omega_i\;.
    \label{jbexpiia}
\ee
The fields $v^i$ now measure the volumes of two-cycles in string units
\be 
v^i = 2\pi M_s^2\;\mathrm{Vol\;} {\cal C}_i \;.
\label{volcisu}
\ee 
The scalars in (\ref{jbexpiia}) combine into the complex fields 
\be 
t^i = v^i + i b^i \;.
\ee 
These are related to the $Z^i$ of section \ref{sec:4dsugra} by
\be 
Z^i = i \frac{t^i}{2\pi} = \frac{1}{2\pi}\left(-b^i+iv^i\right) \;.
\ee 
The $Z^i$ match the usual notation for the scalar components of the vector multiplets in the type IIA literature (see, for example, \cite{Gurrieri:2003st,Grimm:2004ua,Palti:2008mg}).

The gauge fields in the vector multiplets arise from the expansion of the Ramond-Ramond three-form $C_3$ , and one-form $C_1$ , as
\be 
    C_3= \sum_{i}A^i \wedge \omega_i \;\;\;, \;\;\;\; C_1 = A^0 \;.
\ee
Then the $A^{I}_{\mu}$ in section \ref{sec:4dsugra}, with $I \in \left\{ 0 , i \right\}$ , are given by 
\be 
A^I = \left(A^0,A^i+\frac{b^i}{2\pi}A^0\right) \;.
\label{gaugeiiared}
\ee

\subsubsection*{Direct reduction on the Calabi-Yau}

We can now dimensionally reduce the type IIA action
\be 
\frac{2\pi M_s^8}{g_s^2} \int_{\mathbb{R}^{1,3}\times Y} d^{10}x\sqrt{-g^{S}}\; R^{S} = \frac{2\pi M_s^2}{g_4^2} \int_{\mathbb{R}^{1,3}} d^{4}x\sqrt{-g^{S}}\; R^{S} \;,
\ee 
where we define the four-dimensional dilaton
\be 
g_4 = g_s \left({\cal V}_Y \right)^{-\frac12} \;,
\ee 
with ${\cal V}_Y$ the Calabi-Yau volume measured in units of $M_s$. It can be written as
\be 
{\cal V}_Y = \frac16 \left(\frac{1}{2\pi}\right)^3\kappa_{ijk} v^i v^j v^k \;.
\ee 
To go to the Einstein frame we take
\be 
g^S_{\mu\nu} = g_4^2\;g^E_{\mu\nu} \;,
\label{stru10to4f}
\ee 
which leads to
\be 
2\pi M_s^2 \int_{\mathbb{R}^{1,3}} d^{4}x\sqrt{-g^{E}}\; R^{E} \;.
\label{4deinsac}
\ee 
We therefore obtain the expression for $M_p$ in (\ref{IIA}) as
\be 
M_p^2 = 4\pi M_s^2 \;.
\label{mpstr4dfr}
\ee

\subsubsection*{Relation to the five-dimensional solution}
 
To understand the relation to the five dimensional solution, we need to do the reduction of M-theory to five dimensions, and then again on a circle, but unlike in section \ref{sec:embmt}, we need to do it in string units $M_s$. 

Starting from the eleven-dimensional action we have
\be 
\frac{2\pi M_s^9}{g_s^3} \int_{\mathbb{R}^{1,3}\times S^1 \times Y} d^{11}x\;\sqrt{-g^{(11)}} R^{(11)} = \frac{2\pi M_s^3}{g_s^3} {\cal V}_Y \int_{\mathbb{R}^{1,3}\times S^1} d^{5}x\;\sqrt{-g^{(11)}} R^{(11)} \;.
\ee 
To go to the Einstein frame we set
\be 
g^{(11)}_{MN} = g_s^2\left({\cal V}_Y \right)^{-\frac23} g^{E,5}_{MN} \;,
\label{welresmstr}
\ee 
which gives
\be 
2\pi M_s^3 \int_{\mathbb{R}^{1,3}\times S^1} d^{5}x\sqrt{-g^{(E,5)}} R^{(E,5)} = \left(2\pi\right)^2 M_s^3 R_5 \int_{\mathbb{R}^{1,3}} d^{4}x\sqrt{-g^{(E,5)}} R^{(E,5)}\;.
\ee 
Then the four-dimensional Einstein frame is given by
\be 
g^{E,5}_{\mu\nu} = \left(2\pi R_5 M_s\right)^{-1} g^E_{\mu\nu} \;,
\label{einto4ds}
\ee 
yielding (\ref{4deinsac}). From (\ref{einto4ds}) we therefore read off
\be 
e^{\sigma_{M_s}} = 2\pi R_5 M_s = \left({\cal V}_Y \right)^{\frac13}\frac{2\pi R_M M_s}{g_s} = \left({\cal V}_Y \right)^{\frac13} \;,
\label{sigmsdef}
\ee 
where we used that $R_5$ and $R_M$ are related by the metric rescaling (\ref{welresmstr}), and the relation (\ref{gsrmrel}). Comparing with (\ref{esigr5m5de}), we have the identification
\be 
M_5 = 2\pi M_s \;.
\ee 
In (\ref{sigmsdef}) we place an $M_s$ subscript on $\sigma$ to remind us that it is $\sigma$ in string units $M_s$.

\subsubsection*{BPS masses and wrapped branes}

The mass of the BPS states, so the D2-D0 branes wrapping the two cycles, can be determined either by direct evaluation or by writing the results of section \ref{sec:embmt} in string units (rather than M2 units). Converting the results requires using the relation for the four-dimensional Einstein frame metrics
\be 
g^{E,M_2}_{\mu\nu} =  \left(2\pi\right)^2 \left(\frac{g_s}{2\pi} \right)^{\frac23} g^{E,M_s}_{\mu\nu} \;.
\ee
The other quantities are related by
\be 
{\cal V}^{M_2}_Y = \left(\frac{2\pi}{g_s}\right)^2{\cal V}_Y \;,\;\; h^i = \frac{v^i}{2\pi} \left( {\cal V}_Y\right)^{-\frac13} \;,\;\; e^{\sigma_{M2}} = \left( {\cal V}_Y\right)^{\frac13}\;.
\ee 
We then obtain that (\ref{mbpsfrom5dm2}) takes the form in string units of
\be 
M_{BPS}^{M_s} = 2\pi \left(\frac{g_s}{2\pi} \right)^{\frac13} M_{BPS}^{M_2} = \frac{2\pi}{\left({\cal V}_Y\right)^{\frac12}} \left|\beta_i Z^i + n \right|M_s\;,
\label{bpsm4dsun}
\ee 
where we used the usual charge decomposition
\be 
q_I = \left(\beta_i,n\right) \;,
\ee
with respect to the gauge fields (\ref{gaugeiiared}).
We can check that this matches the result of direct dimensional reduction. The BPS state mass in ten-dimensions is, using (\ref{tenvolcyc}) and (\ref{dbranten}), given by
\be 
M_{BPS}^{(S)} = \frac{1}{g_s} \beta_i v^i M_s \;.
\ee 
Then performing the rescaling (\ref{stru10to4f}), we obtain
\be 
M_{BPS}^{M_s} = \frac{1}{\left({\cal V}_Y\right)^{\frac12}} \beta_i v^i M_s \;.
\ee 
This matches the full expression (\ref{bpsm4dsun}). Comparing with (\ref{emmapse}) we then obtain the relations
\be 
\left|2X^0 \right| =  \left({\cal V}_Y\right)^{-\frac12} \;\;,\;\;\;M_Y = 2\pi M_s \;.
\label{mytomsconv}
\ee 
It is worth noting that using these, the central charge (\ref{cencxomod}) takes the form
\be 
Z\left(q\right) =-i \left({\cal V}_Y\right)^{-\frac12}\left(\beta_i Z^i + n \right)  = \frac{1}{2\pi\left({\cal V}_Y\right)^{\frac12}}\left(\beta_i t^i -2\pi i n \right)\;,
\label{cenchargstemb}
\ee 
We have then completed the embedding of the general integrating out calculation into the type IIA setting, in string units.

\subsubsection*{$X^0$ and the conformal gauge choice}

Recall that in the supergravity formulation, there is a rescaling symmetry (\ref{resX}). This symmetry is fixed by the constraint (\ref{constraint}). The constraint fixes a gauge for the symmetry, but it is possible to choose other gauges. The specific gauge (\ref{constraint}) is most appropriate for the superfield notation. It ensures that the action can be written in superspace in the form (\ref{f0ftermac}), and in this form leads to the Einstein frame (so no fields multiplying the Ricci scalar). From the topological string perspective, it is the most appropriate gauge. 

Typically, in studying dimensional reduction of type IIA string theory, for example in \cite{Gurrieri:2003st,Grimm:2004ua,Palti:2008mg}, a different gauge is used. This is because one does not typically utilize the superspace formulation, but works with component fields. The gauge that is usually used is one where $X^0$ is a constant (usually one). In our case, we can write 
\be 
X^{0,\mathrm{com}} = \frac12 X^0\;\;, \;\;\; X^0 = 1 \;.
\label{iiacompgague}
\ee 
where the superscript on $X^{0,\mathrm{com}}$ denotes that it is in component fields gauge, and we have introduced $X^0=1$ in (\ref{iiacompgague}) so as to match the usual conventions.
In this gauge, we can write the Kahler potential (\ref{kahlpot}) as
\be 
e^{-K} = i \left(f_I \bar{X}^I - \bar{f}_I X^I \right) \;,
\label{kapotga}
\ee 
and the prepotential (at large $v^i$) as
\be 
F_0^{(S)} = -\frac{\kappa_{ijk}}{6}\frac{X^iX^jX^k}{2X^{0,\mathrm{com}}} = - \frac{\kappa_{ijk}}{6}\frac{X^iX^jX^k}{X^{0}} \;.
\label{prepoxog}
\ee 
This yields the familiar relation 
\be 
e^{-K} = 8 {\cal V}_Y \;.
\ee 
Note that the $Z^i$ are gauge independent, which is expected since they measure physical volumes. So any expression written in terms of the $Z^i$ holds in any gauge.

%%%%%%%%%%%%%%%%%%%%%%%%%%%%%%%%%%%%%%%%%%%%%%%%%%%%%%%%%%%%%%%%%%%%%%%%%%%%%%%%%%%%
\section{The map to the topological string}
\label{sec:map2top}
%%%%%%%%%%%%%%%%%%%%%%%%%%%%%%%%%%%%%%%%%%%%%%%%%%%%%%%%%%%%%%%%%%%%%%%%%%%%%%%%%%%%

Putting together the final result (\ref{fullhyper}) of section \ref{sec:intm2}, with the embedding into M-theory and type IIA string theory developed in section \ref{sec:embiia}, we arrive at the contribution of the integrating out to the four-dimensional supergravity arising from dimensional reduction of the higher dimensional theories. 

We should fix our units of measurement now. We would like to write the effective action as a superspace integral over the prepotential (\ref{clstFgv}). This leads to an action of the form (\ref{IIA}) as long as we utilize appropriate units for factoring out $M_p^2$ from the matter Lagrangian. The simplest way to fix these units is to demand that the mass of BPS states be given by the central charge of the supersymmetry algebra. So we choose our units such that
\be 
M_{BPS} = \left|Z\right| \;.
\ee 
In the language of section \ref{sec:embiia}, this corresponds to choosing $M_Y=1$. In M-theory units this means $M_2=1$ (\ref{myinm2unit}), while in string theory units it means $M_s=\frac{1}{2\pi}$ (\ref{mytomsconv}). Both of these choices fix the four-dimensional Planck scale, see (\ref{mpinm2unt}) and (\ref{mpstr4dfr}), to be
\be 
M_p^2 = \frac{1}{\pi} \;.
\label{4dplanckscun}
\ee 

Using these units, in (\ref{fullhyper}), we obtain the contribution from a hypermultiplet to the effective action of 
\be 
\Delta \Gamma^{M_2=1}_H = \frac{1}{64\left(2\pi\right)^2} \int d^4x \;d^4\theta \;\sqrt{g^E}\int^{\infty}_{\epsilon} \frac{ds}{s} \;e^{-s \left|Z(q)\right|^2}\; \frac{\left(-W^2\right) }{\left(2 \sin \left(\frac{s \bar{Z}(q) \sqrt{ - W^2 }}{8}\right)\right)^{2-2r}} \;.
\label{fullhyperm2}
\ee 
Note that this matches precisely the relevant expression in \cite{Dedushenko:2014nya}.
We can work with type IIA quantities, so the $t^i$, and bring this to a form that is more convenient for comparison with the topological string by rescaling
\be 
s \rightarrow s \left(\left(2\pi\right)^2 {\cal V}_Y \right) \;,
\ee 
and defining 
\be 
z_{\beta,n} = \beta_i t^i - 2 \pi i n \;.
\label{zbndefto}
\ee 
We also introduce the topological string coupling
\be 
\lambda = \frac{\pi}{2} \left({\cal V}_Y \right)^{\frac12} \sqrt{ - W^2 } = \frac{\pi}{4} \left(\frac{W^2}{\left(X^0\right)^2} \right)^{\frac12}\;.
\label{topstrcpdef}
\ee 
Then, using (\ref{cenchargstemb}) and (\ref{mytomsconv}) , (\ref{fullhyperm2}) can be written as
\be 
\Delta \Gamma_H^{\lambda} = -\frac{1}{\left(2\pi\right)^4} \int d^4x \;d^4\theta \;\sqrt{g^E}\;\left(X^0\right)^2 \lambda^2\;\int^{\infty}_{\epsilon} \frac{ds}{s} \frac{\;e^{-s\left|z_{\beta,n} \right|^2}}{\left(2 \sin \left(\frac{s \bar{z}_{\beta,n} \lambda}{2}\right)\right)^{2-2r}} \;.
\label{delgamlamH}
\ee 
We recall that this expression should be understood as to be evaluated with the full superfields, so $X^I \rightarrow \mathcal{X}^I$ as in (\ref{n2boschmul}) , and $W \rightarrow \mathcal{W}$ as in (\ref{gravsupf}). 

Let us compare this with the effective action (\ref{clstFgv}). We can write (\ref{clstFgv}) as\footnote{Note that we should rotate to Euclidean time ${\cal L}\rightarrow -{\cal L}^E$. However, we can keep the same superspace action (\ref{clstFgv}).}
\bea
{\cal L}^E_{\mathrm{eff}} &=& -\frac{i}{\pi} F^{(S)}\left(X^I,W \right)\nonumber \\
&=& -\frac{i}{\pi}\sum_{g=0}^{\infty} F^{(S)}_g\left(X^I\right) \left(\frac{\pi^2}{16}W^{2}\right)^g \nonumber \\
&=& -\frac{i}{\pi}\sum_{g=0}^{\infty} F^{(S)}_g\left(Z^I\right) \left(X^0\right)^{2-2g} \left(\frac{\pi}{4}\sqrt{W^{2}}\right)^{2g} \nonumber \\
&=& -\frac{i}{\pi}\left(X^0\right)^{2} \lambda^2\sum_{g=0}^{\infty} F^{(S)}_g\left(Z^I\right)  \lambda^{2g-2} \;,
\label{pertexpaninw}
\eea 
where we have utilized that $F^{(S)}_g$ are homogeneous of degree $2-2g$ in the $X^I$. We have also replaced the superfields by their bosonic representatives for clarity of notation.
So the effective Euclidean action, as an expansion in $\lambda$, can be written as
\be 
S^E_{\mathrm{eff}} = -\frac{1}{\left(2\pi\right)^4}\int d^4x \;d^4\theta \;\sqrt{g^E}\;\left(X^0\right)^2 \lambda^2\; \left(2\left(2\pi\right)^3 i \sum_{g=0}^{\infty} F^{(S)}_g(Z^I)\lambda^{2g-2} \right) \;.
\label{eucactmapfor}
\ee 

The action (\ref{eucactmapfor}) gives a perturbative expansion in $\lambda$, but we can define the full non-perturbative topological string free energy $F(t,\lambda)$ , as in (\ref{tsfe}) , through it by sending 
\be 
2\left(2\pi\right)^3 i\sum_{g=0}^{\infty} F^{(S)}_g(Z^I)\lambda^{2g-2} \rightarrow F(t,\lambda) \;.
\ee  
Equivalently, we can write the relation
\be 
\frac{2\left(2\pi\right)^3 i}{\left(X^0\right)^{2}}\;F^{(S)}\left(Z^I,\lambda\right) = \lambda^2\;F(t,\lambda) \;.
\ee 
With this definition, we see that integrating out a hypermultiplet, as in (\ref{delgamlamH}), gives a contribution to the topological string free energy of
\be 
F(t,\lambda)_{H_{\beta,n,r}} = \int^{\infty}_{\epsilon} \frac{ds}{s} \frac{\;e^{-s\left|z_{\beta,n} \right|^2}}{\left(2 \sin \left(\frac{s \bar{z}_{\beta,n} \lambda}{2}\right)\right)^{2-2r}} \;.
\label{tpfreerbn}
\ee 
Each wrapped M2 brane gives one contribution of the form (\ref{tpfreerbn}). The full contribution from integrating out M2 branes is therefore the sum over these. For each value of $\left\{\beta,n,r\right\}$, we say that we have $\alpha^{\beta}_r$ branes (the multiplicity with respect to $n$ is always one). Then the full contribution is 
\be 
F(t,\lambda) = \sum_{\beta,r\geq0,n\in\mathbb{Z}} \alpha^{\beta}_r \int^{\infty}_{\epsilon} \frac{ds}{s} \frac{\;e^{-s\left|z_{\beta,n} \right|^2}}{\left(2 \sin \left(\frac{s \bar{z}_{\beta,n} \lambda}{2}\right)\right)^{2-2r}} \;.
\label{tpfreefullm2}
\ee 
This is the main result of the notes so far: it gives the form of the topological string free energy, defined through integrating out M2 branes. The $\alpha_r^{\beta}$ are the famous Gopakumar-Vafa invariants \cite{Gopakumar:1998ii,Gopakumar:1998jq}. However, the expression (\ref{tpfreefullm2}) differs in an important way from the results of  \cite{Gopakumar:1998ii,Gopakumar:1998jq}, and this is discussed in more detail in section \ref{sec:evalfor}.

\subsubsection*{String frame couplings}

Before proceeding, it is worth noting an important fact about the perturbative expansion (\ref{pertexpaninw}). We see that there is no explicit dependence on the string coupling. This means that all the dependence on the string coupling comes from transforming the metric in the contraction $W^2$ in the topological string coupling (\ref{topstrcpdef}). Therefore, the dependence on the string coupling at each order in $\lambda$ is fixed exactly. This allows us to go to strong string coupling, while maintaining the topological string expansion. 

To see the dependence on the string coupling we need to transform back to the four-dimensional string frame by undoing the transformation (\ref{stru10to4f}). This means that we write
\be 
W^2 = W_{\mu\nu}^-W_{\rho\sigma}^- g^{E,\mu\rho} g^{E,\nu\sigma} =  W_{\mu\nu}^-W_{\rho\sigma}^- g^{S,\mu\rho} g^{S,\nu\sigma} \left(g_4\right)^4 \;.
\label{w2gscolu}
\ee  
It is informative to define also
\be
W^{S,-}_{\mu\nu} = g_s W_{\mu\nu}^- \;\;,\;\;\; \left(W^S\right)^2 = W_{\mu\nu}^{S,-}W_{\rho\sigma}^{S,-} g^{S,\mu\rho} g^{S,\nu\sigma} \;,
\label{Wsdef}
\ee
which allows us to write the topological string coupling as
\be 
\lambda = \frac{\pi}{2} \;g_4 \;\sqrt{ - \left(W^S\right)^2 } \;\;\;.
\label{lamstfrm}
\ee 
We can now see the string perturbation expansion explicitly: each term in the expansion (\ref{pertexpaninw}) comes with a power of $\left(g_4\right)^{2g}$, which is the string coupling associated to the genus expansion of string theory. There is also the factor of $g_s$ in (\ref{Wsdef}). This is the correct assignment for the vertex operator of the graviphoton, because it is a Ramond-Ramond field which comes with an extra power of $g_s$. 

\subsubsection*{Imaginary couplings}

It is worth noting that in the final result (\ref{tpfreefullm2}), the string coupling $g_s$ appears only through (\ref{w2gscolu}), and so it also makes sense to take it imaginary. In that case, we would have  $\left(W^S\right)^2 > 0$ , and so we should write (\ref{lamstfrm}) as
\be 
\lambda = i\frac{\pi}{2} \;g_4 \;\sqrt{\left(W^S\right)^2 } \;\;\;.
\label{lamstfrmimg}
\ee 
Sometimes the relation between the topological string coupling and the string coupling is written with an $i$, and the string coupling taken imaginary. We can have similar factors of $i$ appearing if we rotate to Minkowski space, rather than Euclidean space. Or if we take the graviphoton field strength as real rather than imaginary. However, with all these possibilities, one should really go back to the starting solution (\ref{5dmetim}), and repeat all the analysis of these notes accordingly. It is likely that there are various subtleties which arise in such cases. On the other hand, we also had to deal with an imaginary graviphoton field strength, so some subtlety seems unavoidable. The analysis in these notes, in Euclidean signature and real string coupling, seems to us the most feasible route to the topological string free energy.

\section{Evaluating the Free Energy}
\label{sec:evalfor}
%%%%%%%%%%%%%%%%%%%%%%%%%%%%%%%%%%%%%%%%%%%%%%%%%%%%%%%%%%%%%%%%%%%%%%%%%%%%%%%%%%%%

In section \ref{sec:map2top}, we calculated the expression for the topological string free energy (\ref{tpfreefullm2}), coming from integrating out M2 branes. In this section, we describe how to evaluate that expression. For convenience, let us reproduce the result here:  
\be 
F\left(t,\lambda\right) = \sum_{\beta,r\geq0,n\in\mathbb{Z}} \alpha_r^{\beta}\int^{\infty}_{\epsilon} \frac{ds}{s} \; \frac{e^{-s |z_{\beta,n}|^2}}{\left(2 \sin \left( \frac{s \bar{z}_{\beta,n} \lambda}{2}\right)\right)^{2-2r}} \;.
\label{prefreeeneg}
\ee 
For completeness, we can summarise the meaning of the different quantities in (\ref{prefreeeneg}). The topological string free energy is a function of the Kahler moduli $t^i$ of the Calabi-Yau manifold $Y$, which are counted by $h^{(1,1)}(Y)$. The topological string coupling is $\lambda$, and is taken real and positive $\lambda \in \mathbb{R}^+$. The sum is over different wrapped M2 brane states. The $\beta_i$ are the non-negative integer wrapping numbers of the M2 branes on the two-cycles associated to the Kahler moduli $t^i$.\footnote{The $\beta=0$ case corresponds to pure $D0$ states, rather than $D2-D0$ bound states.} The non-negative integer $r$ is the genus of the cycle wrapped by the brane. The M2 branes have Kaluza-Klein momentum $n$ along the M-theory circle, which can be positive or negative (or zero) depending on the momentum direction. For the given quantum numbers, $\alpha_r^{\beta}$ are the number of such BPS states (more precisely the index), these are the Gopakumar-Vafa invariants. There is a Schwinger integral, with real Schwinger proper time parameter $s$, and an ultraviolet cutoff $\epsilon$. The quantity $z_{\beta,n}$ is related to the central charge, it is defined in (\ref{zbndefto}), and we reproduce it here for convenience
\be 
z_{\beta,n} =  \beta_i t^i - 2\pi i n \;.
\label{zbndefno2}
\ee 

The idea of integrating out the M2 branes to generate the topological string free energy was introduced by Gopakumar and Vafa in their seminal paper \cite{Gopakumar:1998jq}. The expression they obtained differed from (\ref{prefreeeneg}). We now go on to explain, in section \ref{sec:GVasy}, how their result relates to (\ref{prefreeeneg}). We show that by taking the free energy as an asymptotic series in $\lambda$, one can recover the result of \cite{Gopakumar:1998jq}. Following this, in section \ref{sec:GVasyfull}, we show how (\ref{prefreeeneg}) can actually be evaluated exactly, and yields a full function, rather than only an asymptotic series \cite{Hattab:2024ewk}.

The important parts of the analysis will not be affected by the sums over $\beta$ and $r$ in (\ref{prefreeeneg}). We are primarily concerned with the sum over $n$. We therefore restrict at this point to a simple setting where there is only one $t^i$, denoted as $t$, and the only non-vanishing Gopakumar-Vafa invariant is $\alpha_{0}^{1} = 1$ , with all other $\alpha_{r}^{\beta}=0$.\footnote{There is also the $\alpha^0_0$ contribution, which corresponds to pure $D0$ states. We will ignore this contribution for now, and return to it at the end of section \ref{sec:GVasy}.} This is actually the case for a non-compact Calabi-Yau manifold called the resolved conifold. We therefore label the free energy in this simple case as $F^{rc}$ , and write it as
\be 
F^{rc} = \sum_{n \in \mathbb{Z}} \;\int^{\infty}_{\epsilon} \frac{ds}{s} \; \frac{e^{-s |z_{n}|^2}}{\left(2 \sin \left( \frac{s \bar{z}_{n} \lambda}{2}\right)\right)^{2-2r}} \;,
\label{prefreecon}
\ee 
with
\be 
z_{n} =  t - 2\pi i n \;\;, \;\;\; t= v+ ib\;.
\label{zrcsim}
\ee 
We proceed to work with the free energy $F^{rc}$ in (\ref{prefreecon}), and will reinstate the sum over $\beta$ and $r$, as well as the general multi-moduli case, at the end.

%%%%%%%%%%%%%%%%%%%%%%%%%%%%%%%%%%%%%%%%%%%%%%%%%%%%%%%%%%%%%%%%%%%%%%%%%%%%%%%%%%%%
\subsection{The Gopakumar-Vafa asymptotic series}
\label{sec:GVasy}
%%%%%%%%%%%%%%%%%%%%%%%%%%%%%%%%%%%%%%%%%%%%%%%%%%%%%%%%%%%%%%%%%%%%%%%%%%%%%%%%%%%%

The free energy should be holomorphic in the $t^i$. The expression (\ref{prefreecon}) is indeed holomorphic, but not manifestly so since it depends on $\bar{z}_{n}$. To make it manifestly holomorphic we can rescale 
\be 
s \rightarrow \frac{s}{\bar{z}_{n}} \;.
\label{ressz}
\ee 
This would make the integrand manifestly holomorphic. However, it also makes the integration variable complex, and within this complex plane the integration path depends on $n$. To keep track of this carefully, instead of (\ref{ressz}), we introduce a complex variable 
\be 
u_n = s \bar{z}_{n}  \;.
\label{defu}
\ee   
Then (\ref{prefreecon}) can be written as 
\be
\label{angles}
    F^{rc} = \sum_{n\in\mathbb{Z}}\int_{0^+}^{\infty e^{i\theta_n}}\frac{\text{d}u}{u}\frac{e^{-uz_n}}{\left(2\sin\left(\frac{u\lambda}{2}\right)\right)^2}\;.
\ee
Because $u_n$ is a dummy variable in the integral, we drop the $n$ subscript and just write $u_n \rightarrow u$. We also set the ultraviolet cutoff to be arbitrarily close to, but not including, the origin
\be 
\epsilon = 0^+ \;.
\label{uvepcut}
\ee
The integral needs to be performed over a line originating at the origin of the complex plane, or more precisely at $0^+$, and going out to infinity at an angle 
\be 
\theta_n =\tan^{-1}\left(\frac{2\pi n-b}{v}\right) \;,
\ee 
which is the argument of $\bar{z}_n$.

\begin{figure}[h]
\centering
\includegraphics[width=0.8\textwidth]{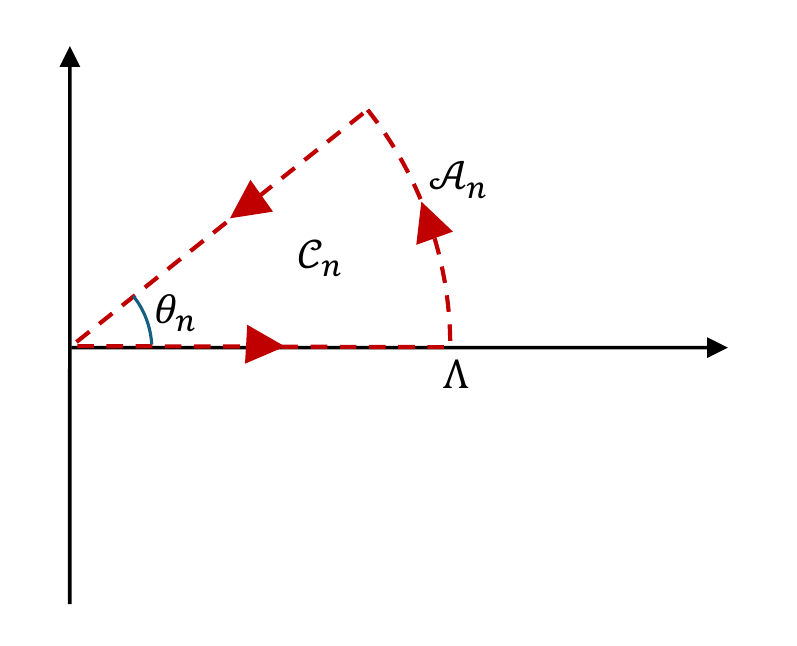}
\caption{Figure illustrating the decomposition of the integration path in (\ref{intpathslice}).}
\label{fig:llslice}
\end{figure}

Next, we would like to perform the sum over $n$ in (\ref{angles}). What we would like to do is bring the sum inside the integral and sum the integrands. However, this is not possible while each of the integrals is over a different integration path in the $u$ plane. Still, since this is complex integration of a meromorphic function, it is possible to deform the paths in the plane under certain conditions. To study these deformations, let us introduce an infrared cutoff to the Schwinger integral. So instead of integrating out to infinity in $s$, or equivalently in $|u|$, we integrate only in the range 
\be 
0 < |u| \leq \Lambda \;.
\ee
Then we can relate an integration line at angle $\theta_n$ , to one along the real line by writing it as 
\be 
\oint_{{\cal C}_{n}} = -\int_{0^+}^{\Lambda e^{i\theta_n}} + \int_{0^+}^{\Lambda} + \; \int_{{\cal A}_{n}} \;\;\;,
\label{intpathslice}
\ee 
where $\oint_{{\cal C}_{n}}$ is a contour integral over the slice, with angle $\theta_n$ , and the two integration lines as boundaries. The piece $\int_{{\cal A}_{n}}$ is the integration along the arc connecting the two lines at radius $\Lambda$. This is illustrated in figure \ref{fig:llslice}. 

\begin{figure}[h]
\centering
 \includegraphics[width=0.8\textwidth]{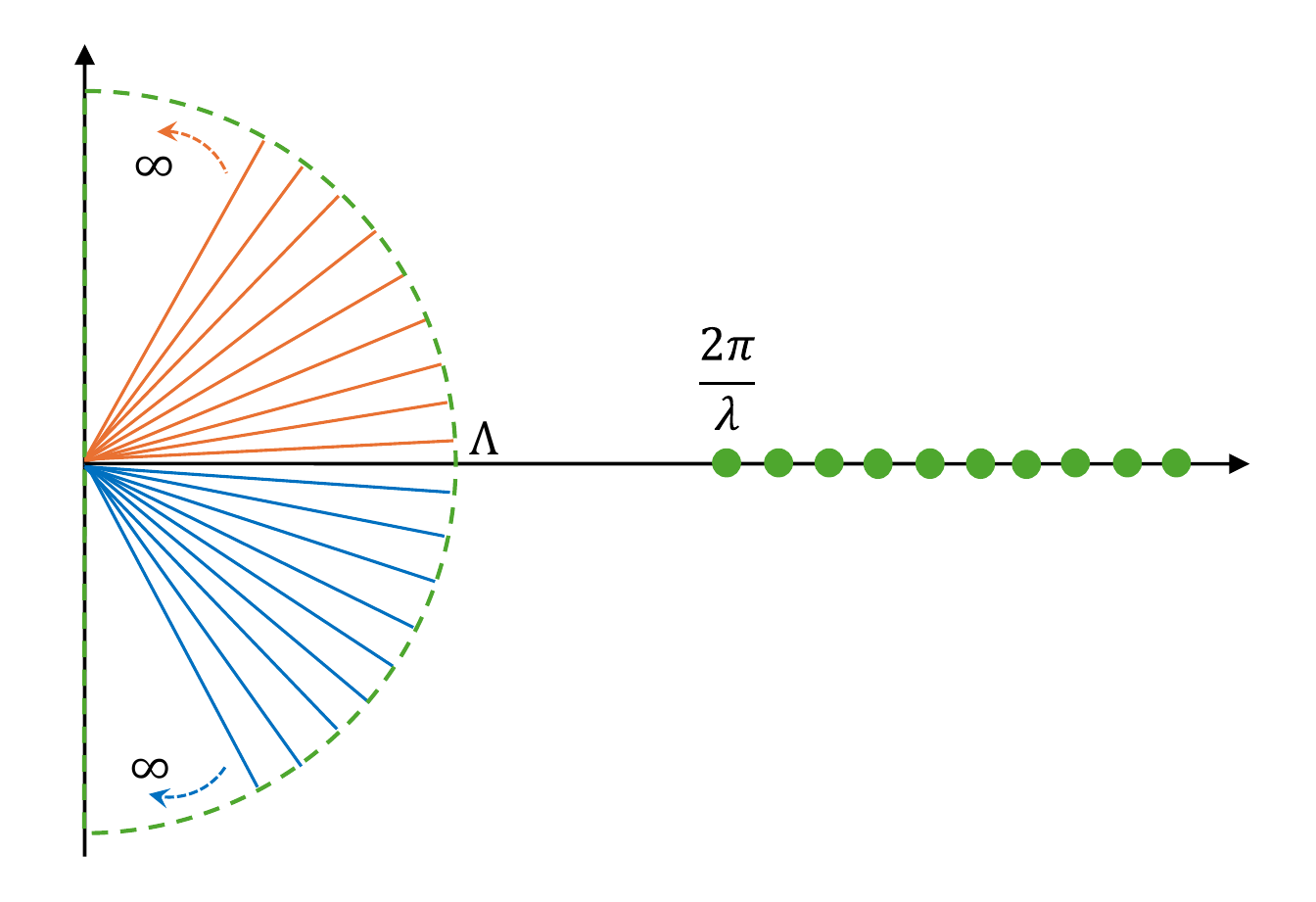}
\caption{Figure showing how the free energy (\ref{angles}) is a sum over different integration lines in the complex $u$ plane. We introduce an infrared cutoff in the Schwinger integral, so that each line only reaches a distance $\Lambda$. The poles (\ref{nppoles}) are located on the real axis and, as long as (\ref{llpolescon}) is satisfied, the integration lines can be rotated to coincide on the real axis without hitting the poles. The contributions from the boundary of the semi-circle are of order $e^{-t \Lambda}$.}
\label{fig:llpoles}
\end{figure}

The closed contour integral $\oint_{{\cal C}_{n}}$ yields the residues at the poles. Therefore, if there are no poles inside the contour, it vanishes. 
The poles of (\ref{angles}) are located on the real axis at
\be 
u \in \left(\frac{2\pi}{\lambda} \right)\mathbb{N}^* \;.
\label{nppoles}
\ee 
If we consider $\lambda \rightarrow 0$, then these poles are sent to infinity, and we can avoid them.
More precisely, as long as 
\be 
\Lambda < \frac{2\pi}{\lambda} \;,
\label{llpolescon}
\ee
the integration contour $\oint_{{\cal C}_{n}}$ has no poles inside. This is illustrated in figure \ref{fig:llpoles}.
Using (\ref{intpathslice}), we can therefore write
\be 
\int_{0^+}^{\Lambda e^{i\theta_n}} = \int_{0^+}^{\Lambda} + \; \int_{{\cal A}_{n}} \;.
\ee 
The integral along the arc $\int_{{\cal A}_{n}}$ will yield a contribution of order
\be 
\int_{{\cal A}_{n}} \frac{\text{d}u}{u}\frac{e^{-uz_n}}{\left(2\sin\left(\frac{u\lambda}{2}\right)\right)^2} \sim {\cal O}\left(e^{-t \Lambda}\right) \;.
\label{contfromarc}
\ee 
We can therefore write (\ref{angles}) as a sum of integrals over the real line, up to the contributions from the arc as in (\ref{contfromarc}), 
\bea
\label{anglesLambda}
    F^{rc}_{\Lambda} &=& \sum_{n\in\mathbb{Z}}\int_{0^+}^{\Lambda}\frac{\text{d}s}{s}\frac{e^{-sz_n}}{\left(2\sin\left(\frac{s\lambda}{2}\right)\right)^2} + {\cal O}\left(e^{-t \Lambda}\right) \nonumber \\
    &=&  \int_{0^+}^{\Lambda}\frac{\text{d}s}{s}\frac{e^{-st}}{\left(2\sin\left(\frac{s\lambda}{2}\right)\right)^2} \left( \sum_{n\in\mathbb{Z}} e^{-2\pi i ns} \right) + {\cal O}\left(e^{-t \Lambda}\right)\;,
\eea
where we sent $u \rightarrow s$, since now we are integrating on the real line. Using Poisson resummation
\be 
\sum_{n\in\mathbb{Z}}e^{2\pi i ns} = \sum_{k\in\mathbb{Z}}\delta(s-k) \;,
\ee 
we can write this as
\bea
F^{rc}_{\Lambda} &=& \sum_{k\in\mathbb{Z}}\int_{0^+}^{\Lambda}\frac{\text{d}s}{s}\frac{\delta(s-k)\;e^{-st}}{\left(2\sin\left(\frac{s\lambda}{2}\right)\right)^2} + {\cal O}\left(e^{-t \Lambda}\right) \nonumber \\
&=& \sum_{k=1}^{k_{\mathrm{max}}} \frac{e^{-k t}}{k\left(2\sin\left(\frac{k\lambda}{2}\right)\right)^2} + {\cal O}\left(e^{-t \Lambda }\right)\;,
\label{sumkLam}
\eea
with $k_{\mathrm{max}}$ being the largest positive integer which is smaller than $\Lambda$, so 
\be 
k_{\mathrm{max}} = \floor{\Lambda} \;.
\label{kmax}
\ee 
The expression (\ref{sumkLam}), with $k_{\mathrm{max}} \rightarrow \infty$ , is the Gopakumar-Vafa result for the resolved conifold \cite{Gopakumar:1998ii,Gopakumar:1998jq}. 

The expression (\ref{sumkLam}) should be understood as an asymptotic series in $\lambda$. To see this, let us take $\lambda \;k_{\mathrm{max}} \ll 1$ , we expand (\ref{sumkLam}) as
\be 
F^{rc}_{\Lambda} = \frac{1}{\lambda^2}\sum_{k=1}^{k_{\mathrm{max}}} \frac{e^{-k t}}{k^3} \left( 1 + \frac{1}{12} \left(\lambda \;k\right)^2  +  \frac{1}{240}\left(\lambda\; k\right)^4 + ...  \right) \;.
\ee 
Then the genus $g$ free energy coefficients in (\ref{tsfe}) are read off as
\be 
F^{rc}_0(t) = \sum_{k=1}^{k_{\mathrm{max}}} \frac{e^{-k t}}{k^3} \;,\;\; F^{rc}_1(t) = \sum_{k=1}^{k_{\mathrm{max}}} \frac{1}{12} \frac{e^{-k t}}{k} \;,\;\; ... \;.
\ee
The usual presentation of the Gopakumar-Vafa prepotentials $F_g(t)$ is an infinite series, so with $k_{\mathrm{max}} \rightarrow \infty$. Using (\ref{kmax}) and (\ref{llpolescon}), we see that this means sending $\lambda \rightarrow 0$. 
Therefore, (\ref{sumkLam}) is an asymptotic series. We also see that corrections to the series behave as
\be 
{\cal O}\left(e^{-t \Lambda}\right) \sim {\cal O}\left(e^{-\frac{t}{\lambda}}\right) \;,
\ee  
which are of precisely the expected magnitude for non-perturbative effects, as in (\ref{tsfe}). 

For completeness, we present here the full Gopakumar-Vafa (asymptotic series) formula, which is just (\ref{sumkLam}) with $k_{\mathrm{max}}=\infty$ and the sum over $\beta$ and $r$ reinstated \cite{Gopakumar:1998ii,Gopakumar:1998jq}
\be 
F^{GV} = \sum_{r,\beta \geq 0,k \geq 1} \alpha_{r}^{\beta}\frac{e^{-k\beta_i t^i}}{k\left(2\sin\left(\frac{k\lambda}{2}\right)\right)^{2-2r}} \;.
\ee 
We also note that this does not yield the asymptotic formula for the free energy precisely because it does not include the polynomial, in $t$, terms in the free energy (which feature in $F_0$ and $F_1$ only). We discuss these missing terms in section \ref{sec:sum}. 

It is worth noting that historically, this is not how the results were first presented. Rather, the result of \cite{Gopakumar:1998ii,Gopakumar:1998jq} was presented directly as (\ref{anglesLambda}) with $\Lambda = \infty$. The analysis in \cite{Dedushenko:2014nya} is more closely related to the one in this section. It was understood that one needs to perform a rotation to the real line, and also that there are poles. However, it was stated that the Gopakumar-Vafa result is an asymptotic series, and therefore one can avoid the poles, as discussed in this section.

%%%%%%%%%%%%%%%%%%%%%%%%%%%%%%%%%%%%%%%%%%%%%%%%%%%%%%%%%%%%%%%%%%%%%%%%%%%%%%%%%%%%
\subsection{The full result}
\label{sec:GVasyfull}
%%%%%%%%%%%%%%%%%%%%%%%%%%%%%%%%%%%%%%%%%%%%%%%%%%%%%%%%%%%%%%%%%%%%%%%%%%%%%%%%%%%%

\begin{figure}[h]
\centering
 \includegraphics[width=0.8\textwidth]{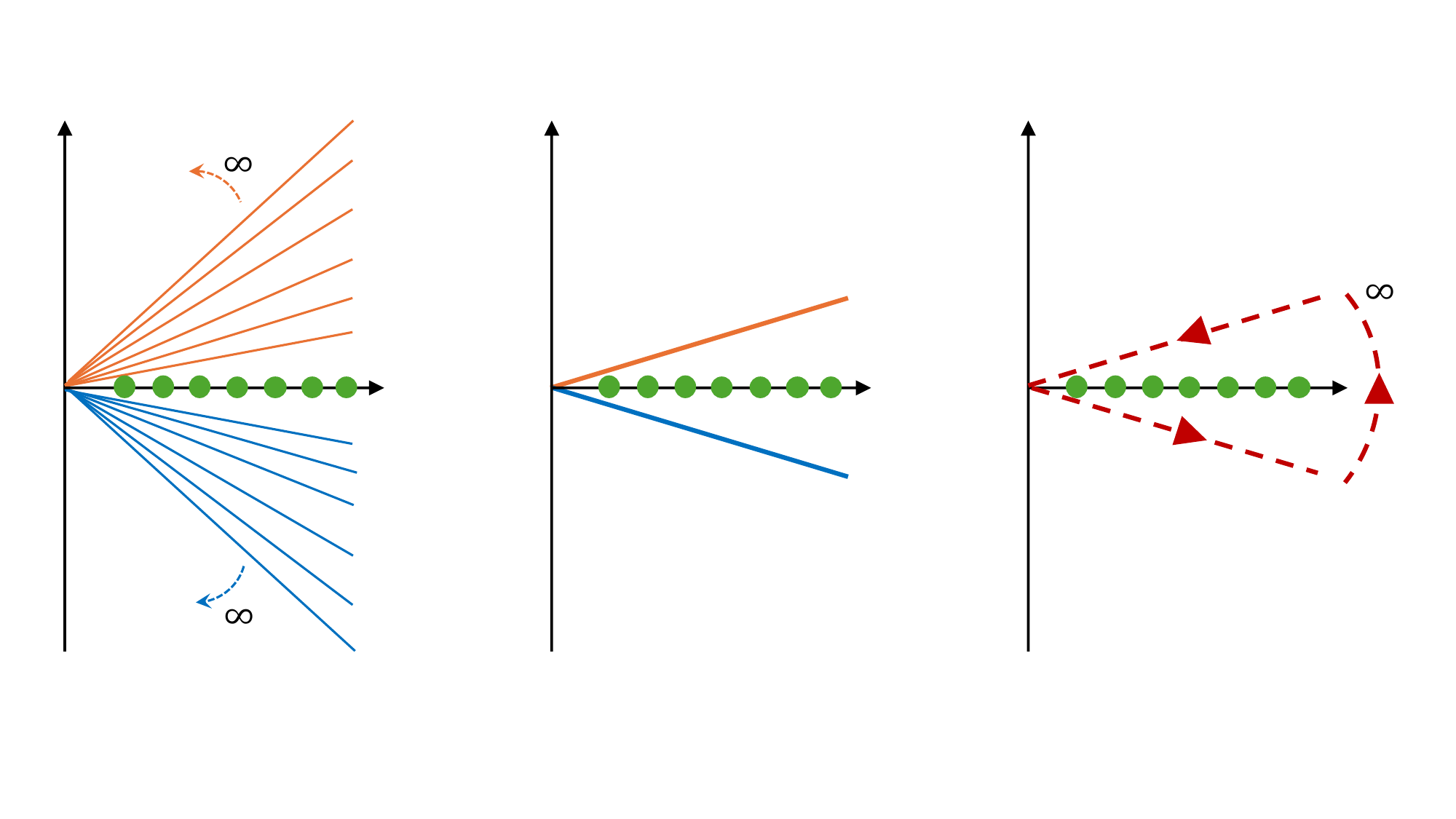}
\caption{Figure, taken from \cite{Hattab:2024ewk}, showing how it is possible to deform all the lines in the upper plane to coincide at some angle $\theta^+$, and the same for the lower plane with angle $\theta^-$. The integration path can then be closed off at infinity yielding a contour integral which picks up all the poles.}
\label{fig:lines}
\end{figure}

If we do not want the free energy only as an asymptotic series, then we need to consider the exact expression (\ref{angles}) for finite values of $\lambda$. To do this, we follow the analysis in \cite{Hattab:2024ewk}. In order to evaluate the sum over $n$, we would like again to make the integration lines coincide. However, now the lines stretch to infinity. This implies two significant differences from the cases studied in section \ref{sec:GVasy}. The first is that upon rotating the lines using contour integrals, as in (\ref{intpathslice}), the contribution from the arc now vanishes. This is simple to see because we can write the exponential factor in the integrand of (\ref{angles}) as
\be 
e^{-u z_n} = e^{- s |z_n|^2} \;,
\ee 
and since $|u|\rightarrow \infty$ implies $s \rightarrow \infty$, this vanishes. 
The second difference is that we can no longer rotate the integration lines to, or through, the real axis because we would hit the poles (\ref{nppoles}). So we need to split the integration lines into two sets, lying in the upper and lower half planes. We can bring all integrals with $2\pi n-b>0$ to coincide on the same line, at angle $\theta_+\in (0,\pi/2)$. Similarly, those with $2\pi n-b<0$ can be brought to coincide on another line with angle $\theta_-\in (0,-\pi/2)$. This is illustrated in figure \ref{fig:lines} (in the second picture).

After this rotation, we can write the sum over the integrals as
\begin{eqnarray}
    F^{rc} &=& \sum_{\{n\;:\;\theta_n>0\}}\int_{0^+}^{\infty e^{i\theta_+}}\frac{\text{d}u}{u}\frac{e^{-uz_n}}{\left(2\sin\left(\frac{u\lambda}{2}\right)\right)^2} \nonumber \\
    &+&\sum_{\{n\;:\;\theta_n<0\}}\int_{0^+}^{\infty e^{i\theta_-}}\frac{\text{d}u}{u}\frac{e^{-uz_n}}{\left(2\sin\left(\frac{u\lambda}{2}\right)\right)^2} \;.
    \label{frcspsum}
\end{eqnarray}
We can now evaluate the (partial) sums over $n$. It is convenient to split $b$ as
\be
b = 2\pi n_b + \tilde{b} \;,
\ee
where 
\be 
n_b = \floor{\frac{b}{2\pi}} \in \mathbb{Z} \;\;,\;\; \tilde{b} \in (0,2\pi) \;.
\ee 
Note that shifting $n_b$ by an integer is a gauge symmetry of the theory, and so the choice of $n_b$ is essentially a gauge choice (it fixes a principle branch).
Let us consider first the sum over the lines in the upper half plane, so with $\theta_n > 0$. We have
\bea 
\sum_{\{n\;:\;\theta_n>0\}} e^{-u z_n} &=& e^{-u\left(v + i \tilde{b}\right)} \left[\sum_{\{n\;:\;n-n_b>0\}} e^{-u 2\pi i \left( n_b - n\right)} \right]
\nonumber \\
&=& e^{-u\left(v + i \tilde{b}\right)} \left[\sum_{m=1}^{\infty} e^{2\pi i um} \right]
\nonumber \\
&=& e^{-\left(t-2\pi i n_{b} \right)u} \left[\sum_{m=1}^{\infty} \left(e^{2\pi i u}\right)^m \right]
\nonumber \\
&=& e^{-\left(t-2\pi i n_{b} \right)u}\frac{e^{2\pi i u}}{1-e^{2\pi i u}} 
\nonumber \\
&=& -\frac{e^{-\left(t-2\pi i n_{b} \right)u}}{1-e^{-2\pi i u}} \;.
\label{uphalfsum}
\eea 
Where we used the fact that the geometric sum is convergent because for $\theta_n > 0$ we have $\mathrm{Im\;}u>0$. Similarly, for the sum over the lower half plane we have 
\bea 
\sum_{\{n\;:\;\theta_n<0\}} e^{-u z_n} &=& e^{-u\left(v + i \tilde{b}\right)} \left[\sum_{\{n\;:\;n-n_b \leq 0\}} e^{-u 2\pi i \left( n_b - n\right)} \right]
\nonumber \\
&=& e^{-u\left(v + i \tilde{b}\right)} \left[\sum_{m=0}^{\infty} e^{-2\pi i um} \right]
\nonumber \\
&=& e^{-\left(t-2\pi i n_{b} \right)u} \left[\sum_{m=0}^{\infty} \left(e^{-2\pi i u}\right)^m \right]
\nonumber \\
&=& e^{-\left(t-2\pi i n_{b} \right)u}\frac{1}{1-e^{-2\pi i u}} 
\nonumber \\
&=& \frac{e^{-\left(t-2\pi i n_{b} \right)u}}{1-e^{-2\pi i u}} \;.
\label{lowhalfsum}
\eea 
Again, the sum is convergent because for $\theta_n < 0$ we have $\mathrm{Im\;}u < 0$.
Using (\ref{uphalfsum}) and (\ref{lowhalfsum}), the free energy (\ref{frcspsum}) can be written as
\begin{eqnarray}
     F^{rc} &=&-\int_{0^+}^{\infty e^{i\theta_+}}\frac{\text{d}u}{u}\frac{1}{1-e^{-2\pi i u}}\frac{e^{-\left(t-2\pi i n_{b} \right)u}}{\left(2\sin\left(\frac{u\lambda}{2}\right)\right)^2} \nonumber \\
    &+&\int_{0^+}^{\infty e^{i\theta_-}}\frac{\text{d}u}{u}\frac{1}{1-e^{-2\pi i u}}\frac{e^{-\left(t-2\pi i n_{b} \right)u}}{\left(2\sin\left(\frac{u\lambda}{2}\right)\right)^2} \;.
\end{eqnarray}
We can now combine these two integrals into a single closed contour integral 
\be
-\int_{0^+}^{\infty e^{i\theta_+}} + \int_{0^+}^{\infty e^{i\theta_-}} = \oint_C \;, 
\ee
where $C$ is a closed contour starting at $0^+$ and encircling all the poles on the real axis, in an anti-clockwise manner. Here, we have used the fact that the arc connecting the two integration lines does not contribute because the integrand vanishes at infinity. This is illustrated in figure \ref{fig:lines} (in the third picture).

We then arrive at the expression for the free energy of the resolved conifold
\begin{eqnarray}
\label{rcres}
F^{rc}=\oint_C\frac{\text{d}u}{u}\frac{1}{1-e^{-2\pi i u}}\frac{e^{-\left(t-2\pi i n_{b} \right)u}}{\left(2\sin\left(\frac{u\lambda}{2}\right)\right)^2} \;.
\end{eqnarray}
This is the contribution to the topological string free energy from integrating out a single M2 brane (over a genus zero surface). 

Given the expression (\ref{rcres}), the general formula for an arbitrary Calabi-Yau now follows directly by summing over the different M2 branes
\be
\label{full}
    F(t,\lambda) = \sum_{\beta,r\geq 0}\alpha_r^{\beta}\oint_C\frac{\text{d}u}{u}\frac{1}{1-e^{-2\pi i u}}\frac{e^{-\left(\beta\cdot t-2\pi i n_{\beta\cdot b} \right)u}}{\left(2\sin\left(\frac{u\lambda}{2}\right)\right)^{2-2r}} \;,
\ee
where we now have $n_{\beta \cdot b} = \floor{\frac{\beta \cdot b}{2\pi}} \in \mathbb{Z}$. We propose that (\ref{full}) gives the full non-perturbative topological string free energy (in the holomorphic limit). In some sense, as discussed in section \ref{sec:int}, this is the answer to the question: "What is topological string theory?"

\subsubsection*{The D0 contributions}

Note that from (\ref{full}) we can write the $\beta=0$ contribution, which we have ignored so far, as
\be
\label{fulld0}
    F(t,\lambda) = - \frac{\chi}{2} \oint_C\frac{\text{d}u}{u}\frac{1}{1-e^{-2\pi i u}}\frac{1}{\left(2\sin\left(\frac{u\lambda}{2}\right)\right)^{2-2r}} \;,
\ee
where $\chi$ is the Euler character of the Calabi-Yau manifold $\chi= 2\left(h^{(2,1)}-h^{(1,1)}\right)$. This is the full pure $D0$ contribution to the free energy.

\subsubsection*{The zero pole}

Finally, as mentioned already in section \ref{sec:GVasy}, the expressions presented for the free energy do not include the polynomial, in the $t^i$, contributions to the free energy. The absence of these contributions was attributed in \cite{Blumenhagen:2023tev,Hattab:2023moj,Hattab:2024thi,Hattab:2024ewk,Hattab:2024chf,Hattab:2024yol} to the ultraviolet cutoff of the Schwinger integral (\ref{uvepcut}). Indeed, for the resolved conifold expression (\ref{rcres}), we can remove this cutoff by simply extending the integration contour to include the pole at the origin. This pole contributes
\be 
\oint_{u=0}\frac{\text{d}u}{u}\frac{1}{1-e^{-2\pi i u}}\frac{e^{-tu}}{\left(2\sin\left(\frac{u\lambda}{2}\right)\right)^2} = \frac{1}{\lambda^2}\left(-\frac{1}{6}t^3 + \frac{\pi i}{2}t^2 + \frac{\pi^2}{3}t\right) + \left( -\frac{1}{12}t + \frac{\pi i}{12} \right)\;,
\label{zeropole}
\ee 
which are exactly the missing polynomial terms \cite{Gopakumar:1998ki,Hattab:2023moj,Hattab:2024ewk}. For the more general cases, the polynomial terms can still be associated to removing the ultraviolet cutoff, but in a more complicated way, as discussed in \cite{Hattab:2023moj,Hattab:2024thi,Hattab:2024ewk,Hattab:2024chf,Hattab:2024yol}.

\subsection{Evaluating the contour integral}

Let us study how the contour integral can be evaluated explicitly. Consider first the resolved conifold (\ref{rcres}). It is convenient to introduce 
\be 
\lambdabar = \frac{\lambda}{2\pi} \;,
\ee
in terms of which (\ref{rcres}) reads
\be
\label{rcreslbar}
F^{rc}=\oint_C\frac{\text{d}u}{u}\frac{1}{1-e^{-2\pi i u}}\frac{e^{-tu}}{\left(2\sin\left(\pi u \lambdabar\right)\right)^2} \;,
\ee
where we set $n_b=0$. 

The contour integral is of a meromorphic function, and therefore localizes onto the poles inside the contour. There are three different types of poles in (\ref{rcres}). The first set are the poles at positive integer values
\be 
u \in \mathbb{N}^* \;.
\label{ppoles}
\ee 
They arise from the factor 
\be
\frac{1}{1-e^{-2\pi i u}} \;,
\ee
which leads to first order poles. The contribution to the free energy from these poles is exactly the Gopakumar-Vafa asymptotic series form of the free energy
\be 
\sum_{k=1}^{\infty} \frac{e^{-k t}}{k\left(2\sin\left(\pi k \lambdabar\right)\right)^2} \;,
\ee 
where we have set $n_b=0$ for simplicity. We have therefore reproduced these results.

The second type of poles in (\ref{rcres}) arise at values of $u$ 
\be 
u \in \frac{1}{\lambdabar}\;\mathbb{N}^* \;.
\label{nppoles2t}
\ee
and come from the factor
\be 
\frac{1}{\left(2\sin\left(\pi u \lambdabar\right)\right)^2} \;.
\ee 
They are second order poles. They contribute to the free energy as (for $n_b=0$)
\be 
\sum_{k=1}^{\infty} c_k(t)\;e^{-\frac{ k t}{\lambdabar}} \;,
\ee 
with
\be
c_k(t) = -\frac{e^{\frac{\pi i k}{\lambdabar }}\left( k t+ \lambdabar \right)}{4\pi k^2 \lambdabar \;\sin \left( \frac{\pi k}{\lambdabar }\right) } - \frac{1}{4 k \lambdabar \;\left(\sin \left( \frac{\pi k}{\lambdabar }\right)\right)^2 } 
   \;\;\;.
   \label{ckedef}
\ee 

It is possible to write the sum of the contributions of the two types of poles as
\be 
\sum_{k=1}^{\infty} \left[\frac{e^{-k t}}{k\left(2\sin\left(\pi k \lambdabar\right)\right)^2} + c_k(t)\;e^{-\frac{ k t}{\lambdabar}} \right]\;.
\label{freesumk}
\ee 
This expression for the free energy takes the form (\ref{tsfe}). It is starting to look like a full function. In fact, the form (\ref{freesumk}) was already known for the resolved conifold case. It can be calculated by the NS limit \cite{Nekrasov:2009rc} of the refined topological string on the resolved conifold, by studying matrix model dualities, Borel resummation techniques or by solving difference equations \cite{Pasquetti:2010bps,Hatsuda:2013oxa,Krefl:2015vna,Hatsuda:2015owa,Alim:2021ukq}. However, our derivation of it from integrating out M2 branes is new. And, more importantly, our result generalizes simply to other cases, since it is just an exact version of the Gopakumar-Vafa integrating out approach, which holds for all Calabi-Yau manifolds.

Some of the coefficients in each of the two sums in (\ref{freesumk}) diverge if $\lambdabar$ is rational. This divergence actually corresponds to a certain rearrangement of the sums such that (\ref{freesumk}) remains finite. The way to understand this is to note that the divergence is associated to third order poles in (\ref{rcreslbar}). Such poles arise whenever 
\be
u \in \mathbb{N}^* \;\;\;\mathrm{and}\;\;\;  u \in \frac{1}{\lambdabar}\;\mathbb{N}^* \;.
\ee
To account for these, we first need to subtract from (\ref{freesumk}) these contributions, so we write (\ref{freesumk}) as
\bea 
\sum_{\left\{k \in \mathbb{N}^* \;:\; k \lambdabar \;\notin\; \mathbb{N}^*\right\}} \frac{e^{-k t}}{k\left(2\sin\left(\pi k \lambdabar\right)\right)^2} \;\;\;+\sum_{\left\{k \in \mathbb{N}^* \;:\;  \frac{k}{\lambdabar} \;\notin \;\mathbb{N}^* \right\}} c_k(t)\;e^{-\frac{ k t}{\lambdabar}}\;\; \;.
\label{freesumknorat}
\eea 
Then we need to add the contributions from the third order poles. These can be written as a sum over all pairs of integers $(p,q)$ under the constraint 
\be 
\lambdabar = \frac{p}{q} \;.
\ee 
For each choice of a pair $(p,q)$, we get a contribution $d_{(p,q)}(t) \; e^{-q \;t} $ with 
\be 
d_{(p,q)}(t) = \frac{1}{6\left(2\pi p\right)^2q} \left[2 \left(\pi ^2 p^2-\pi ^2 q^2-3 i \pi  q+3 \right) + 6\;  \left(1-i \pi  q \right) \;q\;t + 3\left(q\;t \right)^2 \right] \;.
\label{dpqdef}
\ee 
Therefore, the contour integral (\ref{rcreslbar}) can be written as
\bea
F^{rc} &=& \sum_{\left\{k \in \mathbb{N}^* \;:\; k \lambdabar \;\notin\; \mathbb{N}^* \right\}} \frac{e^{-k t}}{k\left(2\sin\left(\pi k \lambdabar\right)\right)^2} \nonumber \\
&+& \sum_{\left\{k \in \mathbb{N}^* \;:\; \frac{k}{\lambdabar} \;\notin \;\mathbb{N}^* \right\}} c_k(t)\;e^{-\frac{ k t}{\lambdabar}}\;\; \nonumber \\
&+& \sum_{\left\{(p,q) \in \mathbb{N}^* \;:\; \lambdabar = \frac{p}{q} \right\}} d_{(p,q)}(t) \; e^{-q \;t}\;,
\eea 
with $c_k(t)$ and $d_{(p,q)}(t)$ defined in (\ref{ckedef}) and (\ref{dpqdef}) respectively. This defines a function that is perfectly well-behaved for any value of $\lambdabar$.\footnote{In fact, this procedure is equivalent to taking (\ref{freesumk}) and performing a careful analysis of the limit approaching rational values of $\lambdabar$. Although this involves subtleties to do with the rationals being dense in the reals.} 

This form of the function for the resolved conifold was also found using resurgence techniques in \cite{Hatsuda:2015owa}.\footnote{Although, the parameter $n_b$, which we have set to zero for simplicity here, is not present in that formulation.} The finite value at rational values of $\lambdabar$ is a property which is also reproduced using techniques which build on geometric transitions duality, and its generalisation to the Fermi Gas of TS/ST correspondence \cite{Drukker:2010nc,Marino:2011eh,Hatsuda:2013oxa,Grassi:2014zfa}. 

It is then simple to write a similar expression for the full general result (\ref{full}), which reads
\bea 
F(t,\lambda) &=& \sum_{\beta\geq 0,\;r\geq 1,\; k \in \mathbb{N}^*\;}  \alpha_r^{\beta}\frac{e^{-k \;\beta\cdot t}}{k\left(2\sin\left(\pi k \lambdabar\right)\right)^{2-2r}}  \nonumber \\ \nonumber
&+& \sum_{\beta \geq 0} \;\alpha_0^{\beta}\;\left[ 
 \sum_{\left\{k \in \mathbb{N}^* \;:\; k \lambdabar \;\notin \; \mathbb{N}^* \right\} }\;\frac{e^{-k \;\beta\cdot t}}{k\left(2\sin\left(\pi k \lambdabar\right)\right)^{2}} \right.
\nonumber \\
&+&\;\;\sum_{\left\{k \in \mathbb{N}^* \;:\; \frac{k}{\lambdabar} \;\notin \;\mathbb{N}^* \right\}} c_k\Big(\beta\cdot t-2\pi i n_{\beta\cdot b}\Big)\;e^{-\frac{ k \left(\beta\cdot t-2\pi i n_{\beta\cdot b} \right)}{\lambdabar}}\; \nonumber \\
&+& \left.\; \sum_{\left\{(p,q) \in \mathbb{N}^* \;:\; \lambdabar = \frac{p}{q} \right\}} d_{(p,q)}\Big(\beta\cdot t-2\pi i n_{\beta\cdot b}\Big) \; e^{-q\; \beta\cdot t} \;\;\right]\;.
\label{genrssumtg}
\eea 
We have reinstated $n_{\beta\cdot b}$ , because when summing over $\beta$ to infinity it only vanishes for $b=0$.

%%%%%%%%%%%%%%%%%%%%%%%%%%%%%%%%%%%%%%%%%%%%%%%%%%%%%%%%%%%%%%%%%%%%%%%%%%%%%%%%%%%%
\section{Summary and Discussion}
\label{sec:sum}
%%%%%%%%%%%%%%%%%%%%%%%%%%%%%%%%%%%%%%%%%%%%%%%%%%%%%%%%%%%%%%%%%%%%%%%%%%%%%%%%%%%%

In these notes, we presented a detailed discussion of integrating out M2 branes on Calabi-Yau manifolds. The calculation is a version of the idea of Gopakumar and Vafa \cite{Gopakumar:1998jq}, and follows closely the paper of Dedushenko and Witten \cite{Dedushenko:2014nya}. There are, however, a number of important new aspects, which were originally reported in \cite{Hattab:2023moj,Hattab:2024thi,Hattab:2024ewk,Hattab:2024chf,Hattab:2024yol}. 

The calculation is framed in the context of topological strings, by identifying the effective action arising from integrating out M2 branes with the topological string free energy. The final result is an expression for this free energy (\ref{full}). The crucial new aspect is that this expression includes non-perturbative topological string contributions to the free energy, which were absent from the original Gopakumar-Vafa formula. As discussed in the introduction, these are analogues of the Schwinger effect in QED. 

It was proposed in \cite{Hattab:2024ewk} that (\ref{full}) not only includes non-perturbative contributions, but is the full answer for the (holomorphic) free energy. In the context of the discussion in the introduction, it can therefore be taken as a non-perturbative definition of topological string theory on Calabi-Yau manifolds. However, as discussed in section \ref{sec:highersp}, the integrating out calculation assumes a restriction on the geometry of the cycle: that it is rigid. Therefore, while the integrating out is exact in terms of the topological string coupling, it does not yet apply to the most general Calabi-Yau geometry. 

There are many directions and open questions that follow from the calculation described in these notes. We will not go into details here, but give a list with small details of some of the most obvious directions:
\begin{itemize}
	\item The most direct open question is whether (\ref{full}) is the full non-perturbative topological string free energy. Certainly it captures non-perturbative physics, but could there be other non-perturbative contributions? There are two aspects to this question. The first is whether integrating out M2 branes yields the full free energy. The second is whether the integrating out calculation performed in these notes, and in \cite{Hattab:2024ewk}, is the full M2 integrating out calculation. We expect that the answer to the first part is yes, that integrating out M2 branes should yield the full answer. The calculations and discussions in these notes are evidence for this. While the answer to the second part is no, this is not the most general integrating out calculation. We have restricted to rigid and smooth cycles only, and have neglected issues such as degenerations in moduli space and possible interactions between M2 branes. This perspective also matches results from the geometric dualities and resurgence approaches. From those approaches one expects further contributions. For example, in the dual to ABJM theory, described using the Fermi Gas method \cite{Drukker:2010nc,Marino:2011eh}, the free energy includes additional terms not present in (\ref{full}). Additional results in resurgence, for example in \cite{Pasquetti:2010bps,Hatsuda:2013oxa,Couso-Santamaria:2013kmu,Couso-Santamaria:2014iia,Gu:2023mgf,Alexandrov:2023wdj,Alim:2024dyi}, suggest that the exponents of the exponentials in the free energy should correspond to various wrapped D-branes. This also connects to the non-holomorphic part of the free energy, controlled by the holomorphic anomaly equation, which captures certain non-local non-Wilsonian parts of the effective action from the integrating out perspective \cite{Antoniadis:1993ze,Bershadsky:1993cx}. It is important to understand these effects, and to try and match the results in the literature which solve the holomorphic anomaly equation. We expect that a full understanding of the integrating out calculation, for general cycle geometries, should reproduce (at least some of) these additional terms.  The calculation performed in these notes therefore serves as a starting base for more involved integrating out calculations.
	\item The form of the free energy (\ref{genrssumtg}) is still as an infinite sum. It is not clear if this sum converges for any value of $t$. It would be good to understand this, perhaps by studying examples. One technical aspect of this question is whether formulating the answer as a sum over BPS states misses some physics, in particular could there be additional poles if we move the sum inside the integral: $\sum \alpha_{r}^{\beta} \oint_C = \oint_C \sum \alpha_{r}^{\beta}$ ?
	\item Our analysis took the topological string coupling $\lambda$ as real. It is possible to consider what happens when $\lambda$ is complex. In that case we expect to capture some of the physics associated to the spectrum of BPS states, such as Wall Crossing and Stokes jumps (see \cite{Hattab:2024ewk} for more details). However, complex $\lambda$ implies a complex spacetime background in the solution (\ref{5dmetim}), and it is not clear what the implications of this are. It is natural to expect that this would affect the integrating out calculation, and therefore the final result. This is certainly an interesting avenue to explore. 
	\item The claim is that (\ref{full}) is exact, and therefore we should be able to evaluate it at strong coupling. In particular, the values $\lambda \rightarrow \infty$ and $\lambda = 2\pi$ are especially interesting.\footnote{Note that we expect that both of these limits are captured by the rigid and smooth cycle calculations performed here.} The former manifests a certain weak-strong coupling duality \cite{Hattab:2024yol}, and the latter is the associated self-dual point. At the self-dual point the free energy takes a very simple form
	\begin{eqnarray}
\nonumber    F(t,2\pi)|_{\text{Exp}} &=& \sum_{\beta}\frac{\alpha_0^{\beta}}{(2\pi)^2}\Big[\text{Li}_3(e^{-\beta\cdot t})+(\beta\cdot t -2\pi i n_{\beta\cdot b}-i\pi)\text{Li}_2(e^{-\beta\cdot t})\\
&+&(\beta\cdot t-2\pi i n_{\beta\cdot b})\left(\frac{\beta\cdot t-2\pi i n_{\beta\cdot b}}{2}-i\pi\right)\text{Li}_1(e^{-\beta\cdot t})\Big] \nonumber \\
&+&\sum_{\beta}\alpha_{1}^{\beta}\text{Li}_1(e^{-\beta\cdot t}) \;,
\end{eqnarray}
where the subscript Exp refers to only the exponential (in $t$) parts.
	In particular, all the higher genus contributions vanish. It would be good to understand the physics of this in more detail. This is a generalization of the maximally supersymmetric points found in the Fermi Gas or TS/ST correspondence setting \cite{Drukker:2010nc,Marino:2011eh,Hatsuda:2013oxa,Grassi:2014zfa}.
	\item The topological string free energy is proposed to count physical black hole microstates. This is the Ooguri-Strominger-Vafa (OSV) conjecture \cite{Ooguri:2004zv}. So far, for strict $SU(3)$ holonomy Calabi-Yau manifolds, fully non-perturbative expressions for the free energy were known only for non-compact cases, for which the four-dimensional Planck scale is infinite and there are no black holes. The expression (\ref{full}) is therefore the first proposal for the exact black hole entropy in Calabi-Yau compactifications. It would be very interesting to test and understand this better. 
	\item The Ooguri-Vafa formula \cite{Ooguri:1999bv} is an open string extension of the Gopakumar-Vafa formula. It also can be derived from integrating out M2 branes, but now ending on M5 branes. One therefore expects that it similarly has a non-perturbative extension from performing the integrating out calculation, as proposed in \cite{Hattab:2024chf}. It would be good to understand the physics of this, and the integrating out procedure in more detail. 
	\item As (\ref{zeropole}) shows, the polynomial terms in the free energy can be associated to the ultraviolet, the zero pole, of the integrating out calculation. The polynomial terms are the leading classical terms in the effective supergravity action. Therefore, the classical dynamics of the fields can be understood as emergent from integrating out the M2 branes. This was actually the original motivation for much of this work, starting from \cite{Hattab:2023moj}. It forms part of the Emergence Proposal \cite{Palti:2019pca} (based on ideas in \cite{Harlow:2015lma,Heidenreich:2017sim,Grimm:2018ohb,Heidenreich:2018kpg,Palti:2019pca}). The relation to emergence, including to open strings and emergent potentials, was developed further in this context in \cite{Hattab:2023moj,Hattab:2024thi,Hattab:2024ewk,Hattab:2024chf,Hattab:2024yol}.\footnote{See also \cite{Blumenhagen:2023tev,Blumenhagen:2023xmk,Blumenhagen:2024ydy,Blumenhagen:2024lmo} for a closely related approach to emergence.} This is certainly an important and fascinating direction of study, and may lead to a microscopic understanding behind some of the central Swampland conjectures, such as the Weak Gravity Conjecture \cite{Arkani-Hamed:2006emk}, the Distance Conjecture \cite{Ooguri:2006in}, the Refined de-Sitter Conjecture \cite{Ooguri:2018wrx}, and the AdS Distance Conjecture \cite{Lust:2019zwm}. 
\end{itemize}

%\vspace{10px}
\vspace{0.1cm}
{\bf Acknowledgements}
%\vspace{10px}
\noindent
These lecture notes are based on a series of lectures given as part of the Physics Latam seminar series. The work of JH and EP is supported by the German Research Foundation through a German-Israeli Project Cooperation (DIP) grant ``Holography and the Swampland". The work of EP is supported by the Israel planning and budgeting committee grant for supporting theoretical high energy physics.

%%%%%%%%%%%%%%%%%%%%%%%%%%%%%%%%%%%%%%%%%%%%%%%%%%%%%%%%%%%%%%%%%%%%%%%%%%%%%%%%%%%%
%%%%%%%%%%%%%%%%%%%%%%%%%%%%%%%%%%%%%%%%%%%%%%%%%%%%%%%%%%%%%%%%%%%%%%%%%%%%%%%%%%%%
\appendix
%%%%%%%%%%%%%%%%%%%%%%%%%%%%%%%%%%%%%%%%%%%%%%%%%%%%%%%%%%%%%%%%%%%%%%%%%%%%%%%%%%%%
%%%%%%%%%%%%%%%%%%%%%%%%%%%%%%%%%%%%%%%%%%%%%%%%%%%%%%%%%%%%%%%%%%%%%%%%%%%%%%%%%%%%

%%%%%%%%%%%%%%%%%%%%%%%%%%%%%%%%%%%%%%%%%%%%%%%%%%%%%%%%%%%%%%%%%%%%%%%%%%%%%%%%%%%%
\section{The Hamiltonian for higher genus contributions}
\label{sec:fertra}
%%%%%%%%%%%%%%%%%%%%%%%%%%%%%%%%%%%%%%%%%%%%%%%%%%%%%%%%%%%%%%%%%%%%%%%%%%%%%%%%%%%%

In this appendix we present the details of how tracing over the additional fermionic zero modes, associated to higher genus surfaces, yields the expression (\ref{trhyrsp}).
Half of the extra fermionic zero modes $\rho_{A\sigma}$ come from (1,0)-forms on $\mathcal{C}_2$ with $A = 1,2$ and $\sigma = 1,\cdots,r$ and the other $\tilde{\rho}_{A\sigma}$ from (0,1)-forms on $\mathcal{C}_2$. The minimal action is given by
\be
S_{r}=\int\text{d}t \sum_{\sigma = 1}^{r}\left(i\tilde{\rho}_{A\sigma}\frac{d}{dt}\rho^{A}_{\sigma}+\frac{i}{8}\bar{Z}(q)W_{\mu\nu}^-\gamma^{\mu\nu}_{AB}\tilde{\rho}_{\sigma}^A\rho_{\sigma}^B\right)\;.
\ee
where the exact normalisation of the magnetic moment coupling can be determined by expanding the M2 action wrapping $\mathcal{C}_2$ as in \cite{Dedushenko:2014nya}.
The Hamiltonian is identified as
\be 
\tilde{H}= -\sum_{\sigma =1}^{r}\frac{i}{8}\bar{Z}(q)W_{\mu\nu}^-\gamma^{\mu\nu}_{AB}\tilde{\rho}_{\sigma}^A\rho_{\sigma}^B \;.
\label{tildHdef}
\ee 
Quantization of the Hamiltonian indeed gives $r$ copies of the hypermultiplet spin content where $\rho_{A\sigma}$ and $\tilde{\rho}_{A\sigma}$ naturally act on the finite dimensional Hilbert space $\mathcal{H}_r = \otimes_{\sigma=1}^r\mathcal{H}_{\sigma}$ where $\mathcal{H}_{\sigma}$ is spanned by $|\sigma \rangle$, $\tilde{\rho}_{1\sigma}|\sigma\rangle$, $\tilde{\rho}_{2\sigma}|\sigma\rangle$, $\tilde{\rho}_{1\sigma}\tilde{\rho}_{2\sigma}|\sigma\rangle$ with $\rho_{A\sigma}|\sigma\rangle = 0$.

We want to perform the trace for the Hamiltonian (\ref{tildHdef}). Since we have $r$ copies of the same system we can focus on
\be
H = -\frac{i}{8}\bar{Z}(q)W_{\mu\nu}^-\gamma^{\mu\nu}_{AB}\Tilde{\rho}^{A}\rho^{B} \;.
\ee
The curved $\gamma^{\mu}$ gamma matrices are given in terms of the flat ones $\gamma^i$ by 
\be 
\gamma^{\mu} = e^{-\sigma/2}\delta^{\mu}_i\gamma^i \;.
\ee
We recall that capital Latin letters $A, B, C, D$ refer to negative chirality indices. We raise or lower indices as $\rho_A = \rho^B\epsilon_{BA}$ and $\rho^A = \epsilon^{AB}\rho_B$ with $\epsilon_{12} = \epsilon^{12} = 1$, so in particular 
\be 
\gamma^{\mu\nu}_{AB} = \frac{1}{2}[\gamma^{\mu},\gamma^{\nu}]_A^{\;\;C}\epsilon_{CB} \;.
\ee 
We take the following representation for the flat gamma matrices
\be
\gamma^i = \begin{pmatrix}
0 & \sigma^i\\
\sigma^i & 0
\end{pmatrix}\;\;\;\text{and}\;\;\;  \gamma^{4} = \begin{pmatrix}
0 & -i\mathbb{1}\\
i\mathbb{1} & 0
\end{pmatrix}\,,
\ee
where $\sigma^i$ with $i = 1,2,3$ are the Pauli matrices. This gives
\be
\frac12[\gamma^i,\gamma^4] = \begin{pmatrix}
i\sigma^i & 0\\
0 & -i\sigma^i
\end{pmatrix}\;\;\;\text{and}\;\;\;  \frac12[\gamma^i,\gamma^j] = \begin{pmatrix}
i\epsilon^{ijk}\sigma^k & 0\\
0 & i\epsilon^{ijk}\sigma^k
\end{pmatrix}\,.
\ee
Contracting the anti-self dual field strength $W_{12}^- = -W_{34}^- $ yields
\be
W_{\mu\nu}^-\gamma^{\mu\nu} = 2W_{12}^-\gamma^{12}+ 2W_{34}^-\gamma^{34}= 4iW_{12}^-e^{-\sigma}\begin{pmatrix}
0 & 0\\
0 & \sigma^3
\end{pmatrix}\;.
\ee
The chirality matrix is taken as
\be
\gamma^5 = \gamma^4 \gamma^1 \gamma^2 \gamma^3 = \begin{pmatrix}
\mathbb{1} & 0\\
0 & -\mathbb{1}
\end{pmatrix}\;.
\ee
%{\bf JH: This seems to be in contradiction with the fact that we need to take negative chirality spinors since this gives $W_{\mu\nu}^-(\gamma^{\mu\nu})_A^{\;\;B} = 0$ and $W_{\mu\nu}^-(\gamma^{\mu\nu})_{\dot{A}}^{\;\;\dot{B}} = 4iW_{12}^-e^{-\sigma}(\sigma_3)_{\dot{A}}^{\;\;\dot{B}}$}\\
%It is possible that $\mu = 1$ in $W_{12}$ should be identified with the time direction and so 2,3,4 with 1,2,3) such that: 
%\be
%W_{\mu\nu}^-\gamma^{\mu\nu} = 2W_{12}^-\gamma^{41}+ 2W_{34}^-\gamma^{23}= 4iW_{12}^-e^{-\sigma}\begin{pmatrix}
%-\sigma^1 & 0\\
%0 & 0
%\end{pmatrix}\;.
%\ee
Finally (using $\epsilon = i\sigma^2$)
\be
W_{\mu\nu}^-\gamma^{\mu\nu}_{AB} = 4iW_{12}^-e^{-\sigma}(\sigma^{3})_A^{\;\;C}\epsilon_{CB} = -4W_{12}^-e^{-\sigma}(\sigma^{3}\sigma^2)_{AB} = 4iW_{12}^-e^{-\sigma}(\sigma^1)_{AB}\;,
\ee
and 
\be
H = -\frac{i}{8}\bar{Z}(q)W_{\mu\nu}^-\gamma^{\mu\nu}_{AB}\Tilde{\rho}^A\rho^B =-\frac{1}{2}\bar{Z}(q)W_{12}^-e^{-\sigma}\left(\Tilde{\rho}_1\rho_2+\Tilde{\rho}_2\rho_1\right)\;.
\ee
We now have to compute the trace. The basis of states is $|0\rangle$, $\Tilde{\rho}_1|0\rangle$, $\Tilde{\rho}_2|0\rangle$ and $\Tilde{\rho}_1\Tilde{\rho}_2|0\rangle$ with $\rho_A|0\rangle = 0$ and the spinor coordinates satisfy the anti-commutation relations
\be
\{\Tilde{\rho}_A,\rho_B\} = \epsilon_{AB} \;\;,\;\; \{\rho_A,\rho_B\} = \{\Tilde{\rho}_A,\Tilde{\rho}_B\} = 0\;.
\ee
This means that we get the eigenstates
\begin{eqnarray}
    H|0\rangle &=& 0 \;,\\
    H \Tilde{\rho}_1  |0\rangle &=& -\frac{1}{2}\bar{Z}(q)W_{12}^-e^{-\sigma} \Tilde{\rho}_1 |0\rangle \;,\\
    H \Tilde{\rho}_2|0\rangle &=& \frac{1}{2}\bar{Z}(q)W_{12}^-e^{-\sigma} \Tilde{\rho}_2|0\rangle \;,\\
    H\Tilde{\rho}_1\Tilde{\rho}_2|0\rangle &=& 0\;.
\end{eqnarray}
So the trace is then computed as 
\begin{eqnarray}
    \text{Tr}((-1)^Fe^{-sH}) = 2-e^{\frac{s}{2}\bar{Z}(q)W_{12}^-e^{-\sigma}}-e^{-\frac{s}{2}\bar{Z}(q)W_{12}^-e^{-\sigma}} = \left(2\sin\left(\frac{s\bar{Z}(q)\sqrt{-W^2}}{8}\right)\right)^2 \;,
\end{eqnarray}
which matches (\ref{trhyrsp}).

\bibliographystyle{jhep}
\bibliography{Higuchi}

\end{document}